\documentclass[12pt]{article}
\usepackage[margin=1in]{geometry}
\usepackage{setspace, mathtools, esvect, authblk, cite, amssymb, IEEEtrantools, slashed, graphicx, subcaption}
\usepackage[colorlinks=True, linkcolor=black, urlcolor=blue, citecolor=blue]{hyperref}
\numberwithin{equation}{section}

\graphicspath{ {./Images/} {./tikz_out/} }
\renewcommand{\bar}{\overline}
\renewcommand{\vec}[1]{\vv{#1}}
\newcommand{\ket}[1]{\lvert #1 \rangle}
\newcommand{\bra}[1]{\langle #1 \rvert}
\newcommand{\braket}[2]{\langle #1 \lvert #2 \rangle}
\newcommand{\ketbra}[2]{\lvert #1 \rangle \langle #2 \rvert}
\newcommand{\expect}[1]{\langle #1 \rangle}
\DeclareMathOperator{\trace}{tr}
\DeclareMathOperator{\sgn}{sgn}

\title{Lattice Quantum Chromodynamics \\ for Quantum Simulations}
\author{Luis Hidalgo}
\author{Patrick Draper}
\affil{Department of Physics, University of Illinois Urbana-Champaign}
\date{}

\begin{document}

\maketitle

\begin{abstract}
	We develop a framework for quantum  simulations of lattice SU($N_c$) gauge theory with quarks. Staggered and Wilson lattice fermions are considered in two and three spatial dimensions and a theta angle is included in three dimensions. The physical, gauge-invariant Hilbert space is formulated in a representation basis, where gauge and fermionic degrees of freedom are  encoded by irreducible representations of SU($N_c$) that tensor at each lattice site to contain a singlet. We discuss algorithms for simulating time evolution on quantum computers and carry out noiseless simulations of lattice quantum chromodynamics ($N_c=3$) on small lattices with up to 32 qubits, showcasing theta angle effects, hadronic states, string dynamics, and a baryon chemical potential for the first time in three spatial dimensions.
\end{abstract}

\tableofcontents

\section{Introduction}

Quantum chromodynamics (QCD) is the microscopic theory of the strong interaction that controls the structure of hadrons such as protons and neutrons \cite{EllisStirlingWebber:1996:qcd_collider_physics, CampbellHustonKrauss:2017:black_book_qcd}. Perturbative QCD describes physics in particle collisions with large momentum transfer. Nonperturbative and computational methods are necessary to compare the theory with experiments that, for example, probe the internal structure of hadrons. A common way to study QCD with a computer is to discretize space, possibly time, and truncate the total volume. This forms a cutoff version of the theory called lattice QCD \cite{DeGrandDetar:2006:lattice_methods}.

The theory of lattice QCD continues to be essential for examining the Standard Model of particle physics \cite{FLAG:2026:flag_2024, MUON:2025:anomalous_magnetic_moment}. The common way to use lattice QCD is to discretize spacetime and calculate correlation functions via importance sampling of an imaginary time path integral. By lowering the lattice spacing, and following a renormalization procedure, an extrapolation to the continuum theory can be made. However, this procedure generally fails when a sign problem occurs, for example, in real time simulations or the presence of a theta angle \cite{TroyerWiese:2005:sign_problem, AartsSexty:2026:sign_problem_LFT, BonannoBonatiDElia:2026:strong_cp_problem}.

An alternative way to use lattice QCD is with its Hamiltonian formulation \cite{KogutSusskind:1975:KS_hamiltonian, ZoharBurrello:2015:LGT4QS}. An initial quantum state can be prepared, evolved over time with the time evolution operator, and then sampled. However, the dimension of the Hilbert space grows exponentially with the volume of the lattice. For this reason it is likely that only quantum computers will be able to simulate large-scale lattice QCD processes in the Hamiltonian formulation.

An effort to develop use cases for quantum computers in high energy physics has recently grown \cite{BauerEtal:2023:quantsim4HEP, BauerEtal:2023:quantum_simulating, DiMeglioEtal:2024:QC4HEP, HalimehEtal:2025:out_of_equilibrium, RobinSavage:2026:quantum_complexity}. At this point, several frameworks for Hamiltonian simulations of lattice QCD have been established. Some fundamentally differ by which basis is mapped to the computational basis. In the representation basis, recent proposals formulate the physical Hilbert space by manually constructing gauge-invariant singlets at each site \cite{BalajiEtal:2025:qcirc_for_su3}, including in the large-$N_c$ limit \cite{ModiEtal:2026:large_Nc}, or by application of native gauge-invariant operators \cite{KadamEtal:2025:LSH_part_1, KadamEtal:2025:LSH_part_2}. In the group element basis, the gauge degrees of freedom are the elements of discrete SU($N_c$) subgroups \cite{MarianiPradhanErcolessi:2023:finite_LGT, PerezEtal:2025:primitive_gates_S216}. A mixed basis has also been formulated on top of a full gauge fixing \cite{FrolandGrabowskaLi:2025:fully_gauge_fixed_su2}. Other frameworks differ by which Hamiltonian, such as the orbifold lattice Hamiltonian \cite{BergnerHanadaMendicelli:2026:orbifold}, is used. Ultimately each framework may be useful to answer different questions about QCD, and the continual development of quantum computing hardware and algorithms will guide the applications.

In  this work we use an elaboration of the representation basis. Building on previous works \cite{ByrnesYamamoto:2006:sim_LGT_on_QC, CiavarellaKlcoSavage:2021:trailhead, BalajiEtal:2025:qcirc_for_su3}, the physical Hilbert space is constructed with irreducible representations of SU($N_c$) for gauge and fermionic degrees of freedom, along with a label for the different singlets found at each lattice site. Staggered and Wilson fermions -- being among the simplest lattice fermions to implement -- are used, and we include a theta angle in three spatial dimensions. The purpose of this paper is to present ingredients amenable to designing  quantum simulations of lattice QCD, in this Hamiltonian formulation, in both the near and long terms. Primary theoretical products include formulae for efficiently numerically computable matrix elements of the Hamiltonian, and a proof-of-concept Trotterization algorithm is used for small-scale noiseless quantum simulations.

The paper is organized as follows. Sec.~\ref{sec:continuum_theory} provides a brief review of Hamiltonian SU($N_c$) gauge theory with fermions, and establishes some of the notation used throughout the paper. Sec.~\ref{sec:lattice_theory} presents a Hamiltonian for the lattice gauge theory with both staggered and Wilson fermions. It also describes the Hilbert spaces associated with each link and site of the lattice and formulates the physical Hilbert space of the whole lattice in the basis described above. Sec.~\ref{sec:quantum_simulation} develops Trotterized time evolution quantum circuits. The circuits are used to perform small-scale noiseless simulations of time-dependent expectation values as well as to probe string and meson dynamics. Although our focus is on quantum simulation, the infrastructure can also be used for  exact diagonalization treatments, which we illustrate in Sec.~\ref{sec:exact_diagonalization} by computing the baryon chemical potential on a cube. Sec.~\ref{sec:conclusion} concludes by discussing some directions for future research.

Five appendices supplement the main text with background information and calculations. App.~\ref{app:continuum_hamiltonian} reviews the canonical quantization of SU($N_c$) gauge theory. App.~\ref{app:dirac_fermions} reviews the staggered and Wilson formulations of lattice fermions and the implementations used in the main work. App.~\ref{app:aspects_of_SUNc_LGT} reviews tools used in the formalism, including SU($N_c$) Clebsch-Gordan coefficients, representation matrices, and identities among Hilbert space operators. It also contains an explicit construction of the physical Hilbert space. App.~\ref{app:hamiltonian_matrix_elements} calculates matrix elements of the various operators found in the Hamiltonian and tabulates matrix element counts for different Hilbert space truncations. App.~\ref{app:the_theta_angle} describes an alternative implementation of the theta term in the Hamiltonian and results are shown for quantum simulations in this case.

\section{Continuum Theory}
\label{sec:continuum_theory}

Let $\mathcal{R}$ be an irreducible representation (irrep) of SU($N_c$) for $N_c \geq 2$. In this paper $\mathcal{R}$ will denote the fundamental representation, with $\dim(\mathcal{R}) = N_c$, unless otherwise specified. Among the irreps of SU($N_c$), the fundamental representation will play a variety of privileged roles, including being the representation of the spin-$\frac{1}{2}$ fermion fields and the representation associated with the link operator in the construction of the lattice Hamiltonian.

The basis elements in a general irrep $R$ of the algebra $\mathfrak{su}(N_c)$ are denoted by $T^{Ra}$. When such representation-dependent quantities are written without an irrep label as in $T^a$, the theory's representation $\mathcal{R}$ is assumed, unless otherwise specified. Here, $a=1,\dots,N_c^2-1$ is the gauge index that runs over the dimension of the algebra. Because $\mathfrak{su}(N_c)$ is a vector space, $[T^{Ra},T^{Rb}] = if^{abc} T^{Rc}$ can be written as a linear combination of the basis matrices, where the structure constants $f^{abc}$ form a totally antisymmetric tensor. Furthermore, the basis elements are chosen to be orthogonal such that $\trace[T^a T^b] = I_\mathcal{R} \delta_{ab}$, where $I_\mathcal{R}$ is the Dynkin index of $\mathcal{R}$. It is conventional to choose $I_\mathcal{R}=\frac{1}{2}$ for the fundamental representation, but this substitution will not be made for most of the paper.

Let d be the number of spatial dimensions and let $\text{D}=\text{d}+1$ be the number of spacetime dimensions. Notation such as 3d is used interchangeably with $\text{d}=3$. The focus of this paper is 3d, but sometimes 2d will be considered as well. A suitable Lagrangian density for the theory in $3+1\text{D}$ is
\begin{equation}
	\mathcal{L} = -\frac{1}{4g^2} F_{\mu\nu}^a F^{\mu\nu,a} + \sum_{f=1}^{N_f} \bar{\psi}^{(f)} (i\gamma^\mu \partial_\mu + \gamma^\mu A_\mu^a T^a - m_f) \psi^{(f)} + \frac{\theta}{64\pi^2} \epsilon^{\mu\nu\rho\sigma} F_{\mu\nu}^a F_{\rho\sigma}^a
\end{equation}
Here, $g$ is the coupling. $A_\mu^a(x)$ is the gauge field and $F_{\mu\nu}^a(x)$ is the field strength tensor. $\psi^{(f)}(x)$ is a collection of Dirac fields, and has components $\psi_{\alpha,c}^{(f)}(x)$, where $f=1,\dots,N_f$ is the flavor index, $\alpha=1,\dots,2^{\lfloor \frac{\text{D}}{2} \rfloor}$ is the Dirac/spinor index, and $c=1,\dots,N_c$ is the color index. Each flavor can have a different mass $m_f$.

As reviewed in App.~\ref{app:continuum_hamiltonian}, the Hamiltonian can be derived via canonical quantization:
\begin{IEEEeqnarray*}{rCl}
	H = \int d^3x \bigg[ & - & \sum_{f=1}^{N_f} \bar{\psi}^{(f)} (i\gamma^i \partial_i + \gamma^i A_i^a T^a - m_f) \psi^{(f)} - A_0^a \bigg( D_i \Pi^{i,a} + \sum_{f=1}^{N_f} \psi^{\dag(f)} T^a \psi^{(f)} \bigg) \\
	& + & \frac{g^2}{2} \Pi^{i,a} \Pi^{i,a} + \frac{1}{4g^2} \left( 1 + \frac{\theta^2 g^4}{64\pi^4} \right) F_{ij}^a F^{ij,a} + \frac{\theta g^2}{16\pi^2} \epsilon_{ijk} \Pi^{i,a} F_{jk}^a \bigg] \yesnumber
\end{IEEEeqnarray*}
where now all field quantities are operators on a Hilbert space. In this quantum theory, $\Pi^{0,a}(x) \ket{\text{phys}} = 0$ is a primary constraint from which Gauss's law,
\begin{equation}
	i[H, \Pi^{0,a}(x)] \ket{\text{phys}} = \bigg( D_i \Pi^{i,a}(x) + \sum_{f=1}^{N_f} \psi^{\dag(f)}(x) T^a \psi^{(f)}(x) \bigg) \ket{\text{phys}} \equiv G^a(x) \ket{\text{phys}} = 0
\end{equation}
follows as a secondary constraint \cite{Dirac:1964:lectures}. 

It is convenient to work in the temporal gauge, $A_0^a=0$, achieved with the gauge transformation
\begin{equation}
	\Omega(x) = \mathcal{P} e^{-i \int_{t_0}^{x^0} dt \, A_0^a(t,\vec{x}) T^a}
\end{equation}
Time-dependent gauge transformations are subsequently forbidden because they will not preserve the gauge. Spatial gauge transformations preserve the gauge and are therefore permitted. The Gauss operator $G^a(x)$ generates spatial gauge transformations, so physical states $\ket{\text{phys}}$ are gauge invariant.

\section{Lattice Theory}
\label{sec:lattice_theory}

It is useful to think of the Hamiltonian of the lattice theory as being -- at base level -- a Riemann sum version of the continuum Hamiltonian: $\int d^3x \ f(\vec{x}) \to a^3 \sum_{\vec{s}} f(\vec{s})$. Thus, a spatial lattice is formed, and time remains continuous. However, many different lattice Hamiltonians can recover the continuum Hamiltonian in the $a \to 0$ limit.

This paper restricts attention to a cubic lattice where all neighboring lattices sites are separated by an equal lattice spacing $a$. Other geometries such as the honeycomb or triamond lattices are also of interest \cite{IllaSavageYao:2025:honeycombs, KavakiLewis:2025:triamonds}. Lattice sites have coordinates $\vec{s} = (n_1a,n_2a,n_3a)$, where $n_i \in \mathbb{Z}$. Links $\ell(\vec{s},\vec{e}_i)$ connect sites $\vec{s}$ and $\vec{s}+a\vec{e}_i$, where the Carestian unit vectors $\vec{e}_i$ are
\begin{equation}
	\vec{e}_1 = (1,0,0) \qquad \vec{e}_2 = (0,1,0) \qquad \vec{e}_3 = (0,0,1)
\end{equation}

An issue arises when disretizing the fermionic portion of the theory. The dispersion relation from naively discretized Dirac fermions shows a ``doubling'' of the number of fermions per flavor. While the continuum dispersion relation has an obvious minimum at zero momentum, lattice fermions have a periodic dispersion relation, leading to minima at large momentum. The minima correspond to tastes, which are copies of a fermion of a given flavor. In $3+1\text{D}$, there are eight tastes in the Hamiltonian approach. It is possible to mitigate this fermion doubling problem, but it is impossible to do so in an ideal way \cite{NielsonNinomiya:1981:no_go_theorem}. Common methods are to use staggered or Wilson fermions \cite{ZacheEtal:2018:quantum_simulation_wilson_fermions, AngelidesEtal:2023:mass_shift_wilson_staggered, BharadwajEtal:2026:wilson_qed3}, as will be done in this paper.

Furthermore, it will be convenient to work with projected Dirac field operators using projectors $P_\pm = \frac{1\pm\gamma^0}{2}$ \cite{Wilson:1977:quarks_strings}. In the Dirac basis of gamma matrices, these operators are
\begin{equation}
	P_+ \psi \coloneq \psi = \begin{pmatrix} \psi_1 \\ \psi_2 \\ 0 \\ 0 \end{pmatrix} \qquad \psi^\dag P_- \coloneq \chi = (0, 0, \chi_3, \chi_4)
\end{equation}
See App.~\ref{app:dirac_fermions} for a discussion on this Dirac fermion formalism. The lattice Hamiltonian will be normal-ordered with respect to a zero-point state $\ket{0}$ for which $\psi_\alpha(\vec{s}) \ket{0} = 0$ and $\chi_\alpha(\vec{s}) \ket{0} = 0$. The zero-point state is neither the free fermion vacuum nor ground state. Nevertheless, this scheme grants a useful interpretation of fermionic states. The state $\psi_\alpha^\dag \ket{0}$ has support in a particle at rest, and the state $\chi_\alpha^\dag \ket{0}$ has support in an antiparticle at rest. These states also contain particles and antiparticles with higher momenta, but in the zero-momentum sector, they contain either only a particle or only an antiparticle. Thus, $\ket{0}$ effectively matches the zero-momentum sector of the free fermion ground state.

The lattice theory has discrete (anti)commutation relations
\begin{IEEEeqnarray*}{rCl}
	[A_i^a(\vec{s}), \Pi^{j,b}(\vec{r})] & = & i\eta_i^j \delta_{ab} \delta_{\vec{s},\vec{r}} \\
	\{ \psi^{(f)}_{\alpha,c}(\vec{s}), \psi_{\beta,c'}^{\dag(f')}(\vec{r}) \} & = & (P_+)_{\alpha\beta} \delta_{cc'} \delta_{ff'} \delta_{\vec{s},\vec{r}} \\
	\{ \chi_{\alpha,c}^{\dag(f)}(\vec{s}), \chi^{(f')}_{\beta,c'}(\vec{r}) \} & = & (P_-)_{\alpha\beta} \delta_{cc'} \delta_{ff'} \delta_{\vec{s},\vec{r}} \yesnumber
\end{IEEEeqnarray*}
Written in this form, the lattice operators are dimensionless, which means that in passing from the continuum to the lattice they have been rescaled as follows:
\begin{equation}
	A_i^a \to \frac{1}{a}A_i^a \qquad \Pi^{i,a} \to \frac{1}{a^2}\Pi^{i,a} \qquad \psi_{\alpha,c}^{(f)} \to \frac{1}{\sqrt{a^3}} \psi_{\alpha,c}^{(f)} \qquad \chi_{\alpha,c}^{(f)} \to \frac{1}{\sqrt{a^3}} \chi_{\alpha,c}^{(f)}
\end{equation}
Then for the continuum limit, for instance, the discrete relations can be multiplied by $1=\frac{a^3}{a^3}$ to recover the continuum relations.

Gauge transformations of $A_\mu \equiv A_\mu^a T^a$ become nontrivial on the lattice due to derivative terms. Fortunately, Wilson line and loop operators \cite{Wilson:1974:confinement_of_quarks}, which are functions of the gauge field, follow straightforward transformation rules. The Wilson line (link) operator is
\begin{equation}
	U(\vec{s},\vec{e}_i) = e^{iA_i(\vec{s})} \qquad U(\vec{s}+a\vec{e}_i,-\vec{e}_i) = U^\dag(\vec{s},\vec{e}_i) = e^{-iA_i(\vec{s})}
\end{equation}
The Wilson loop (plaquette) operator is
\begin{equation}
	\Box(\vec{s},\vec{e}_i,\vec{e}_j) = U^\dag(\vec{s}, \vec{e}_j) U^\dag(\vec{s}+a\vec{e}_j, \vec{e}_i) U(\vec{s}+a\vec{e}_i, \vec{e}_j) U(\vec{s}, \vec{e}_i)
\end{equation}
for which this paper uses the convention that $i<j$. To expose the structure of the plaquette operator, Taylor expand the gauge fields via $A_i(\vec{s} + a\vec{e}_j) \approx A_i(\vec{s}) + a \partial_j A_i(\vec{s})$, and use the BCH formula $e^X e^Y \approx e^{X+Y+\frac{1}{2}[X,Y]}$ to keep an exponent of order $\mathcal{O}(aA,AA)$:
\begin{IEEEeqnarray*}{rCl}
	\Box(\vec{s},\vec{e}_i,\vec{e}_j) & = & e^{-iA_j(\vec{s})} e^{-iA_i(\vec{s} + a\vec{e}_j)} e^{iA_j(\vec{s} + a\vec{e}_i)} e^{iA_i(\vec{s})} \\
	& \approx & e^{-iA_j(\vec{s})} e^{-i(A_i(\vec{s}) + a\partial_j A_i(\vec{s}))} e^{i(A_j(\vec{s}) + a\partial_i A_j(\vec{s}))} e^{iA_i(\vec{s})} \\
	& \approx & e^{-i( A_j + A_i + a\partial_j A_i ) + \frac{(-i)^2}{2}[ A_j, A_i + a\partial_j A_i ]} e^{i( A_j + a\partial_i A_j + A_i ) + \frac{i^2}{2}[ A_j + a\partial_i A_j, A_i ]} \\
	& \approx & e^{-i( A_j + A_i + a\partial_j A_i ) + \frac{(-i)^2}{2}[ A_j, A_i ]} e^{i( A_j + a\partial_i A_j + A_i ) + \frac{i^2}{2}[ A_j, A_i ]} \\
	& \approx & e^{-ia \partial_j A_i + ia \partial_i A_j - [A_j,A_i]} \\
	& = & e^{ia^2 F_{ij}} \yesnumber
\end{IEEEeqnarray*}
where $F_{ij} \equiv F^a_{ij} T^a$. Hence,
\begin{IEEEeqnarray*}{rCl}
	\Box(\vec{s},\vec{e}_i,\vec{e}_j) & \approx & 1 + ia^2F_{ij}(\vec{s}) - \frac{a^4}{2} F_{ij}(\vec{s})^2 \\
	F_{ij}(\vec{s}) & \approx & -\frac{i}{2a^2} \left( \Box(\vec{s},\vec{e}_i,\vec{e}_j) - \Box^\dag(\vec{s},\vec{e}_i,\vec{e}_j) \right) \\
	F_{ij}(\vec{s})^2 & \approx & \frac{2}{a^4} - \frac{1}{a^4} \left( \Box(\vec{s},\vec{e}_i,\vec{e}_j) + \Box^\dag(\vec{s},\vec{e}_i,\vec{e}_j) \right) \yesnumber
\end{IEEEeqnarray*}
Note that $i,j$ in $F_{ij}^2$ are not summed over. Because the plaquette operator contains the field strength tensor in its matrix form, it will be necessary to rewrite the following Hamiltonian terms:
\begin{IEEEeqnarray*}{rCl}
	\frac{1}{4g^2} \left( 1 + \frac{\theta^2 g^4}{64\pi^4} \right) F_{ij}^a F^{ij,a} & = & \frac{1}{4 I_\mathcal{R} g^2} \left( 1 + \frac{\theta^2 g^4}{64\pi^4} \right) \trace[ F_{ij} F_{ij} ] \\
	& = & \frac{1}{2 I_\mathcal{R} g^2} \left( 1 + \frac{\theta^2 g^4}{64\pi^4} \right) \sum_{i<j} \trace[F_{ij}^2] \\
	\frac{\theta g^2}{16\pi^2} \epsilon_{ijk} \Pi^{i,a} F_{jk}^a & = & \frac{\theta g^2}{16 I_\mathcal{R} \pi^2} \epsilon_{ijk} \trace[ \Pi^i F_{jk} ] \\
	& = & \frac{\theta g^2}{8 I_\mathcal{R} \pi^2} \trace[ \Pi^1 F_{23} - \Pi^2 F_{13} + \Pi^3 F_{12} ] \\
	& = & - \frac{\theta g^2}{8 I_\mathcal{R} \pi^2} \sum_{\substack{ i \neq j,k \\ j<k }} (-1)^i \trace[\Pi^i F_{jk}] \yesnumber
\end{IEEEeqnarray*}
Thus, the pure-gauge Hamiltonian is comprised of three operators,
\begin{IEEEeqnarray*}{rCl}
	H_E & = & \frac{g^2}{2a} \sum_{\vec{s}} \sum_{i} \Pi^{i,a}(\vec{s}) \Pi^{i,a}(\vec{s}) \\
	H_B & = & \frac{1}{2 I_\mathcal{R} g^2 a} \left( 1 + \frac{\theta^2 g^4}{64 \pi^4} \right) \sum_{\vec{s}} \sum_{i < j} \trace\left[ 2 - \Box(\vec{s},\vec{e}_i,\vec{e}_j) - \Box^\dag(\vec{s},\vec{e}_i,\vec{e}_j) \right] \\
	H_\theta & = & \frac{i\theta g^2}{16 I_\mathcal{R} \pi^2 a} \sum_{\vec{s}} \sum_{\substack{i \neq j,k \\ j<k}} (-1)^i \trace\left[ \Pi^i(\vec{s}) \left( \Box(\vec{s},\vec{e}_j,\vec{e}_k) - \Box^\dag(\vec{s},\vec{e}_j,\vec{e}_k) \right) \right] \yesnumber
\end{IEEEeqnarray*}

Meanwhile, link operators can be inserted into the free fermion Hamiltonian to make it gauge invariant. The fermion Hamiltonian will be split into the kinetic term $H_\text{kin}$ and the mass term $H_\text{mass}$. (They are shown here with a neglected flavor index for brevity.) The kinetic term for Wilson fermions is
\begin{IEEEeqnarray*}{rCl}
	H^\text{W}_\text{kin} = \sum_{\vec{s}} \bigg[ & - & \frac{i}{2a} \sum_{i=1}^{3} \left[ \psi^\dag(\vec{s}) \gamma^i U^\dag(\vec{s},\vec{e}_i) \chi^\dag(\vec{s}+a\vec{e}_i) - \psi^\dag(\vec{s}) \gamma^i U(\vec{s}-a\vec{e}_i, \vec{e}_i) \chi^\dag(\vec{s}-a\vec{e}_i) \right] \\
	& + & \frac{i}{2a} \sum_{i=1}^{3} \left[ \chi(\vec{s}) \gamma^i U^\dag(\vec{s},\vec{e}_i) \psi(\vec{s}+a\vec{e}_i) - \chi(\vec{s}) \gamma^i U(\vec{s}-a\vec{e}_i, \vec{e}_i) \psi(\vec{s}-a\vec{e}_i) \right] \\
	& - & \frac{r}{2a} \sum_{i=1}^{3} \left[ \psi^\dag(\vec{s}) U^\dag(\vec{s},\vec{e}_i) \psi(\vec{s}+a\vec{e}_i) + \psi^\dag(\vec{s}) U(\vec{s}-a\vec{e}_i, \vec{e}_i) \psi(\vec{s}-a\vec{e}_i) \right] \\
	& - & \frac{r}{2a} \sum_{i=1}^{3} \left[ U_{cc'}(\vec{s},\vec{e}_i) \chi_{c'}^\dag(\vec{s}) \chi_c(\vec{s}+a\vec{e}_i) + \chi_c^\dag(\vec{s}) \chi_{c'}(\vec{s}-a\vec{e}_i) U_{c'c}^\dag(\vec{s}-a\vec{e}_i, \vec{e}_i) \right] \bigg] \\ \yesnumber
\end{IEEEeqnarray*}
In the last line, the explicit contraction of the color indices was shown for clarity; all other operators have the usual contraction as in $\psi^\dag_c U_{cc'} \psi_{c'}$. The kinetic term for staggered fermions is
\begin{IEEEeqnarray*}{rCl}
	H^\text{stag}_\text{kin} = & - & \frac{i}{2a} \sum_{\vec{s} \text{even}} \sum_{i=1}^{3} \gamma^i(\vec{s}) \left[ \psi^\dag(\vec{s}) U^\dag(\vec{s},\vec{e}_i) \chi^\dag(\vec{s}+a\vec{e}_i) - \psi^\dag(\vec{s}) U(\vec{s}-a\vec{e}_i, \vec{e}_i) \chi^\dag(\vec{s}-a\vec{e}_i) \right] \\
	& + & \frac{i}{2a} \sum_{\vec{s} \text{odd}} \sum_{i=1}^{3} \gamma^i(\vec{s}) \left[ \chi(\vec{s}) U^\dag(\vec{s},\vec{e}_i) \psi(\vec{s}+a\vec{e}_i) - \chi(\vec{s}) U(\vec{s}-a\vec{e}_i, \vec{e}_i) \psi(\vec{s}-a\vec{e}_i) \right] \\ \yesnumber
\end{IEEEeqnarray*}
This is written in a condensed manner, where the specific Dirac index is inferred from the site. The phases $\gamma^i(\vec{s}) = \pm 1, \pm i$ are gamma matrix elements. (See App.~\ref{app:dirac_fermions:lattice_fermions} for explicit conventions.) Finally, the mass terms are
\begin{IEEEeqnarray*}{rCl}
	H^\text{W}_\text{mass} & = & (m + \tfrac{3r}{a}) \sum_{\vec{s}} \psi^\dag(\vec{s}) \psi(\vec{s}) + \chi^\dag(\vec{s}) \chi(\vec{s}) \\
	H^\text{stag}_\text{mass} & = & m \sum_{\vec{s} \text{even}} \psi^\dag(\vec{s}) \psi(\vec{s}) + m \sum_{\vec{s} \text{odd}} \chi^\dag(\vec{s}) \chi(\vec{s}) \yesnumber
\end{IEEEeqnarray*}

The complete Hamiltonian is the sum of all terms: $H = H_E + H_B + H_\theta + H_\text{kin} + H_\text{mass}$. This incarnation of the lattice theory is usually referred to as the Kogut-Susskind Hamiltonian \cite{KogutSusskind:1975:KS_hamiltonian}.

\subsection{Link Hilbert Space}

A link operator is defined on each link, and each link hosts a bosonic Hilbert space. The link operator $e^{iA_i(\vec{s})}$ carries the form of a matrix in the representation $\mathcal{R}$ for a group element $g \in \text{SU}(N_c)$. A suitable basis for the link Hilbert space is therefore the set of all states labeled by group elements, $\{ \ket{g} \}$. This group element basis is defined to diagonalize the link operator, with representation matrices $D(g)$ as its eigenvalues. For an arbitrary SU($N_c$) irrep, the link operator is, in component form,
\begin{equation}
	U_{mn}^R = \int dg \ D_{mn}^R(g) \ketbra{g}{g}
\end{equation}
Refer to App.~\ref{app:aspects:representation_matrices} for more information on representation matrices and App.~\ref{app:aspects:link_hilbert_space} for an extended discussion on the link Hilbert space.

Gauge transformations are expressed with representation matrices; throughout this work, the shorthand $\Omega \equiv D(\Omega)$ and $\Omega^\dag \equiv D(\Omega^{-1})$ is used. Define translation operators $\Theta_{\mathsf{L}\Omega}(\vec{s},\vec{e}_i)$ and $\Theta_{\mathsf{R}\Omega}(\vec{s},\vec{e}_i)$ that together enact a gauge transformation $\Omega(\vec{s}+a\vec{e}_i) U(\vec{s},\vec{e}_i) \Omega^\dag(\vec{s})$ on a link operator. This is possible with
\begin{equation}
	\Theta_{\mathsf{L}\Omega} \ket{h} = \ket{h\Omega^{-1}} \qquad \Theta_{\mathsf{R}\Omega} \ket{h} = \ket{\Omega^{-1}h}
\end{equation}
In particular, let $\Theta_{\mathsf{L}\Omega} \equiv \Theta_{\mathsf{L}\Omega(\vec{s})}$ and let $\Theta_{\mathsf{R}\Omega} \equiv \Theta_{\mathsf{R}\Omega(\vec{s}+a\vec{e}_i)}$. Then
\begin{equation}
	\Theta^\dag_{\mathsf{L}\Omega} U \Theta_{\mathsf{L}\Omega} = U \Omega^\dag(\vec{s}) \qquad \Theta_{\mathsf{R}\Omega} U \Theta^\dag_{\mathsf{R}\Omega} = \Omega(\vec{s}+a\vec{e}_i) U
\end{equation}
Hence, $\Theta_{\mathsf{L}\Omega}(\vec{s},\vec{e}_i)$ transforms the link operator from the ``left end'' of the link, and $\Theta_{\mathsf{R}\Omega}(\vec{s},\vec{e}_i)$ transforms the link operator from the ``right end'' of the link. This establishes left and right half-links.

The translation operators are group-valued operators, and can be written in terms of algebra-valued operators via the exponential map:
\begin{equation}
	\Theta_{\mathsf{L}\Omega}(\vec{s},\vec{e}_i) = e^{i \phi^a_\Omega \Pi^{i,a}_\mathsf{L}(\vec{s})} \qquad \Theta_{\mathsf{R}\Omega}(\vec{s},\vec{e}_i) = e^{i \phi^a_\Omega \Pi^{i,a}_\mathsf{R}(\vec{s})}
\end{equation}
where the Hermitian operators $\Pi_{\mathsf{L},\mathsf{R}}^{i,a}(\vec{s})$ have been suggestively written. They satisfy the commutation relations
\begin{equation}
	[\Pi_\mathsf{L}^a, \Pi_\mathsf{L}^b] = if^{abc} \Pi_\mathsf{L}^c \qquad [\Pi_\mathsf{R}^a, \Pi_\mathsf{R}^b] = -if^{abc} \Pi_\mathsf{R}^c \qquad [\Pi_\mathsf{L}^a, \Pi_\mathsf{R}^b] = 0
\end{equation}
This shows there exists two algebras on a link. The quadratic Casimir operators $\Pi^a_\mathsf{L} \Pi^a_\mathsf{L}, \Pi^a_\mathsf{R} \Pi^a_\mathsf{R}$ and the Cartan subalgebra operators $\Pi^k_{\mathsf{L},z}, \Pi^k_{\mathsf{R},z}$ ($k=1,\dots,N_c-1$) form maximal sets of commuting operators for left and right half-link Hilbert spaces. The eigenbases of these operators are suitable bases for these two Hilbert spaces. Eigenstates of the Casimir operator are labeled by an SU($N_c$) irrep $R$. Eigenstates of a Cartan operator are labeled by a basis vector $r$ of $R$. However, $\Pi^a_\mathsf{L} \Pi^a_\mathsf{L} = \Pi^a_\mathsf{R} \Pi^a_\mathsf{R}$, and therefore left and right half-link basis states coalesce into one link basis state labeled by a single irrep $R$ and two independent basis vectors, $r_\mathsf{L}$ and $r_\mathsf{R}$, of $R$. This establishes the representation basis for the link Hilbert space, with representation basis states denoted by $\ket{(R,r_\mathsf{R},r_\mathsf{L})} = \ket{(R,r_\mathsf{R})} \otimes \ket{(R,r_\mathsf{L})}$.

\subsection{Site Hilbert Space}

The fermionic operators $\psi_{\alpha,c}^{(f)}(\vec{s})$ and $\chi_{\alpha,c}^{(f)}(\vec{s})$ change the number of fermions at a site. It is convenient to refer to $\psi^\dag \ket{0} \equiv \ket{1}$ as a single particle state, and $\chi^\dag \ket{0} \equiv \ket{1}$ as a single antiparticle state.\footnote{Strictly speaking, these are only excitations of fundamental and antifundamental charge.} Each particle/antiparticle has a specific flavor, spinor, and color index. This is the notion of the occupation number basis for fermionic states.

To be more explicit, the fermionic occupation number basis is composed of tensor products of basis states, $\ket{0}$ and $\ket{1}$. The annihilation and creation operators ($\psi, \chi$ and $\psi^\dag, \chi^\dag$) have a nontrivial tensor-product structure to ensure anticommutation relations between them. To make this work, the single-particle states are enumerated via some convention to define multi-particle states. In this paper, an occupation number basis state is built according to
\begin{equation}
	\ket{n_{1,1}^{(1)}(s_1), n_{1,2}^{(1)}(s_1), \dots} = \prod_{\vec{s}} \prod_{f=1}^{N_f} \left( \prod_{\alpha=1}^{2} \prod_{c=1}^{N_c} \left( \psi_{\alpha,c}^{(f)\dag}(\vec{s}) \right)^{n^{(f)}_{\alpha,c}(\vec{s})} \times \prod_{\alpha=3}^{4} \prod_{c=1}^{N_c} \left( \chi_{\alpha,c}^{(f)\dag}(\vec{s}) \right)^{\bar{n}^{(f)}_{\alpha,c}(\vec{s})} \right) \ket{0}
\end{equation}
where $n^{(f)}_{\alpha,c}(\vec{s}) \in \{ 0,1 \}$ denotes a particle occupation number and $\bar{n}^{(f)}_{\alpha,c}(\vec{s}) \in \{ 0,1 \}$ denotes an antiparticle occupation number. The sites may be ordered lexicographically, for example, although more optimal enumerations exist \cite{ChiewStrelchuk:2023:fermion_mapping_enumeration}.

Now, for brevity, let $k$ be a multi-index for particle/antiparticle flavor, spinor, color, and site labels. Applying $\psi_k^\dag$ to $\ket{n_1,n_2,\dots,n_k,\dots}$ involves $k-1$ anticommutations until the operator is placed in its appropriate position. This process accumulates a product of negative signs such that an occupation number basis state transforms as
\begin{IEEEeqnarray*}{rCl}
	\psi_k \ket{n_1, \dots, n_k, \dots} & = & (-1)^{\zeta_k} \sqrt{n_k} \ket{n_1, \dots, n_k - 1, \dots} \\ [12pt]
	\psi_k^\dag \ket{n_1, \dots, n_k, \dots} & = & (-1)^{\zeta_k} \sqrt{1-n_k} \ket{n_1, \dots, n_k + 1, \dots} \yesnumber
\end{IEEEeqnarray*}
where $\zeta_k = \sum_{j<k} n_j$. The operators that accomplish this are
\begin{equation}
	\psi_k = \bigotimes_{j<k} Z_j \otimes \sigma^-_k \otimes \bigotimes_{j>k} I_j \qquad \psi_k^\dag = \bigotimes_{j<k} Z_j \otimes \sigma^+_k \otimes \bigotimes_{j>k} I_j
\end{equation}
where
\begin{equation}
	I = \ketbra{0}{0} + \ketbra{1}{1} \qquad Z = \ketbra{0}{0} - \ketbra{1}{1} \qquad \sigma^- = \ketbra{0}{1} \qquad \sigma^+ = \ketbra{1}{0}
\end{equation}
(The same statements and definitions hold for $\chi_k$ and $\chi^\dag_k$.) This is the essence of the Jordan-Wigner transformation, which expresses the annihilation and creation operators in the occupation number basis. The Jordan-Wigner transformation has the drawback that the nonlocal product of $Z$ operators can become large, which comes at a price for numerical simulations. There are alternatives in somewhat less intuitive bases, such as the Bravyi-Kitaev transformation, that tame the locality of these operators \cite{SeeleyRichardLove:2012:fermion_encodings, OBrienStrelchuk:2024:ultrafast_fermion_to_qubit}. It is also possible to eliminate the fermionic matter from the theory, replacing it with bosonic degrees of freedom \cite{Pardo:2023:fermion_elimination, BallarinEtal:2024:local_encoding}.

In its current form, the occupation number basis carries no obvious relation to the charges of the theory. The charge operators are\footnote{The Dirac indices in the charge and translation operators make sense for Wilson fermions. For staggered fermions, the Dirac index can be omitted.}
\begin{equation}
	Q_\alpha^{a(f)}(\vec{s}) = \psi_{\alpha,c}^{\dag(f)}(\vec{s}) T^a_{cc'} \psi_{\alpha,c'}^{(f)}(\vec{s}) \qquad \bar{Q}_\alpha^{a(f)}(\vec{s}) = -\chi_{\alpha,c}^{\dag(f)}(\vec{s}) T^{a*}_{cc'} \chi_{\alpha,c'}^{(f)}(\vec{s})
\end{equation}
where only the color indices $c,c'$ are summed over. Hence, particles are charged in the fundamental representation, and antiparticles are charged in the antifundamental representation. These Hermitian charge operators can be exponentiated to form translation operators
\begin{equation}
	\Theta_{Q_\alpha \Omega}^{(f)}(\vec{s}) = e^{i \phi_\Omega^a Q_\alpha^{a(f)}(\vec{s})} \qquad \bar{\Theta}_{Q_\alpha \Omega}^{(f)}(\vec{s}) = e^{i \phi_\Omega^a \bar{Q}_\alpha^{a(f)}(\vec{s})}
\end{equation}
that enact gauge transformations on the fermions. If $\Theta_{Q_\alpha\Omega} \equiv \Theta_{Q_\alpha\Omega(\vec{s})}$ and $\bar{\Theta}_{Q_\alpha\Omega} \equiv \bar{\Theta}_{Q_\alpha\Omega(\vec{s})}$, then
\begin{equation}
	\Theta^\dag_{Q_\alpha\Omega} \psi_\alpha \Theta_{Q_\alpha\Omega} = \Omega(\vec{s}) \psi_\alpha \qquad \bar{\Theta}^\dag_{Q_\alpha\Omega} \chi_\alpha \bar{\Theta}_{Q_\alpha\Omega} = \bar{\Omega}(\vec{s}) \chi_\alpha
\end{equation}
where $\bar{\Omega} \equiv (D(\Omega))^*$ and $\bar{\Omega}^\dag \equiv (D(\Omega^{-1}))^*$. Note that the charge and translation operators act on a subspace of the occupation number basis -- one of particles or antiparticles of a given flavor and Dirac index at a site. This $2^{N_c}$-dimensional subspace can be called a color subspace, and the inherited basis for this subspace can be called the color occupation number basis.

\begin{figure}[!ht]
	\centering
	\includegraphics[width=0.825\textwidth]{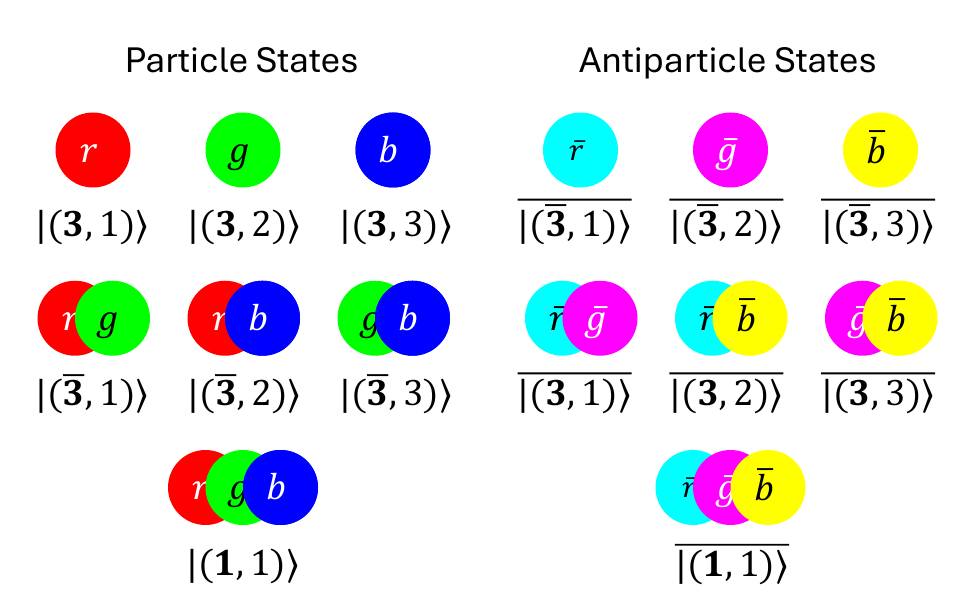}
	\caption{The representation basis for SU(3) fermionic states. Single occupied particle (antiparticle) states correspond to states in the fundamental (antifundamental) representation. Double occupied states form the conjugate representation. Fully occupied states live in a trivial representation, a color singlet. The unoccupied state (not shown) also is a trivial representation state.}
	\label{fig:fermion_states}
\end{figure}

The charge operators form algebras:
\begin{equation}
	[Q^a, Q^b] = if^{abc} Q^c \qquad [\bar{Q}^a, \bar{Q}^b] = if^{abc} \bar{Q}^c \qquad [Q^a, \bar{Q}^b] = 0
\end{equation}
The quadratic Casimir operators $Q^a Q^a, \bar{Q}^a \bar{Q}^a$ and the Cartan subalgebra operators $Q_z^k, \bar{Q}_z^k$ ($k=1,\dots,N_c-1$) form maximal sets of commuting operators for color subspaces. Eigenstates $\ket{(R,r)}$ of these operators are labeled by an SU($N_c$) irrep $R$, and a basis vector $r$ of $R$. These states can be realized as a change of basis\footnote{The zero-point state $\ket{0}$ corresponds to a trivial representation.}
\begin{IEEEeqnarray*}{rCl}
	\ket{(R,r)} & \coloneq & \frac{1}{\sqrt{N_q!}} \sum_{c_1=1}^{N_c} \cdots \sum_{c_{N_q}=1}^{N_c} \braket{(R,r)}{\otimes_{i=1}^{N_q} (\mathcal{R},c_i)} \prod_{i=1}^{N_q} \psi_{c_i}^\dag \ket{0} \\
	\bar{\ket{(R,r)}} & \coloneq & \frac{1}{\sqrt{N_q!}} \sum_{c_1=1}^{N_c} \cdots \sum_{c_{N_q}=1}^{N_c} \braket{\otimes_{i=1}^{N_q} (\mathcal{R},c_i)}{(\bar{R},\tilde{r})} \prod_{i=1}^{N_q} \chi_{c_i}^\dag \ket{0} \yesnumber
\end{IEEEeqnarray*}
Here, $N_q \in \{ 0,\dots,N_c \}$ is the color occupation number -- the number of particles/antiparticles in a color occupation number basis state. The coefficients are Clebsch-Gordan coefficients (CGCs), and the notation $\tilde{r}$ means the basis vector of $\bar{R}$ conjugate to the basis vector $r$ of $R$. This establishes a representation basis for fermionic states, illustrated in Fig.~\ref{fig:fermion_states} for SU(3). See App.~\ref{app:aspects:CGCs} for a review of CGCs and App.~\ref{app:aspects:site_hilbert_space} for an extended discussion on the site Hilbert space.

The representation basis states will be useful to formulate the physical Hilbert space. However, in practice, the occupation number basis remains convenient to perform calculations, due to the straightforward action of the annihilation and creation operators. Fortunately, when $\mathcal{R}$ is the fundamental representation, each representation basis state is equal to a color occupation number basis state. Therefore, the two bases can be readily translated from one to another.

\subsection{Physical Hilbert Space}

The lattice theory inherits Gauss's law as a constraint on physical states of the Hilbert space. As reviewed in App.~\ref{app:aspects:physical_hilbert_space}, the Gauss operator
\begin{equation}
	G^a(\vec{s}) = \sum_{i=1}^{3} \Pi^{i,a}_\mathsf{L}(\vec{s}) - \Pi^{i,a}_\mathsf{R}(\vec{s}-a\vec{e}_i) + \sum_{f=1}^{N_f} \sum_{\alpha=1}^{4} Q_\alpha^{a(f)}(\vec{s}) + \bar{Q}_\alpha^{a(f)}(\vec{s})
\end{equation}
generates spatial gauge transformations. To see the correspondence with the continuum Gauss operator, consider the relation $\Pi^{i,a}_\mathsf{R}(\vec{s}) = U^\text{adj}_{ab}(\vec{s},\vec{e}_i) \Pi_\mathsf{L}^{i,b}(\vec{s})$, where ``adj'' denotes the adjoint representation. The lattice Gauss operator contains (up to factors of $a$) a covariant derivative approximated by a backwards finite difference:
\begin{equation}
	D_i \Pi^{i,a} \to \sum_{i=1}^{3} \frac{\Pi_\mathsf{L}^{i,a}(\vec{s}) - U^\text{adj}_{ab}(\vec{s}-a\vec{e}_i,\vec{e}_i) \Pi_\mathsf{L}^{i,b}(\vec{s}-a\vec{e}_i)}{a}
\end{equation}
It is therefore admissable to equate $\Pi^{i,a}(\vec{s}) = \Pi^{i,a}_\mathsf{L}(\vec{s})$ as the lattice canonical momentum operator, and thus $G^a(\vec{s})$ is a discretization of the continuum Gauss operator.

\begin{figure}[!ht]
	\centering
	\includegraphics[width=0.745\textwidth]{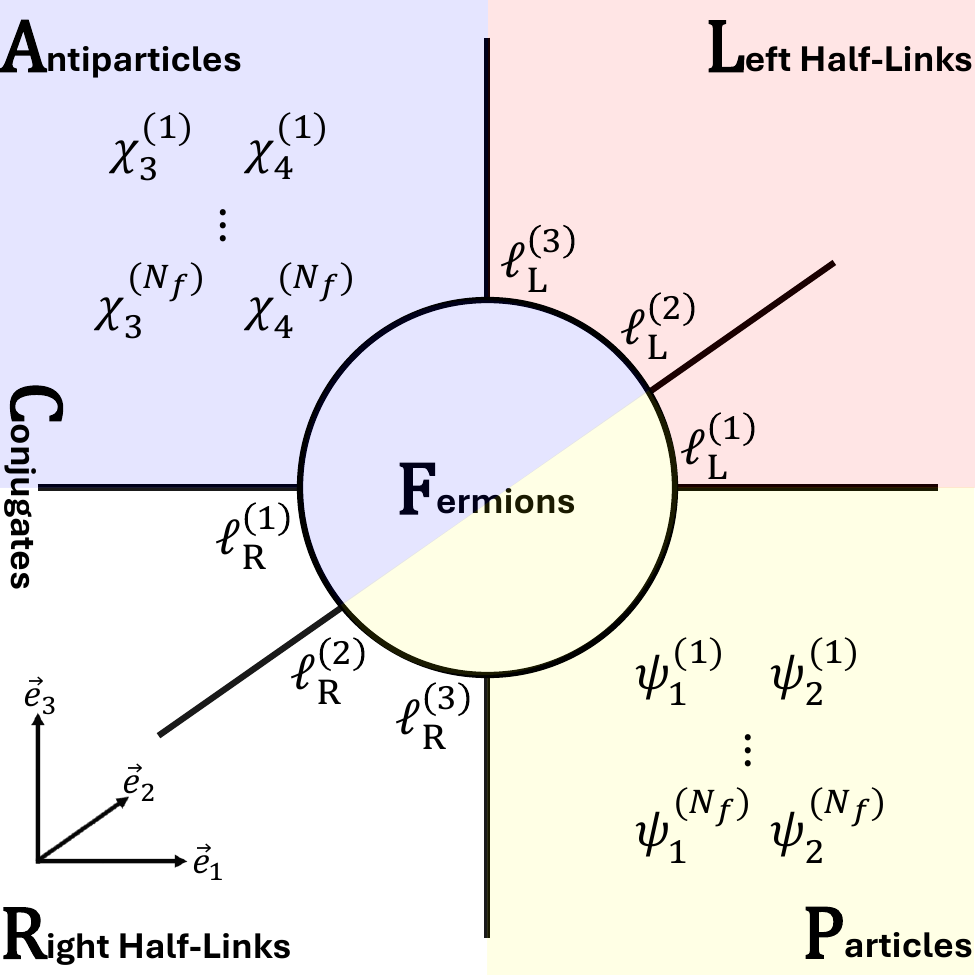}
	\caption{The degrees of freedom meeting at a site. The state of each degree of freedom can be written with representation basis states $\ket{(R,r)}$. To satisfy Gauss's law, the direct product of all SU($N_c$) irreps $R$ must contain a singlet. In 3d, for Wilson-like fermions, there are $6+4N_f$ irreps that meet at a site.}
	\label{fig:site_diagram}
\end{figure}

Physical basis states expressed in the representation basis are\footnote{See App.~\ref{app:hamiltonian_matrix_elements} for calculations of matrix elements in the physical Hilbert space.}
\begin{equation}
	\ket{\Lambda} = \bigotimes_{\vec{s}} \sum_{\{ r_x \}} \Big( {\textstyle\prod\limits_{x \in \mathbf{C}}} \varphi(r_x) \Big) \braket{\mathbf{1},\Gamma_s}{\underset{\ell_\mathsf{L} \in \mathbf{L}}{\otimes} (R_{\ell_\mathsf{L}}, r_{\ell_\mathsf{L}}) \underset{\ell_\mathsf{R} \in \mathbf{R}}{\otimes} (\bar{R}_{\ell_\mathsf{R}}, \tilde{r}_{\ell_\mathsf{R}}) \underset{f \in \mathbf{F}}{\otimes} (R_f, r_f)} \ \ket{\underset{x \in \mathbf{S}}{\otimes} (R_x, r_x)}
\end{equation}
These states form a basis for solutions $\ket{\text{phys}}$ to Gauss's law. For each site $\vec{s}$, $\mathbf{L}$ is the set of all left half-links, $\mathbf{R}$ is the set of all right half-links, $\mathbf{P}$ is the set of all particles, and $\mathbf{A}$ is the set of all antiparticles meeting at the site. This is illustrated in Fig.~\ref{fig:site_diagram}. (Fermions in $\mathbf{F} = \mathbf{P} \cup \mathbf{A}$ are specified by a flavor index, a spinor index, and a particle/antiparticle label, such as $\psi_\alpha^{(f)}$ or $\chi_\alpha^{(f)}$.) The set of all lattice data meeting at a site is $\mathbf{S} = \mathbf{L} \cup \mathbf{R} \cup \mathbf{P} \cup \mathbf{A}$. The label ``$x$'' is used as a catch-all for sets with both link and fermion data, and the sum over $\{ r_x \} = \{ r_x \, | \, x \in \mathbf{S} \}$ runs through all possible basis vectors of the irreps. States from $\mathbf{C} = \mathbf{R} \cup \mathbf{A}$ contribute conjugate phases $\varphi(r) = \pm 1$ that relate $r$ with $\tilde{r}$.

Physical basis states $\ket{\Lambda}$ are partially specified by an assignment of irreps for each link and each fermion. The direct product of irreps meeting at a site must contain a singlet, or trivial representation (denoted by $\mathbf{1}$). The singlet CGCs of $\ket{\Lambda}$, wherein the right half-link states appear conjugated, enforce this. Multiple singlets can exist in a direct product of irreps; different singlets at each site can be labeled by a multiplicity index $\Gamma_s$.

However, irreps and singlet labels are not enough to numerically specify a physical basis state \cite{BalajiEtal:2025:qcirc_for_su3}. The singlet CGCs as written in $\ket{\Lambda}$ contain $(R,r)$ factors strung together in some unspecified order. When the same irrep appears in a CGC multiple times, that CGC takes on nontrivial symmetries under the exchange of same-irrep $(R,r)$ factors. As a toy example, let $x_1$ and $x_2$ be two lattice data meeting at a site such that $R_{x_1}=R_{x_2}=R$, and define some simple objects $C_1 = \braket{\mathbf{1}}{(R,r_{x_1}) \otimes (R,r_{x_2})}$ and $C_2 = \braket{\mathbf{1}}{(R,r_{x_2}) \otimes (R,r_{x_1})}$. It need not be the case that $C_1=C_2$ if $r_{x_1} \neq r_{x_2}$. Different orderings lead to different definitions of the basis state $\ket{\Lambda}$. Which of $C_1$ or $C_2$ should be the coefficient of the state $\ket{(R,r_{x_1}) \otimes (R,r_{x_2})}$? Because $C_1$ and $C_2$ are singlet CGCs, there is no wrong answer, and the choice made defines a convention: if $C_1$ is chosen, then states of $x_1$ appear first in the singlet CGCs, followed by $x_2$. If $C_2$ is chosen, then vice versa. This convention for the order in which lattice data slots into a singlet CGC will be called the site order (also known as an F-order \cite{BalajiEtal:2025:qcirc_for_su3}).

It is convenient and practical to impose the same site order across all sites of the lattice for all physical basis states. This, together with assigning irreps and singlet labels, facilitates an unambiguous complete specification of physical basis states. For calculations done in this work, the site order used runs through the sequence $(\mathbf{L}, \mathbf{R}, \mathbf{P}, \mathbf{A})$. $\mathbf{L}$ is ordered according to $(\vec{e}_1, \vec{e}_2, \vec{e}_3)$, $\mathbf{R}$ is ordered according to $(-\vec{e}_1, -\vec{e}_2, -\vec{e}_3)$, $\mathbf{P}$ is ordered according to $(\psi^{(1)}_1,\psi^{(1)}_2,\dots,\psi^{(N_f)}_1,\psi^{(N_f)}_2)$, and $\mathbf{A}$ is ordered according to $(\chi^{(1)}_3,\chi^{(1)}_4,\dots,\chi^{(N_f)}_3,\chi^{(N_f)}_4)$.

\section{Quantum Simulation}
\label{sec:quantum_simulation}

Simulations of lattice field theory require a finite system size. The lattice volume can be truncated by imposing periodic boundary conditions (PBCs), which preserve translation invariance, or open boundary conditions (OBCs), which terminate the extent of the lattice. What remains to be done is a truncation on the physical Hilbert space.

SU($N_c$) has an infinite number of irreps, so the link Hilbert space is infinite-dimensional. To develop truncations on the link Hilbert space, first note that SU($N_c$) irreps can be labeled by [normalized] i-weights \cite{ArneEtal:2011:cgcs}:
\begin{equation}
	R \equiv (n_1, n_2, \dots, n_{N_c-1}, n_{N_c}) \qquad n_1 \geq n_2 \geq \cdots \geq n_{N_c-1} \geq n_{N_c} = 0 \qquad n_k \in \mathbb{N}
\end{equation}
The dimension and quadratic Casimir eigenvalue\footnote{This formula assumes the Dynkin index of the fundamental representation is one-half.} of an irrep are
\begin{equation}
	\dim(R) = \prod_{k=2}^{N_c} \prod_{\ell=1}^{k-1} \left( 1 + \frac{n_\ell - n_k}{k-\ell} \right) \qquad C_2(R) = \frac{1}{2} \sum_{k=1}^{N_c} n_k(n_k + N_c + 1 - 2k) - \frac{1}{2N_c} \left( \sum_{k=1}^{N_c} n_k \right)^2
\end{equation}
One truncation follows from the form of the i-weights: set a cutoff on $n_1$. Denote this truncation scheme by $T_n$, where i-weights with $n_1 \leq n$ are kept in the finite set of irreps. A more fine-grained scheme can be devised from imposing a cutoff on $C_2(R)$. Let $C_2^{\max}$, the maximum quadratic Casimir eigenvalue allowed for an irrep, be the cutoff in this truncation scheme. A less local cutoff is on the maximum sum of quadratic Casimir eigenvalues meeting at a site. Let $C_2^*$ be this cutoff (also known as $B$ \cite{BalajiEtal:2026:ground_state}). Any of these truncations will yield a finite-dimensional physical Hilbert space.

Although the fermionic Hilbert space is naturally finite, it may be useful to truncate color occupation numbers. Let $M_{N_q}$ denote a truncation where, at a site, each flavor and their spinor components can have at most color occupation number $N_q$. Let $N_{N_q}$ denote a truncation where, summing over all flavors and their spinor components at a site, the total color occupation number is at most $N_q$. Both truncations can be made flavor-dependent. Let $[M^{(1)}_{N_{q,1}}, \dots, M^{(N_f)}_{N_{q,N_f}}]$ denote a truncation where, for a flavor $f$ at a site, each spinor component can have at most color occupation number $N_{q,f}$. Finally, let $[N^{(1)}_{N_{q,1}}, \dots, N^{(N_f)}_{N_{q,N_f}}]$ denote a truncation where, for a flavor $f$ at a site, the total color occupation number across all spinor components is at most $N_{q,f}$.

To simulate the truncated lattice theory on a digital quantum computer, the physical Hilbert space must be encoded with qubits. There are three main encodings that have appeared in the literature so far. The first is a direct encoding of all irreps and their basis vectors. One then prepares the superposition $\ket{\Lambda}$, or more generally, $\ket{\text{phys}}$. This direct, or ``BY,'' encoding was first suggested by Byrnes and Yamamoto \cite{ByrnesYamamoto:2006:sim_LGT_on_QC}. It has been used to make resource estimates for simulations in the fault-tolerant era \cite{KanNam:2022:kannam, RhodesKreshchukPathak:2024:exponential_improvements}. The other two encodings may be classified as global and local \cite{KlcoSavageStryker:2020:one_dim_su2, CiavarellaKlcoSavage:2021:trailhead}. The global encoding is an encoding of the physical basis states themselves. Because this is an encoding of whole-lattice states, it can be useful for devising a simulation contained within a symmetry sector \cite{CiavarellaKlcoSavage:2021:trailhead, RahmanEtal:2021:su2_quantum_annealer}. The local encoding is an encoding of all irreps and multiplicity indices \cite{BalajiEtal:2025:qcirc_for_su3}. This is the encoding used in this paper because it can require less qubits than the BY encoding, and it can require less classical precomputation than the global encoding for a large Hilbert space.

A local encoding can be carried out by devoting qubits to irreps of the fermions and gauge degrees of freedom, as well as to singlet multiplicity indices. For simplicity, this data can be encoded in a dense binary fashion. Enumerate the irreps and multiplicity indices, and assign their decimal number the corresponding bitstring. For example, the tenth representation becomes encoded by $\ket{1010}$. If employing the Jordan-Wigner transformation, it is convenient to enumerate the fermionic irreps according to the parity of their color occupation number. For example, $\ket{01}$ and $\ket{11}$ can encode fermionic states with odd occupation, while $\ket{00}$ and $\ket{10}$ can encode states with even occupation. The number of qubits required per site or link depends logarithmically on the number of irreps kept and the maximum multiplicities of the possible singlets.

\subsection{Trotterization}

Different quantum algorithms exist to perform Hamiltonian simulation on a digital quantum computer \cite{HariprakashEtal:2025:time_evolution_strategies}. This is because there are multiple ways to approximate the time evolution operator $e^{-iHt}$. For instance, there is the Taylor series, but there exist other polynomial approximations that can be implemented on quantum computers through a technique called qubitization \cite{MartynEtal:2021:grand_unification}. Hamiltonian simulation with qubitization has been found to have state-of-the-art optimal quantum resource scalings for fault-tolerant quantum devices \cite{MohseniEtal:2026:quantum_supercomputer}. However, such an algorithm would be difficult to implement on current devices. As a result, this work considers Trotterization, which will enable small-scale quantum simulations that could in principle be ran on near-term devices.

In Trotterization, the time evolution operator is approximated as a product of unitaries made from sub-Hamiltonians $h_k$ in a decomposition $H = \sum_k h_k$. This is done in such a way that $e^{-ih_k}$ should be straightforward to write in terms of quantum gates, compared to $e^{-iH}$. A first-order Trotterization is \cite{NielsenChuang:2012:quantum_computing}
\begin{equation}
	e^{-iHt} \approx U(t) \equiv \left( \prod_{k} e^{-ih_k\Delta t} \right)^{n_t} \qquad n_t \Delta t = t
\end{equation}
where $e^{-iHt} = U(t) + \mathcal{O}(\Delta t^2)$. In this work, the following Trotterization is used:
\begin{equation}
	U(t) = \left( \prod_{k} e^{-iH_{\text{mass},k} \Delta t} \prod_{k} e^{-iH_{E,k} \Delta t} \prod_{k} e^{-iH_{B+\theta,k} \Delta t} \prod_{k} e^{-iH_{\text{kin},k} \Delta t} \right)^{n_t}
\end{equation}
Here, $H_{B+\theta} = H_B + H_\theta$ is the sum of both Hamiltonian operators proportional to the plaquette operator. $H_B$ and $H_\theta$ both induce transitions on a single plaquette, except $H_B$ has real coefficients, and $H_\theta$ has imaginary coefficients. Thus, $H_{B+\theta}$ changes the state of a plaquette, generally, by a complex coefficient.

\begin{figure}[!h]
	\centering
	\begin{subfigure}{0.68\textwidth}
		\centering
		\includegraphics[height=0.4125in]{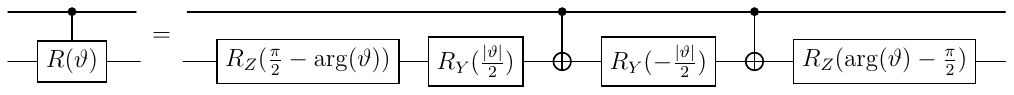}
		\caption{Rotation $e^{-\frac{i}{2}(\vartheta\sigma^+ + \vartheta^*\sigma^-)} = R(\vartheta)$}
	\end{subfigure}
	\begin{subfigure}{0.31\textwidth}
		\centering
		\includegraphics[height=0.4125in]{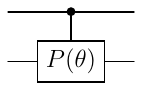}
		\caption{Phase shift $e^{i\theta \pi^1} = P(\theta)$}
	\end{subfigure}
	
	\vspace{12pt}
	
	\begin{subfigure}{0.55\textwidth}
		\centering
		\includegraphics[height=0.4125in]{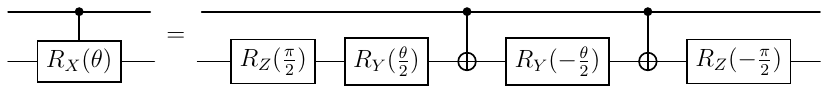}
		\caption{Rotation $e^{-\frac{i}{2}(\theta\sigma^+ + \theta\sigma^-)} = R_X(\theta)$}
	\end{subfigure}
	\begin{subfigure}{0.44\textwidth}
		\centering
		\includegraphics[height=0.4125in]{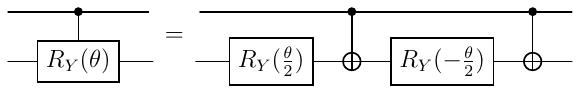}
		\caption{Rotation $e^{-\frac{i}{2}(i\theta\sigma^+ - i\theta\sigma^-)} = R_Y(\theta)$}
	\end{subfigure}
	\caption{Controlled gates used in Trotterization. Each gate can be generalized to a multi-controlled gate by adding more controls. In simulations, the gate angles are Hamiltonian matrix elements, which encode specific, one-to-one physical link or plaquette transitions.}
	\label{fig:controlled_gates}
\end{figure}

\begin{figure}[!h]
	\centering
	\includegraphics{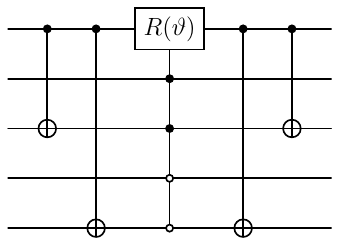}
	\caption{Example quantum circuit for a Givens rotation $e^{-ih}$, where $h = \frac{\vartheta}{2} \ketbra{11001}{01100} + \text{H.c.} = \frac{\vartheta}{2} \sigma^+ \pi^1 \sigma^- \pi^0 \sigma^+ + \text{H.c.}$ for $\vartheta \in \mathbb{C}$. Pick one ``pivot'' qubit that a ladder operator flips. Apply CNOTs controlled on the pivot qubit, targeting all other qubits that get flipped by $\sigma^\pm$. Targeted qubits flipped by the same (opposite) ladder operator as the pivot qubit become projected by $\pi^0$ ($\pi^1$). As shown, the CNOTs partially diagonalize $h$ into $\tilde{h} = \frac{1}{2}(\vartheta \sigma^+ + \vartheta^* \sigma^-) \pi^1 \pi^1 \pi^0 \pi^0$. Hence, $e^{-i\tilde{h}}$ is a controlled $R(\vartheta)$ gate.}
	\label{fig:givens_example}
\end{figure}

The Trotterization is done such that each $e^{-ih\Delta t}$ operator contains a sub-Hamiltonian $h = \vartheta \ketbra{b_1}{b_2} + \text{H.c.}$, where $b_{1,2}$ are bitstrings such as $\ket{b_1} = \ket{011}$, encoding link and fermion irreps, and site multiplicity indices. (Not shown in $h$ are a plethora of identity operators acting all over the lattice.) If $b_1 \neq b_2$, then $e^{-ih\Delta t}$ is referred to as a two-level unitary operator \cite{NielsenChuang:2012:quantum_computing}, or a Givens rotation; otherwise, it is a phase shift. Each operator $\ketbra{b_1}{b_2}$ can be written as a tensor product of ladder operators $\sigma^\pm$ and projector operators $\pi^{0,1}$:
\begin{equation}
	\sigma^+ = \ketbra{1}{0} \qquad \sigma^- = \ketbra{0}{1} \qquad \pi^0 = \ketbra{0}{0} \qquad \pi^1 = \ketbra{1}{1}
\end{equation}
For example, $\ketbra{01}{00} = \pi^0 \otimes \sigma^+$. Fig.~\ref{fig:controlled_gates} shows how to write these $e^{-ih\Delta t}$ operators with quantum gates when at most one tensor-operator in $h$ is a ladder operator. When $\ketbra{b_1}{b_2}$ contains multiple ladder operators, a diagonalization procedure can be performed with CNOTs, as shown in Fig.~\ref{fig:givens_example}.

\begin{figure}[!ht]
	\centering
	\begin{subfigure}{0.495\textwidth}
		\centering
		\includegraphics[width=\textwidth]{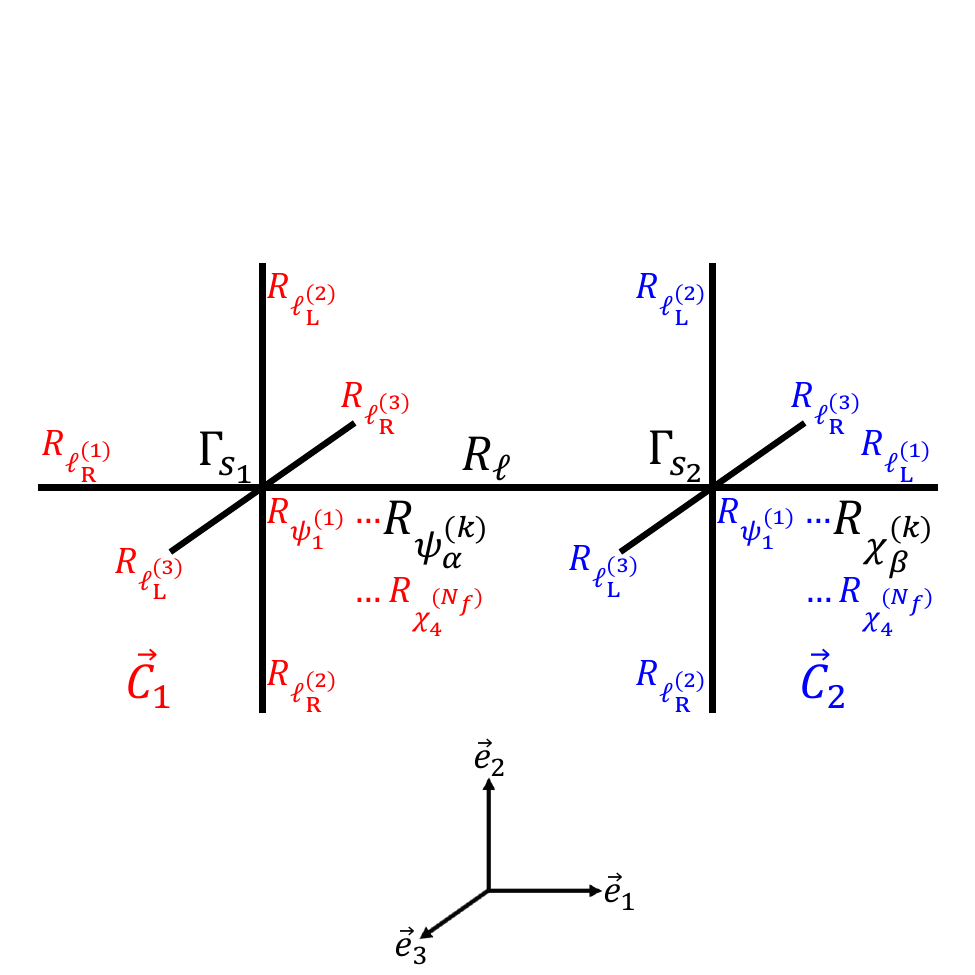}
		\caption{Physical link state}
	\end{subfigure}
	\begin{subfigure}{0.495\textwidth}
		\centering
		\includegraphics[width=\textwidth]{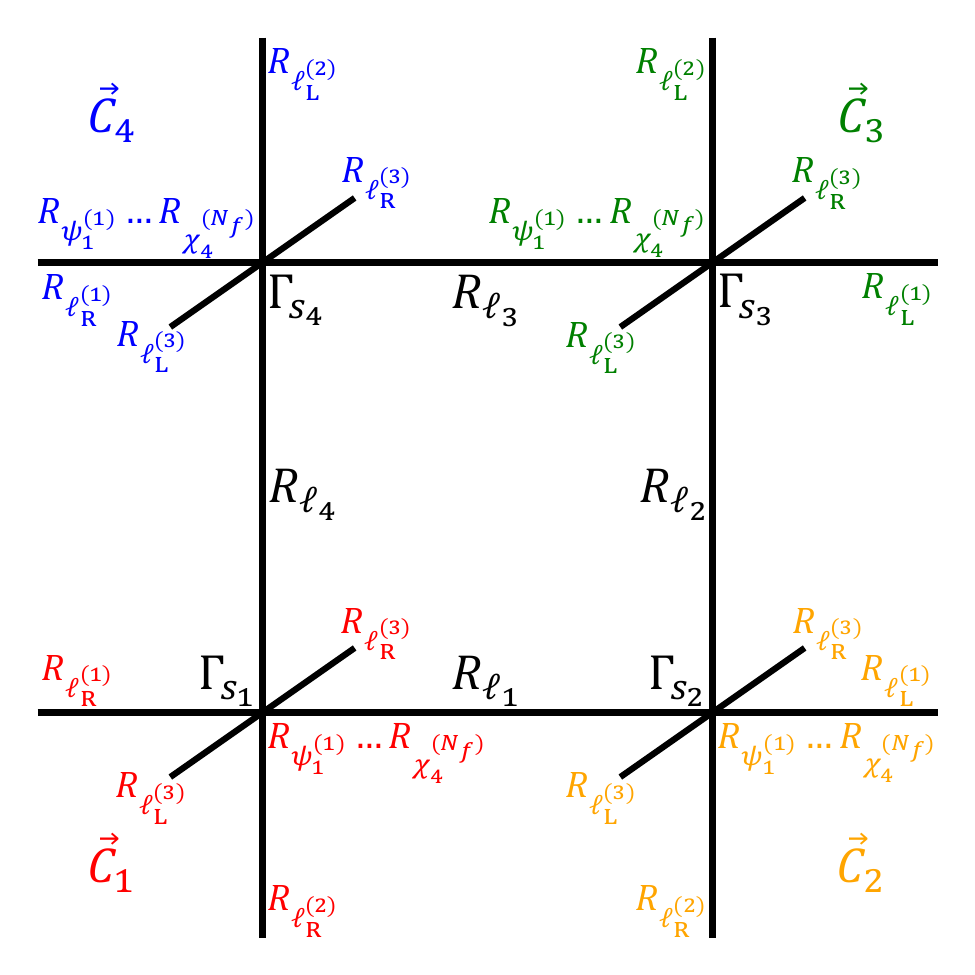}
		\caption{Physical plaquette state}
	\end{subfigure}
	\caption{Physical link and plaquette states. These are specified by assignments of irreps to the degrees of freedom meeting at each site of the link or plaquette, as well as singlet multiplicity indices for each site. $H_\text{kin}$ generates transitions between physical link states, and $H_{B+\theta}$ generates transitions between physical plaquette states. (a)  $H_\text{kin}$ will change the link irrep $R_\ell$, the multiplicity indices $\Gamma_{s_k}$, and two fermion irreps, all written in black font. (b) $H_{B+\theta}$ will change the plaquette link irreps $R_{\ell_k}$ and the multiplicity indices $\Gamma_{s_k}$, written in black font. All other irreps, written in colored font, remain unchanged upon action of the Hamiltonian. These irreps can be grouped site-by-site into vectors $\protect\vec{C}_k$ comprised of ``control links'' and other fermions.}
	\label{fig:link_plaquette_states}
\end{figure}

\subsubsection*{Plaquette Evolution}

The Hamiltonian operator $H_{B+\theta}$ changes a physical plaquette state, shown in Fig.~\ref{fig:link_plaquette_states} -- specifically, the irreps of four plaquette links and site singlet multiplicity indices. The notation $H_{B+\theta,k}$ contains a subscript $k$ that indexes a specific transition on a specific plaquette $P(\vec{s},\vec{e}_i,\vec{e}_j)$. The exponential $e^{-iH_{B+\theta,k} \Delta t}$ is a Givens rotation between the two physical plaquette states associated with this transition. The amplitude for this transition can be real (if only $H_B$ allows the transition), imaginary (if only $H_\theta$ allows the transition), or complex (if both $H_B$ and $H_\theta$ allow the transition).

\subsubsection*{Link Evolution}

\begin{figure}[!h]
	\centering
	\includegraphics{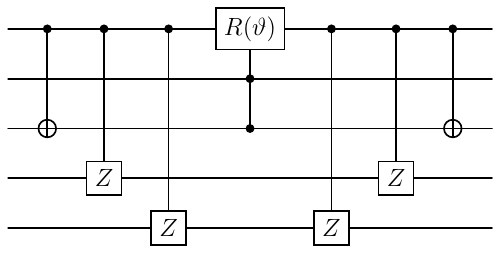}
	\caption{Example quantum circuit for a phased Givens rotation $e^{-ih}$, where $h = \frac{\vartheta}{2} \sigma^+ \pi^1 \sigma^- ZZ + \text{H.c.}$ for $\vartheta \in \mathbb{C}$. $h$ contains $Z$ operators that can be transformed into identity operators with C$Z$s connecting the pivot qubit with qubits phased by $Z$ gates. After applying the C$Z$s and the usual CNOTs of a Givens rotation, $h$ is transformed into $\tilde{h} = \frac{1}{2}(\vartheta \sigma^+ + \vartheta^* \sigma^-) \pi^1 \pi^1 I I$.}
	\label{fig:phased_givens_example}
\end{figure}

The kinetic Hamiltonian $H_\text{kin}$ changes a physical link state, shown in Fig.~\ref{fig:link_plaquette_states} -- specifically, the irreps of a link and fermions of a given spinor and flavor index on the sites connected by the link, together with site singlet multiplicity indices. The notation $H_{\text{kin},k}$ contains a subscript $k$ that indexes a specific transition on a specific link $\ell(\vec{s},\vec{e}_i)$. The exponential $e^{-iH_{\text{kin},k} \Delta t}$ is a Givens rotation between the two physical link states associated with this transition. The angle of the rotation can be either real or imaginary. A consequence of the Jordan-Wigner transformation is the need for phase flips applied to the rotation when fermionic states on sites ``between'' the sites of the link have odd color occupation number. This can be accounted for with C$Z$ gates as shown in Fig.~\ref{fig:phased_givens_example}.

\subsubsection*{Phase Shifts}

Physical basis states are eigenstates of $H_E$ and $H_\text{mass}$. $H_E$ multiplies gauge representation basis states by a quadratic Casimir eigenvalue. The subscript $k$ in $H_{E,k}$ indexes a specific link on the lattice and a specific irrep that link can have. The exponential $e^{-iH_{E,k} \Delta t}$ applies a phase on that link, conditioned on the irrep the link has. $H_\text{mass}$ multiplies fermionic representation basis states by the color occupation number. The subscript $k$ in $H_{\text{mass},k}$ indexes a fermion of a given spinor and flavor index on a specific site, in a particular irrep. The exponential $e^{-iH_{\text{mass},k} \Delta t}$ applies a phase on that fermion, conditioned on the irrep the fermion has.

\subsection{Results}

A full quantum simulation routine would include state preparation, time evolution, and the measurement of observables. A simulation of scattering would require wave packet preparation. A calculation of Wilson loop correlation functions in equilibrium would require ground state preparation. These examples are beyond the scope of this paper, which instead focuses on development of the site factor and matrix element formalism and exploratory noiseless circuit demonstrators, varying parameters and Hilbert space truncations. As such, none of the results shown here attempt to make serious statements of the physics of real-time QCD. Rather, they are intended to qualitatively sketch out what could become larger scale, carefully prepared and parameterized simulations in future work.

Without sophisticated state preparation algorithms and measurement protocols, working with a trivial initial state (all links in the trivial representation and all fermions unoccupied) denoted by $\ket{0}$ is the most convenient. At $g \gtrsim 1$ (a good long-term target for simulations) the overlap between this strong-coupling ground state and the true ground state is modest. All simulations in this paper use $g=2$ and $a=1$, deep in the strong-coupling regime. The remaining parameters $m_f$, $r$, and $\theta$ will be varied.

\begin{figure}
	\centering
	\includegraphics[width=\textwidth]{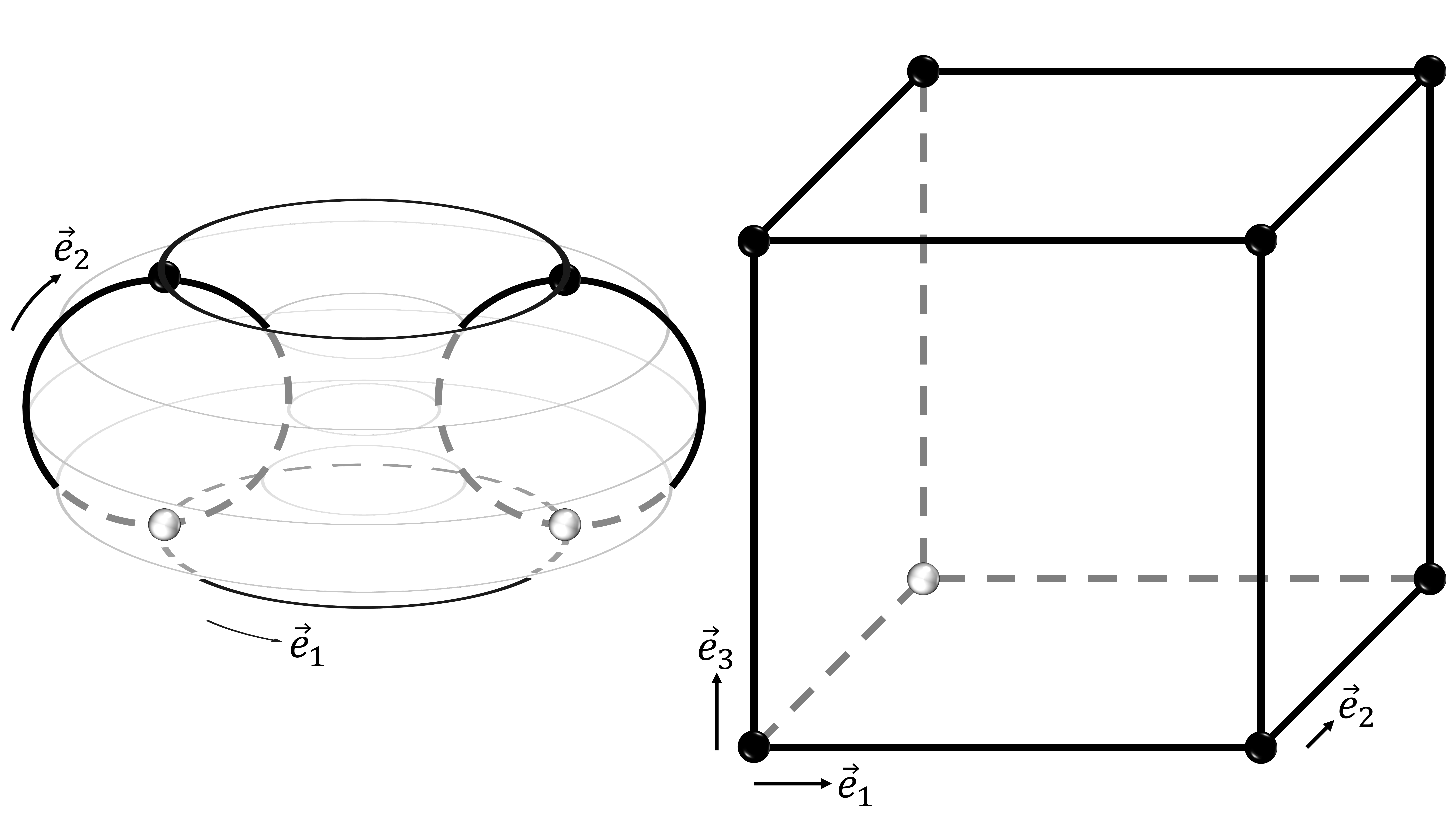}
	\caption{The $2 \times 2$ lattice with PBCs and the OBC cube used for simulations. Thick lines indicate lattice links and spheres correspond to lattice sites.}
	\label{fig:lattices}
\end{figure}

\begin{figure}[!ht]
	\centering
	\begin{subfigure}{0.495\textwidth}
		\centering
		\includegraphics[width=\textwidth]{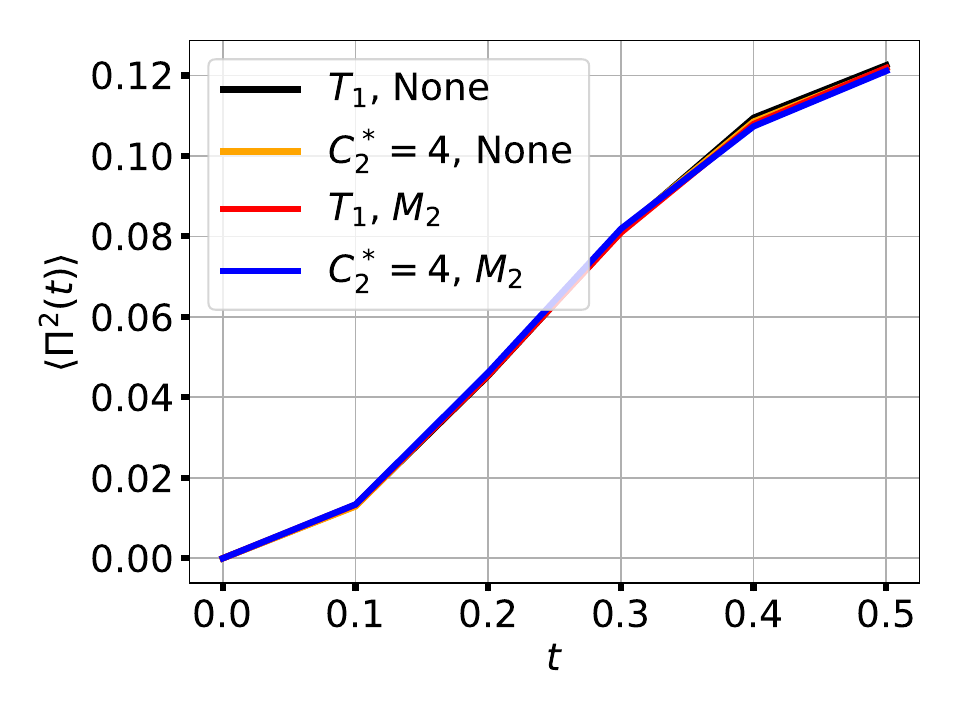}
		\caption{$\expect{\Pi^2(t)}$ with staggered fermions}
	\end{subfigure}
	\begin{subfigure}{0.495\textwidth}
		\centering
		\includegraphics[width=\textwidth]{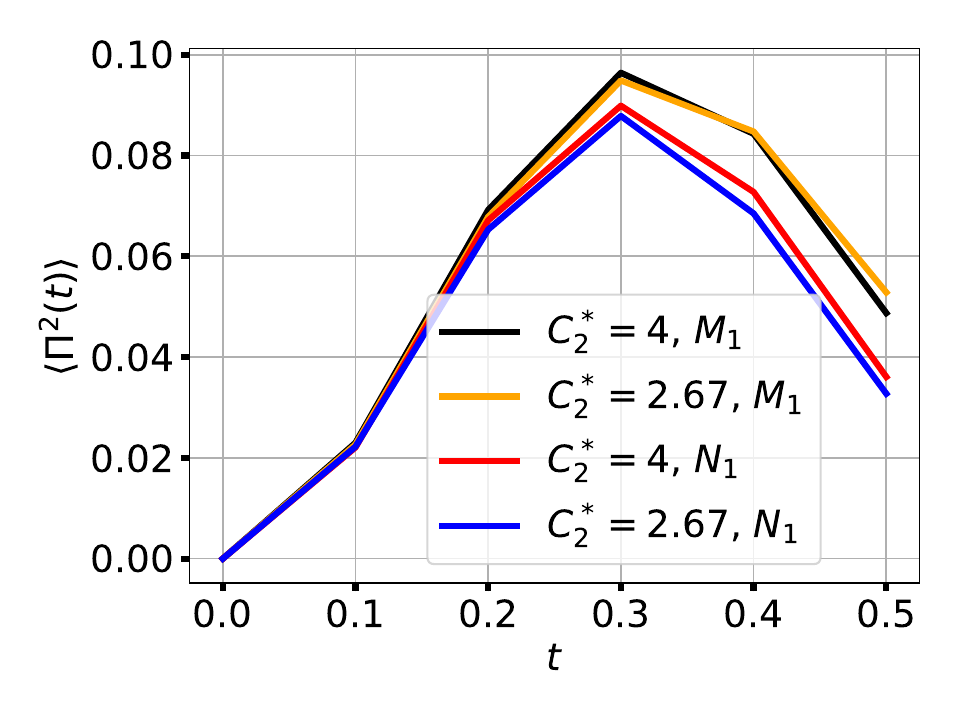}
		\caption{$\expect{\Pi^2(t)}$ with Wilson fermions}
	\end{subfigure}
	
	\begin{subfigure}{0.495\textwidth}
		\centering
		\includegraphics[width=\textwidth]{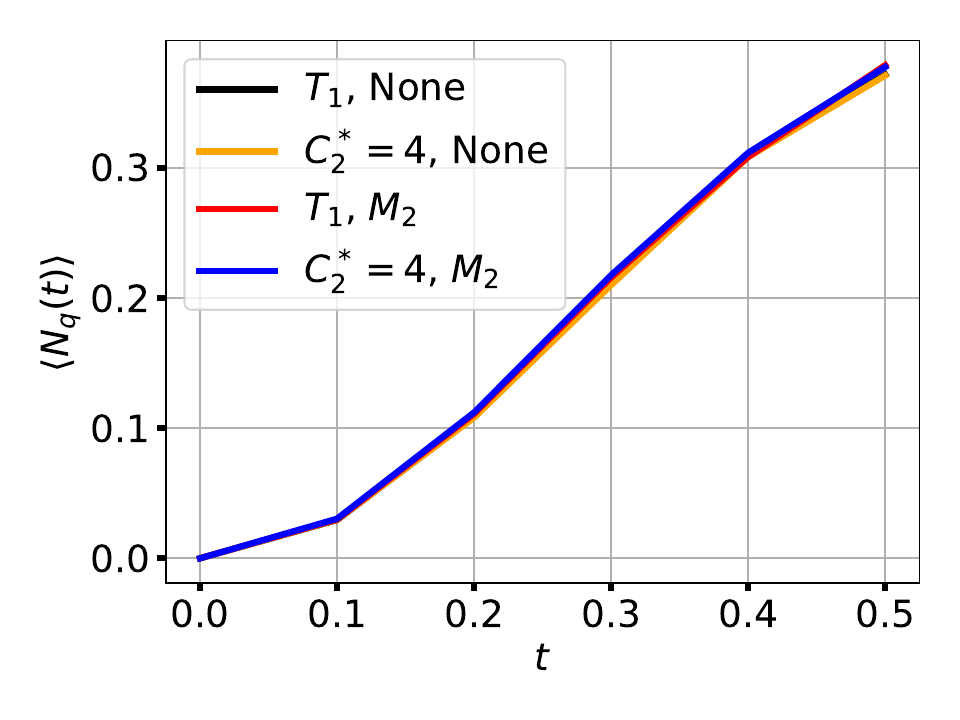}
		\caption{$\expect{N_q(t)}$ with staggered fermions}
	\end{subfigure}
	\begin{subfigure}{0.495\textwidth}
		\centering
		\includegraphics[width=\textwidth]{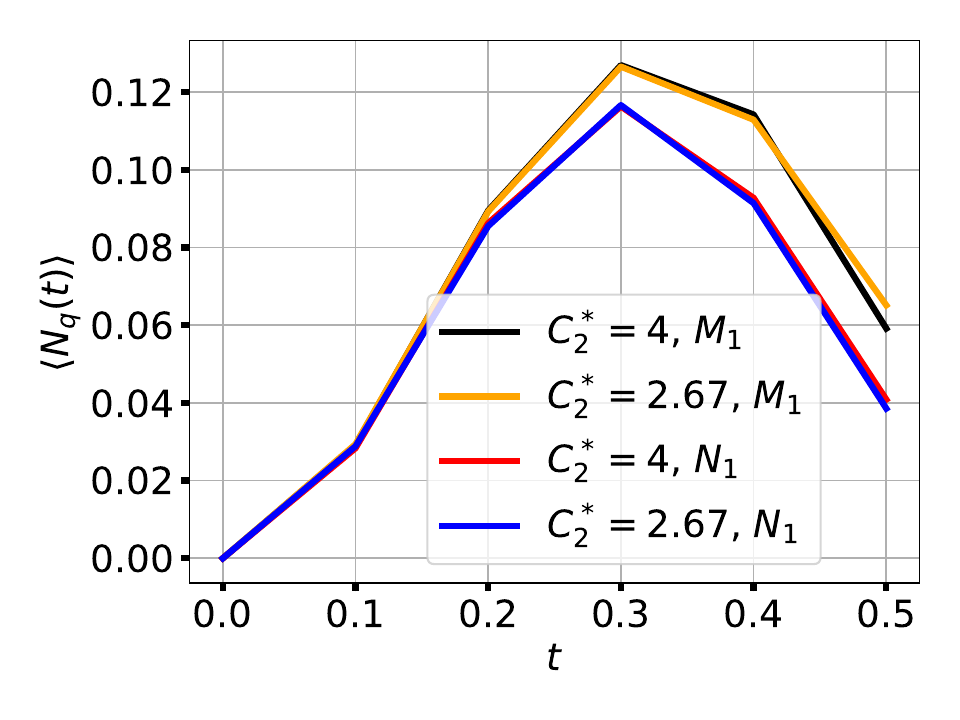}
		\caption{$\expect{N_q(t)}$ with Wilson fermions}
	\end{subfigure}
	\caption{Measurements of $\expect{\Pi^2(t)}$ and $\expect{N_q(t)}$ on the $2 \times 2$ PBC lattice with different Hilbert space truncations. Curves are labeled with ``gauge truncation, fermion truncation.'' For these simulations, $10^5$ shots were ran on Qiskit Aer's statevector simulator \cite{JavadiAbhariEtal:2024:qiskit} with up to 32 qubits and $\mathcal{O}(10^7)$ untranspiled gates. The simulation parameters used were $\Delta t = 0.1$, $N_f=1$, and $r=m_f=1$. (a) and (c) show results for staggered fermions. $T_1$ permits $\{ \mathbf{1}, \mathbf{3}, \bar{\mathbf{3}} \}$ on each link. $C_2^*=4$ forces one of the links meeting at a site to be in the trivial representation. ``None'' means no truncation was made to the fermion Hilbert space. $M_2$ excludes the fully occupied color singlet state from the fermion Hilbert space. (b) and (d) show results for Wilson fermions. $C_2^*=2.67$ forces two of the links meeting at a site to be in trivial representations. $M_1$ allows at most a particle-antiparticle pair at a site. $N_1$ allows only a single particle or antiparticle at a site.}
	\label{fig:2x2_basics}
\end{figure}

What remains to specify and vary are truncations to the link and fermion Hilbert spaces. It is instructive to see the effects of these truncations on observables. In particular, consider the quadratic Casimir eigenvalue of a link, averaged over the lattice:
\begin{equation}
	\expect{\Pi^2(t)} \equiv \bra{\psi(t)} \frac{\sum_{\vec{s}} \sum_{i} \Pi^{i,a}(\vec{s}) \Pi^{i,a}(\vec{s})}{N_\text{links}} \ket{\psi(t)}
\end{equation}
and the color occupation number of a fermion, averaged over the lattice:
\begin{equation}
	\expect{N_q(t)} \equiv \bra{\psi(t)} \frac{\sum_{\vec{s}} \psi^\dag(\vec{s}) \psi(\vec{s}) + \chi^\dag(\vec{s}) \chi(\vec{s})}{N_\text{sites} \times N_\text{flavors} \times N_\text{Dirac}} \ket{\psi(t)}
\end{equation}
where $\ket{\psi(t)} = U(t) \ket{0}$. (For staggered fermions, $\expect{N_q(t)}$ has no factor of $N_\text{Dirac}$ and the sum over sites is appropriately split into even and odd sites.) Fig.~\ref{fig:2x2_basics} shows these expectation values measured for staggered and Wilson fermions on the $2 \times 2$ lattice (see Fig.~\ref{fig:lattices}). For staggered fermions, none of the Hilbert space truncations used have a noticeable effect. The largest relative difference among the results is about 3.5\%, found in $\expect{N_q(t)}$. Using no fermion truncation or the modest $M_2$ truncation has little impact; strong coupling suppresses highly populated states anyway. For Wilson fermions, the truncations have more impact, with relative differences reaching about 50\% at later times. There is a clear division between the $M_1$ fermion truncation data and the more aggressive $N_1$ data.

\clearpage

\begin{figure}[!h]
	\centering
	\begin{subfigure}{0.495\textwidth}
		\centering
		\includegraphics[width=\textwidth]{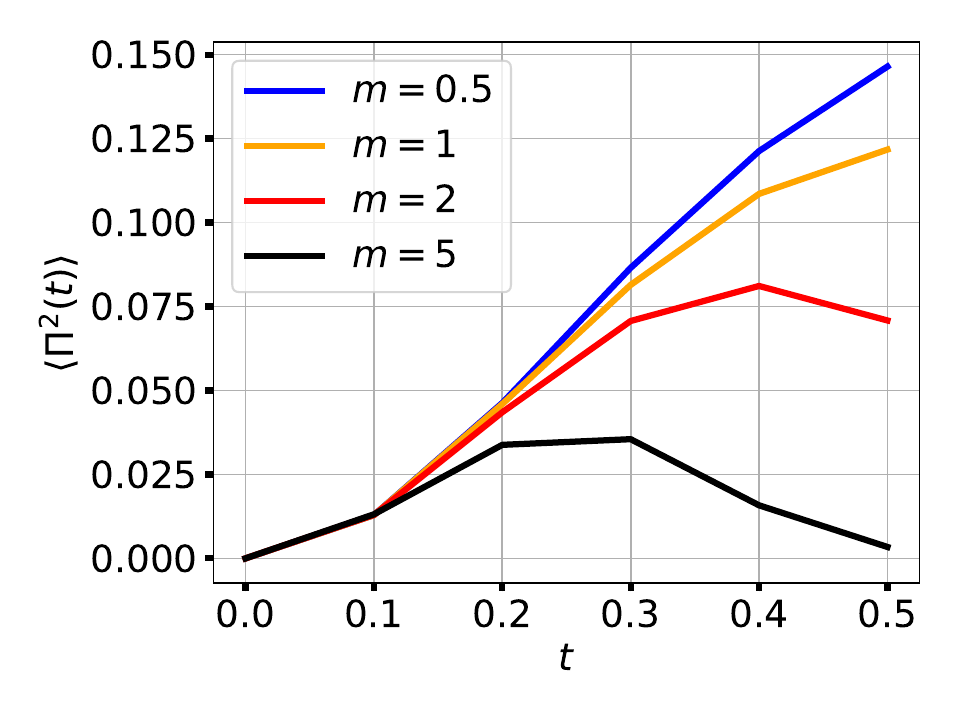}
		\caption{$\expect{\Pi^2(t)}$}
	\end{subfigure}
	\begin{subfigure}{0.495\textwidth}
		\centering
		\includegraphics[width=\textwidth]{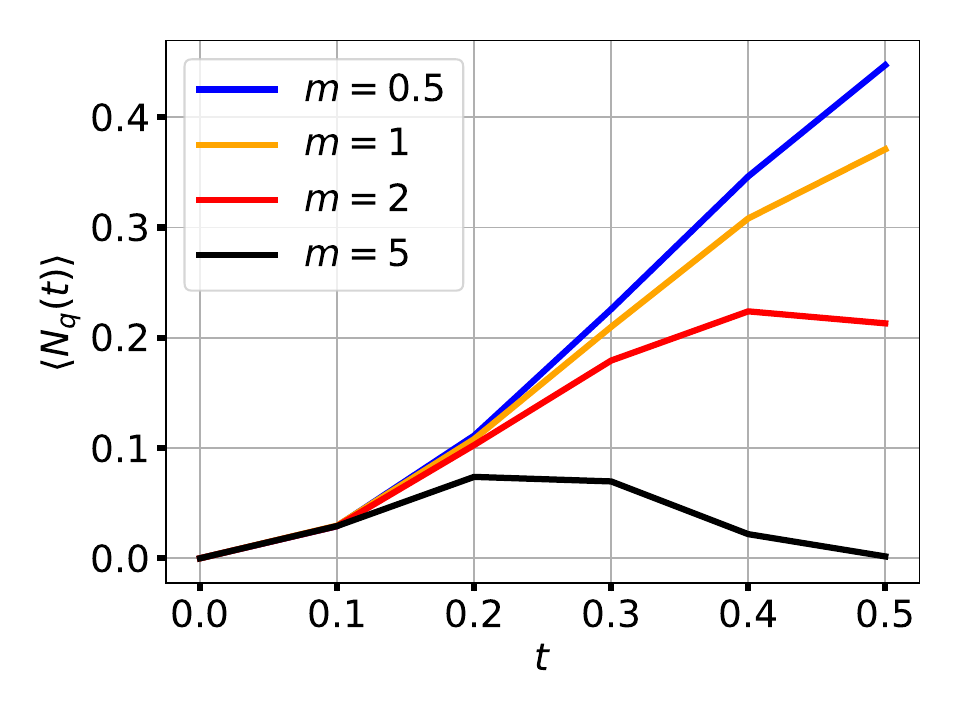}
		\caption{$\expect{N_q(t)}$}
	\end{subfigure}
	\caption{Measurements of $\expect{\Pi^2(t)}$ and $\expect{N_q(t)}$ on the $2 \times 2$ PBC lattice with different staggered fermion masses. The gauge truncation used here is $C_2^*=4$ and no truncation was made on the fermion Hilbert space. For these simulations, $10^5$ shots were ran on Qiskit Aer's statevector simulator with 28 qubits and $\mathcal{O}(10^5)$ untranspiled gates. The simulation parameters used were $\Delta t = 0.1$ and $N_f=1$. Each curve is labeled by the fermion mass parameter.}
	\label{fig:2x2_masses}
\end{figure}

Next we vary the fermion mass, which contributes phase shifts during the time evolution. Fig.~\ref{fig:2x2_masses} shows the effect of varying the mass of a staggered fermion on the $2 \times 2$ lattice. Larger masses are seen to suppress oscillation into fermion-rich states. Intuitively, a lighter fermion has a smaller energy denominator and is easier to produce than a heavier one.

\begin{figure}[!h]
	\centering
	\begin{subfigure}{0.495\textwidth}
		\centering
		\includegraphics[width=\textwidth]{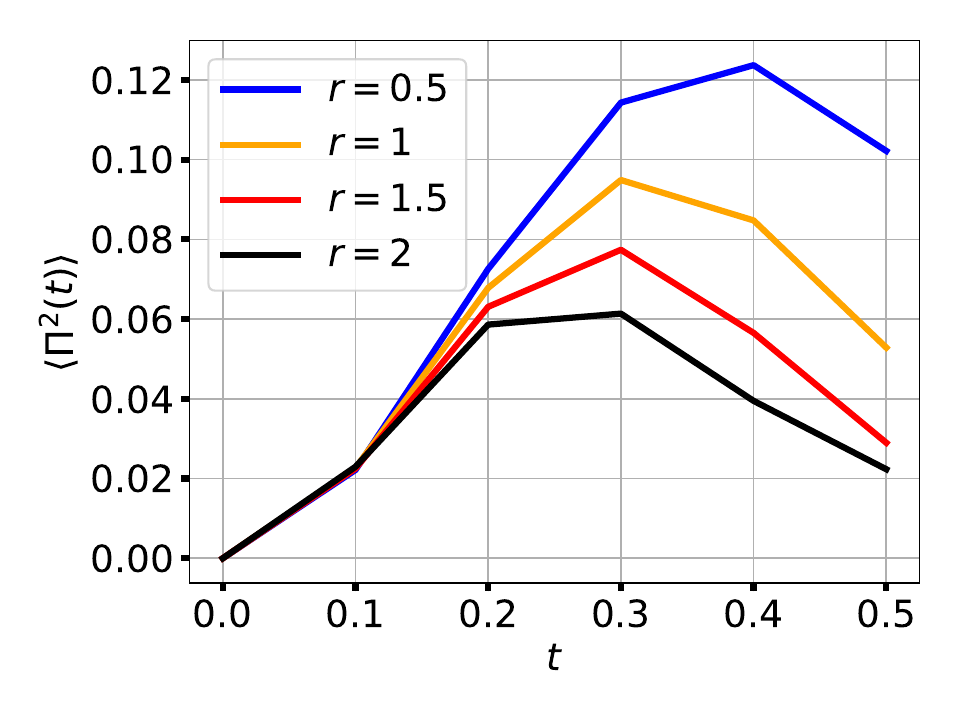}
		\caption{$\expect{\Pi^2(t)}$}
	\end{subfigure}
	\begin{subfigure}{0.495\textwidth}
		\centering
		\includegraphics[width=\textwidth]{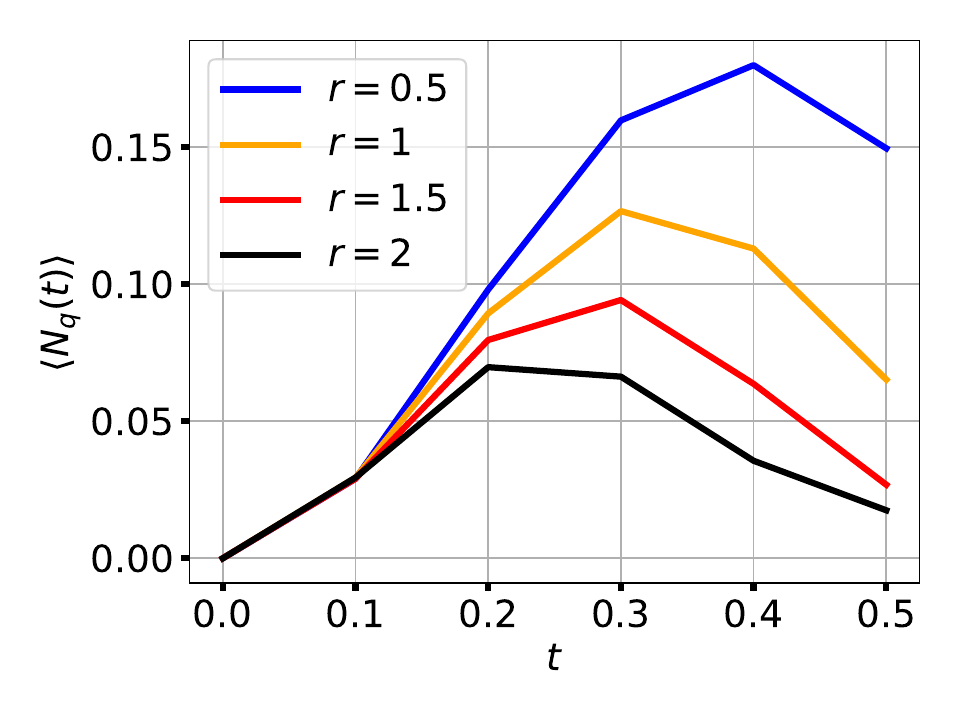}
		\caption{$\expect{N(t)}$}
	\end{subfigure}
	\caption{Measurements of $\expect{\Pi^2(t)}$ and $\expect{N_q(t)}$ on the $2 \times 2$ lattice with different Wilson parameters. The gauge truncation used here is $C_2^*=2.67$ and the fermion truncation was $M_1$. For these simulations, $10^5$ shots were ran on Qiskit Aer's statevector simulator with 28 qubits and up to $\mathcal{O}(10^5)$ untranspiled gates. The simulation parameters used were $\Delta t = 0.1$, $N_f=1$, and $m_f=1$. Each curve is labeled by the Wilson parameter.}
	\label{fig:2x2_wilson_term}
\end{figure}

The Wilson parameter contributes to phase shifts during time evolution in the mass term, but it also parameterizes the Wilson kinetic term. By comparing the coefficients $\frac{r}{2a}$ of the Wilson term and $\frac{i}{2a}$ of the usual kinetic term, it is reasonable to set ${r=1}$. We can also check how different choices of $r$ affect time-dependent observables. This is shown in Fig.~\ref{fig:2x2_wilson_term} for the $2 \times 2$ lattice. The effects of different Wilson parameters are comparable to those of different fermion masses above. Because the fermion mass is effectively $m + 2r$, a larger value of $r$ yields characteristically lower expectation values.

\begin{figure}[!h]
	\centering
	\includegraphics[width=\textwidth]{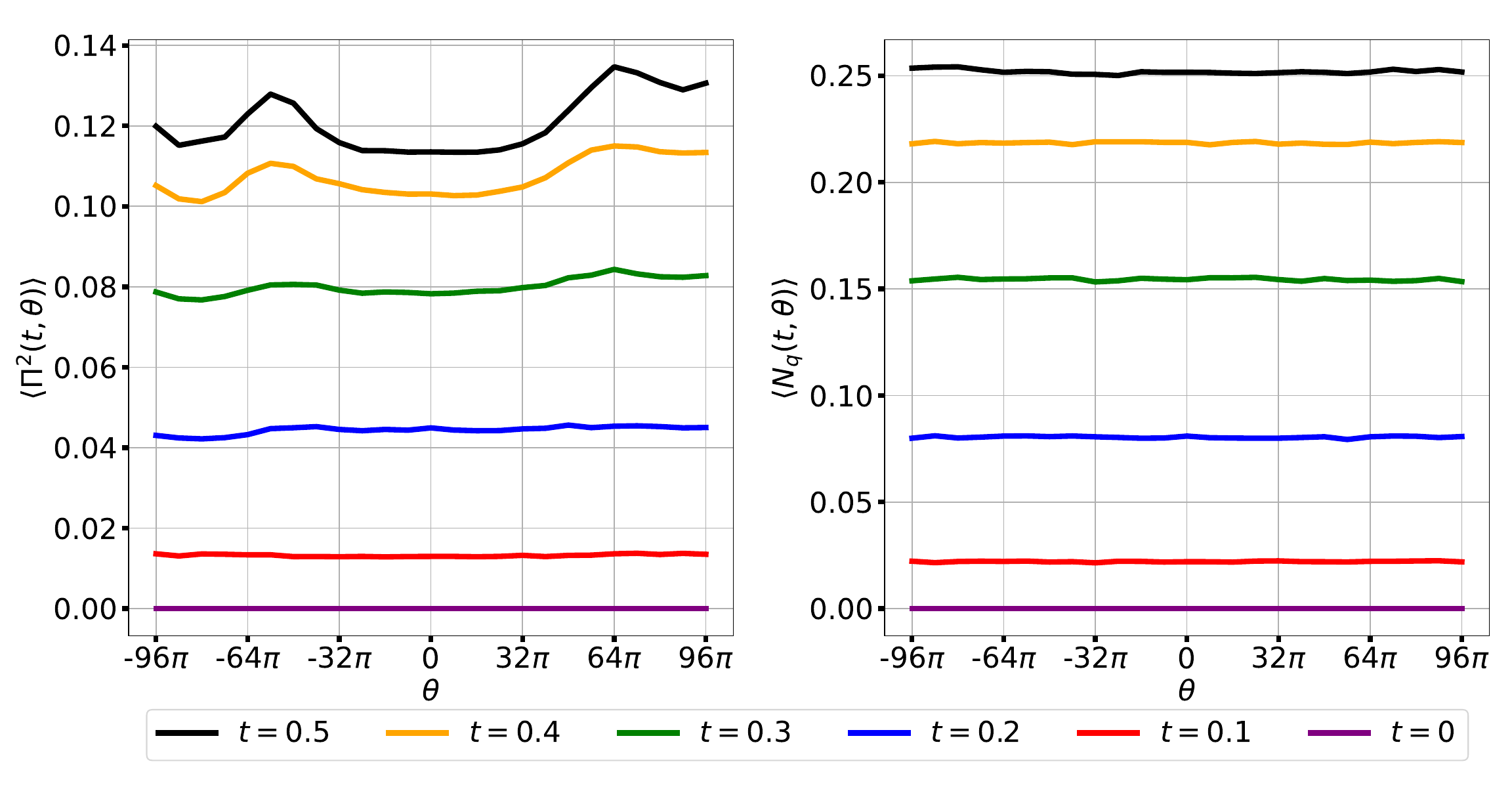}
	\caption{Measurements of $\expect{\Pi^2(t)}$ and $\expect{N_q(t)}$ on the cube with varying theta angle. The gauge truncation used here is $C_2^*=2.67$ and the fermion truncation was $M_1$. For these simulations, $10^5$ shots were ran on CUDA-Q's statevector simulator \cite{Kim:2025:cudaq} with 32 qubits and up to $\mathcal{O}(10^4)$ untranspiled gates. The simulation parameters used were $\Delta t = 0.1$, $N_f=1$, and $m_f=1$. Each curve is labeled by a different time $t$.}
	\label{fig:theta_gauge}
\end{figure}

Finally, we vary the theta angle \cite{KanEtal:2021:theta_term, LewisEtal:2026:su2_cube}. The lattice theory is not necessarily periodic in $\theta$, since topological charge quantization is not preserved, although the time reversal (T) invariance of the Hamiltonian does grant a symmetry $\text{T}H(\theta)\text{T} = H(-\theta)$. Fig.~\ref{fig:theta_gauge} probes this aperiodicity using observables for a staggered fermion on a cube. The aperiodicity is more pronounced for $\expect{\Pi^2(t)}$, whereas $\expect{N_q(t)}$ is mostly independent of $\theta$ in the strong coupling (coarse lattice) regime of the simulation. (See App.~\ref{app:the_theta_angle} for an alternative implementation of the theta angle where both link and fermion dynamics are modified, and a brief discussion on why Fig.~\ref{fig:theta_gauge} and Fig.~\ref{fig:theta_fermion} differ.) Variations with $\theta$ in $\expect{\Pi^2(t)}$ begin to appear at $\theta \approx 16\pi$, which is far from the $2\pi$ periodicity found in the continuum.

Eventually, we would like to prepare hadronic bound states of quarks and gluons in quantum simulations \cite{AtasEtal:2023:tetra_pentaquarks, FarrellEtal:2023:preparations_1+1_qcd, Schuhmacher:2025:observation_hadron_scattering}. Hadronic states are built on the lattice using interpolating operators that share the same quantum numbers as the target states. For example, minimal interpolating operators for charged pions $\pi^\pm$ in $2+1$D may be written as\footnote{For simplicity, the example interpolating operator is for Wilson-like fermions; the treatment for staggered fermions involves accounting for gauge invariance and taste with the spin-taste basis \cite{DeGrandDetar:2006:lattice_methods}.}
\begin{equation}
	\pi^+ = \frac{1}{\sqrt{3}} \psi_c^{\dag(u)} \chi_c^{\dag(d)} \qquad \pi^- = \frac{1}{\sqrt{3}} \chi_c^{\dag(u)} \psi_c^{\dag(d)}
\end{equation}
Applying these operators on the zero-point state creates a superposition of fermionic representation basis states:
\begin{IEEEeqnarray*}{rCl}
	\ket{\pi^+} & \equiv & \pi^+ \ket{0} = \sum_{c_1=1}^{3} \sum_{c_2=1}^{3} \phi(c_2) \braket{\mathbf{1}}{(\mathbf{3}^{(u)}, c_1) \otimes (\mathbf{\bar{3}}^{(\bar{d})}, c_2)} \ \ket{(\mathbf{3}^{(u)}, c_1) \otimes (\mathbf{\bar{3}}^{(\bar{d})}, c_2)} \\
	\ket{\pi^-} & \equiv & \pi^- \ket{0} = \sum_{c_1=1}^{3} \sum_{c_2=1}^{3} \phi(c_1) \braket{\mathbf{1}}{(\mathbf{\bar{3}}^{(\bar{u})}, c_1) \otimes (\mathbf{3}^{(d)}, c_2)} \ \ket{(\mathbf{\bar{3}}^{(\bar{u})}, c_1) \otimes (\mathbf{3}^{(d)}, c_2)} \yesnumber
\end{IEEEeqnarray*}
where the magnitudes of the CGCs are equal to $\frac{1}{\sqrt{3}}$. These are physical basis states -- as expected, because the interpolating operators are gauge invariant. In the local encoding, these mesonic states require simple qubit flips (with $X$ gates) to prepare.

\begin{figure}[!h]
	\centering
	\includegraphics[width=\textwidth]{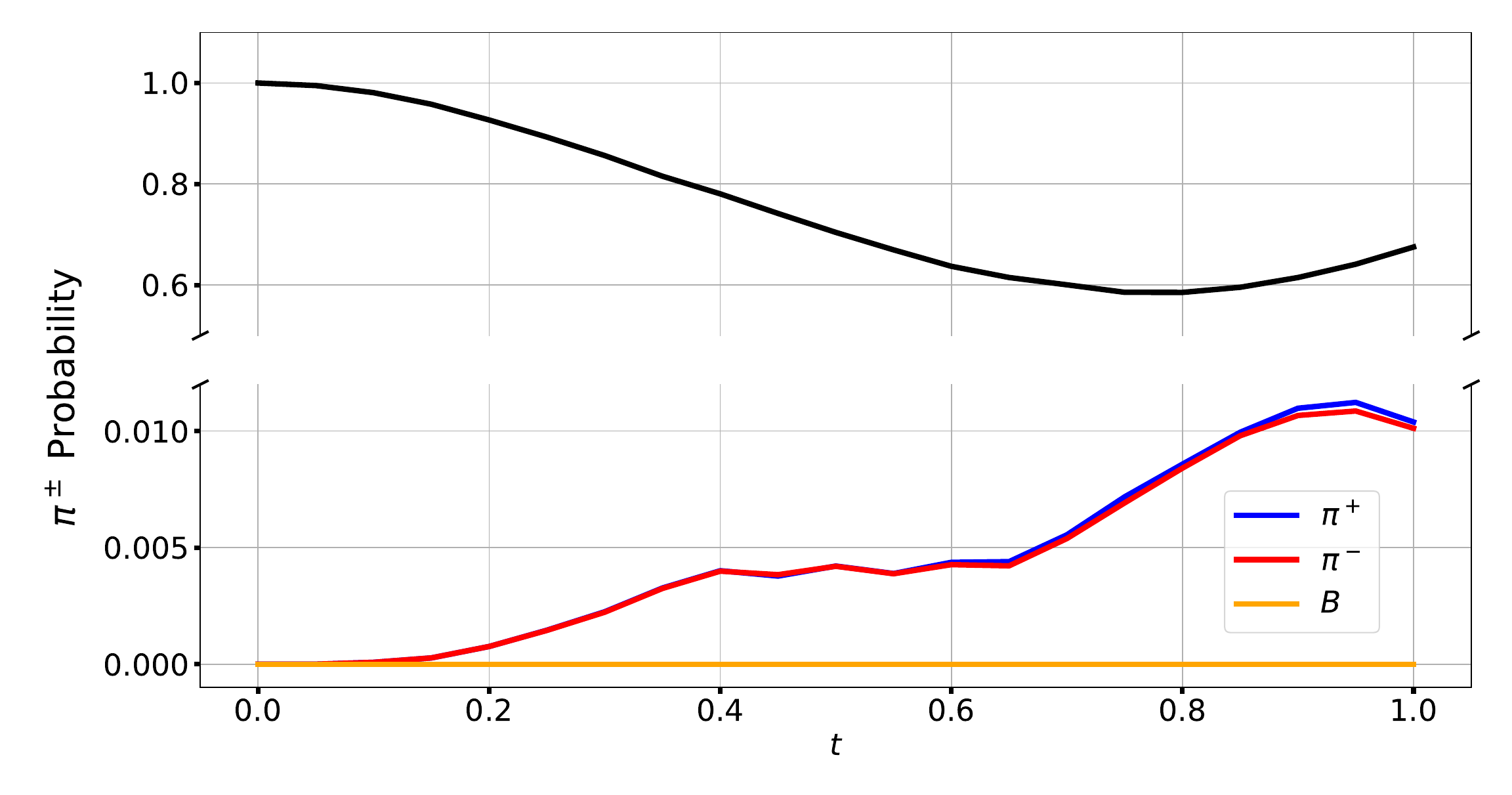}
	\caption{Detections of ``charged pions'' on the $2 \times 2$ lattice with $N_f=2$ Wilson fermions. The gauge truncation used here is $C_2^*=1.33$, which allows at most one link at a site to be in the $\mathbf{3}$ or $\bar{\mathbf{3}}$ representations. The fermion truncation used was $[N_1^{(u)}, N_1^{(d)}]$, where there can be only one particle or antiparticle of a given flavor on a site. $10^5$ shots were ran on CUDA-Q's statevector simulator with 32 qubits and up to $\mathcal{O}(10^5)$ untranspiled gates. The simulation parameters used were $\Delta t=0.05$ and $r=m_u=m_d=1$. Top Half: Probability for a $\pi^+$ on the site $(0,0)$. Bottom Half: Probability for a $\pi^\pm$ on the other three sites, and a constant line that indicates baryon number $B$ was conserved in all the shots data.}
	\label{fig:pi_plus}
\end{figure}

Fig.~\ref{fig:pi_plus} plots the evolution of the $2 \times 2$ lattice initialized with a $\pi^+$ at site $(0,0)$. The probabilities to find a $\pi^+$ on the original site, and $\pi^\pm$ on the other three sites, are shown. Due to statevector qubit constraints, an aggressive, minimal truncation was used. One mode of $\pi^\pm$ creation on other sites is via two applications of the kinetic term on a link, thereby generating a $\pi^\pm$ pair. This is reflected by the probabilities for $\pi^+$ and $\pi^-$ being approximately equal. However, due to strong coupling and truncation, these probabilities grow slowly. Instead states obtained from single applications of a kinetic term on a link dominate in probability, especially at earlier times.

\begin{figure}[!h]
	\centering
	\begin{subfigure}{0.24\textwidth}
		\centering
		\includegraphics[width=\textwidth]{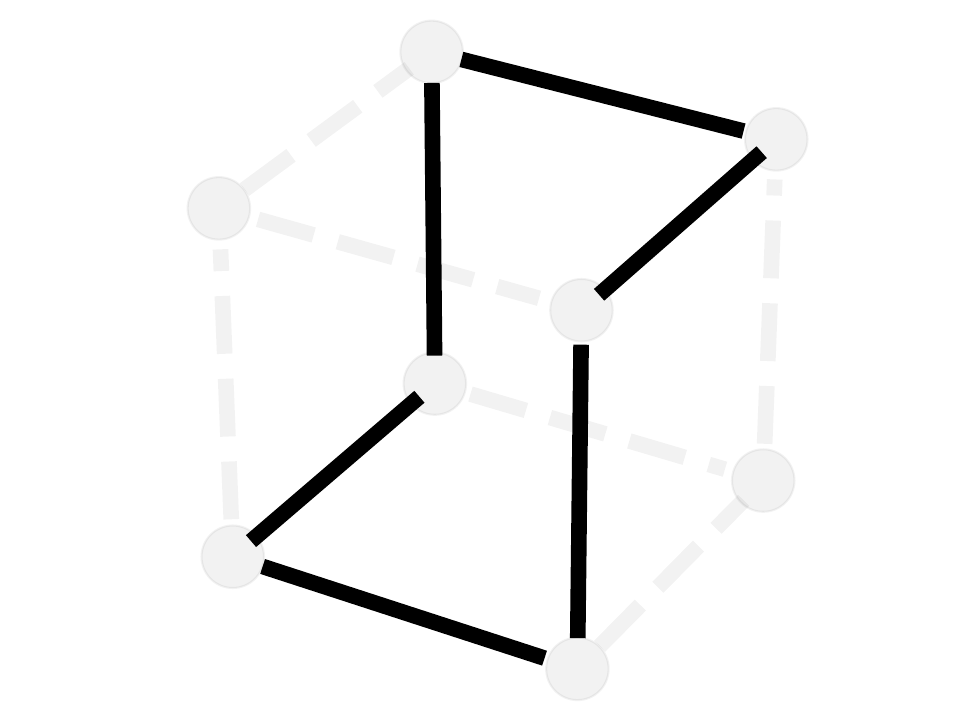}
		\caption{$t=0$}
	\end{subfigure}
	\begin{subfigure}{0.24\textwidth}
		\centering
		\includegraphics[width=\textwidth]{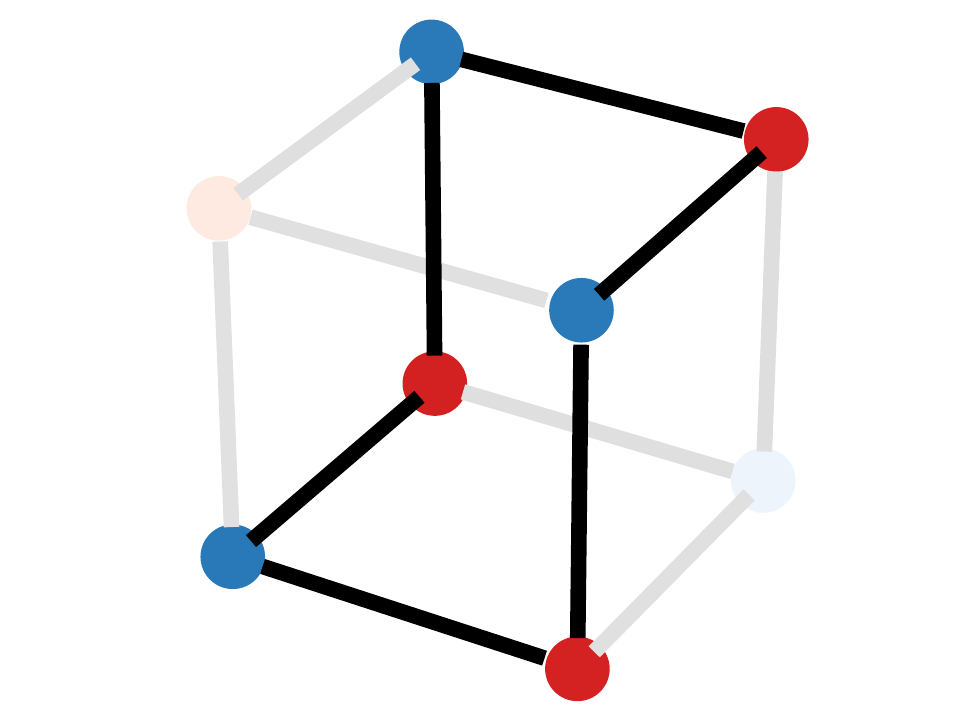}
		\caption{$t=0.5$}
	\end{subfigure}
	\begin{subfigure}{0.24\textwidth}
		\centering
		\includegraphics[width=\textwidth]{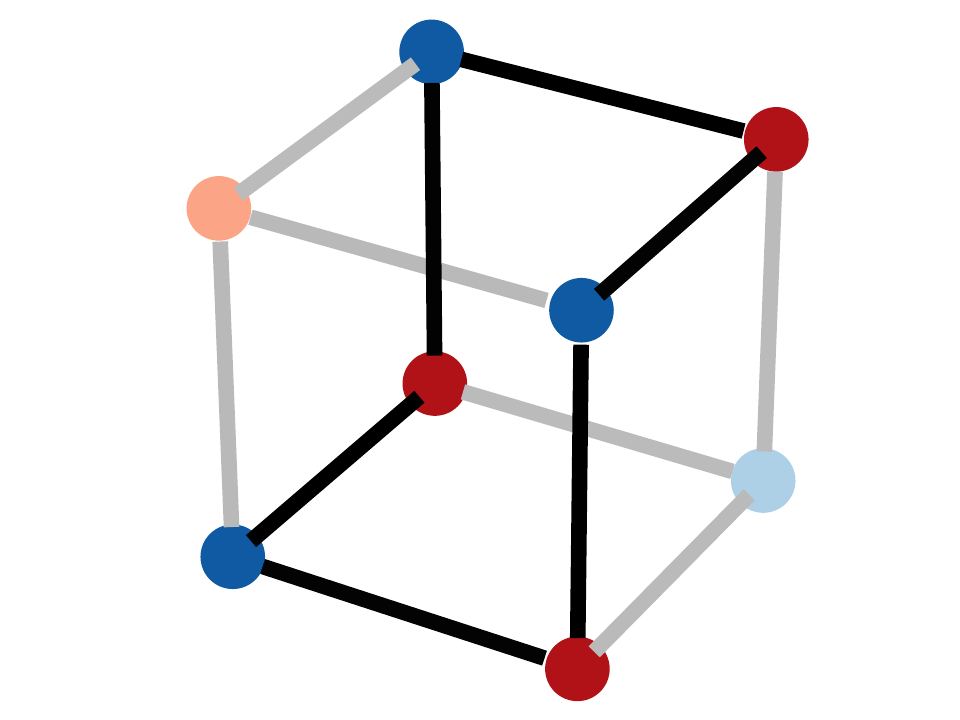}
		\caption{$t=1$}
	\end{subfigure}
	\begin{subfigure}{0.24\textwidth}
		\centering
		\includegraphics[width=\textwidth]{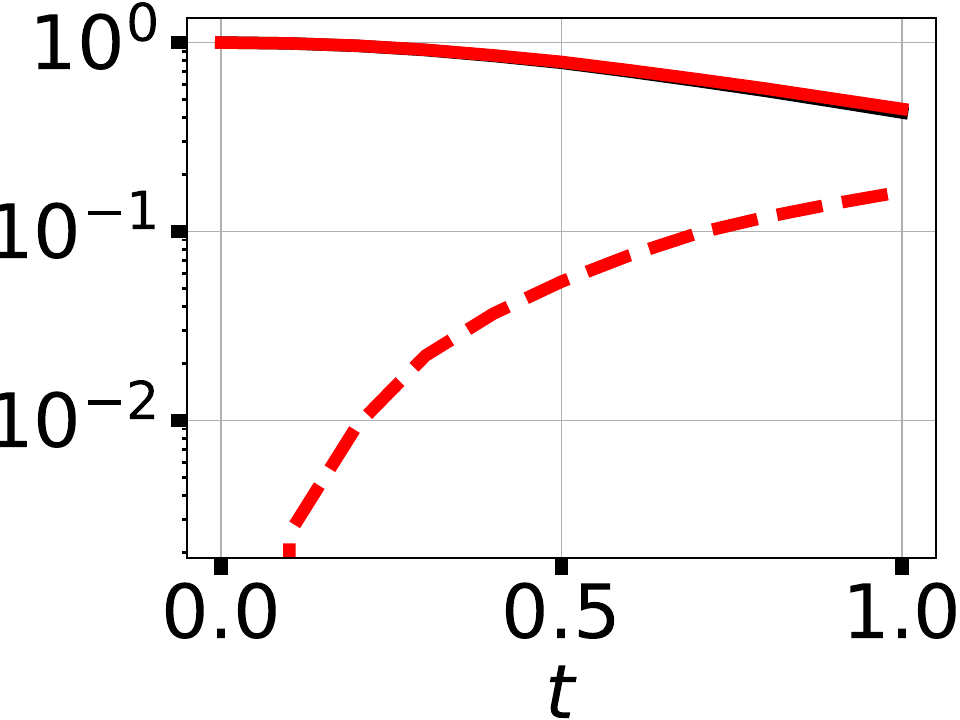}
		\caption{Loop string}
	\end{subfigure}
	
	\begin{subfigure}{0.24\textwidth}
		\centering
		\includegraphics[width=\textwidth]{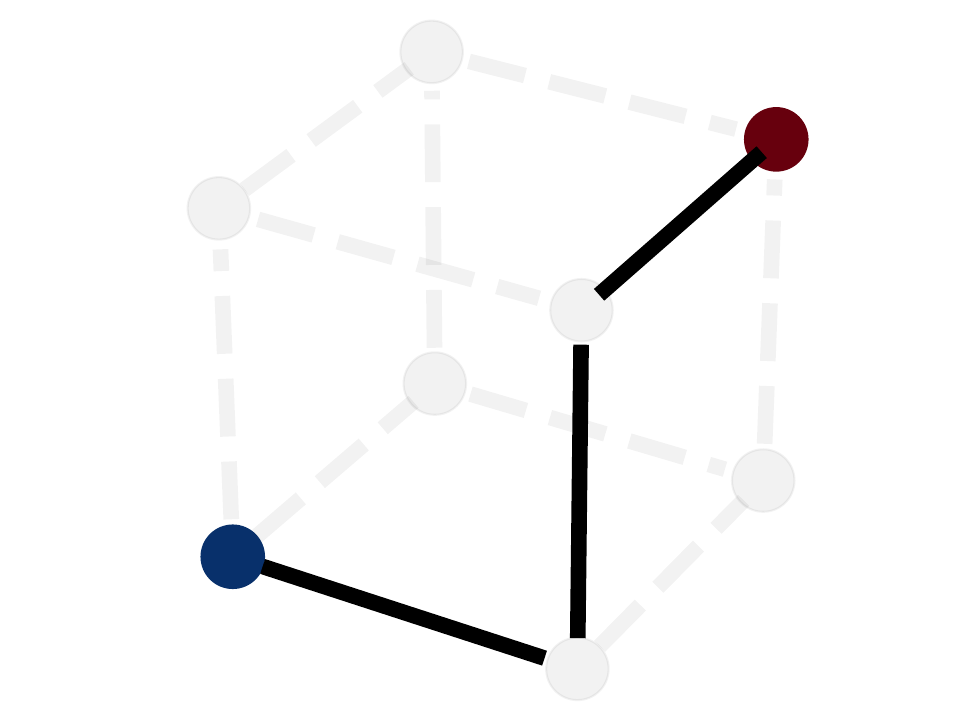}
		\caption{$t=0$}
	\end{subfigure}
	\begin{subfigure}{0.24\textwidth}
		\centering
		\includegraphics[width=\textwidth]{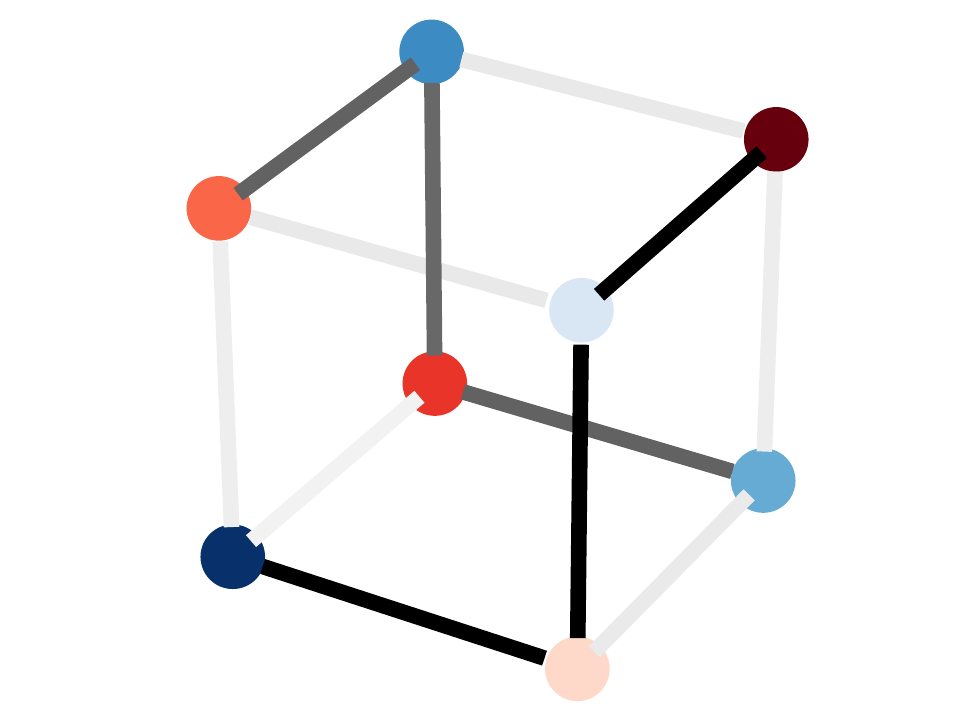}
		\caption{$t=0.5$}
	\end{subfigure}
	\begin{subfigure}{0.24\textwidth}
		\centering
		\includegraphics[width=\textwidth]{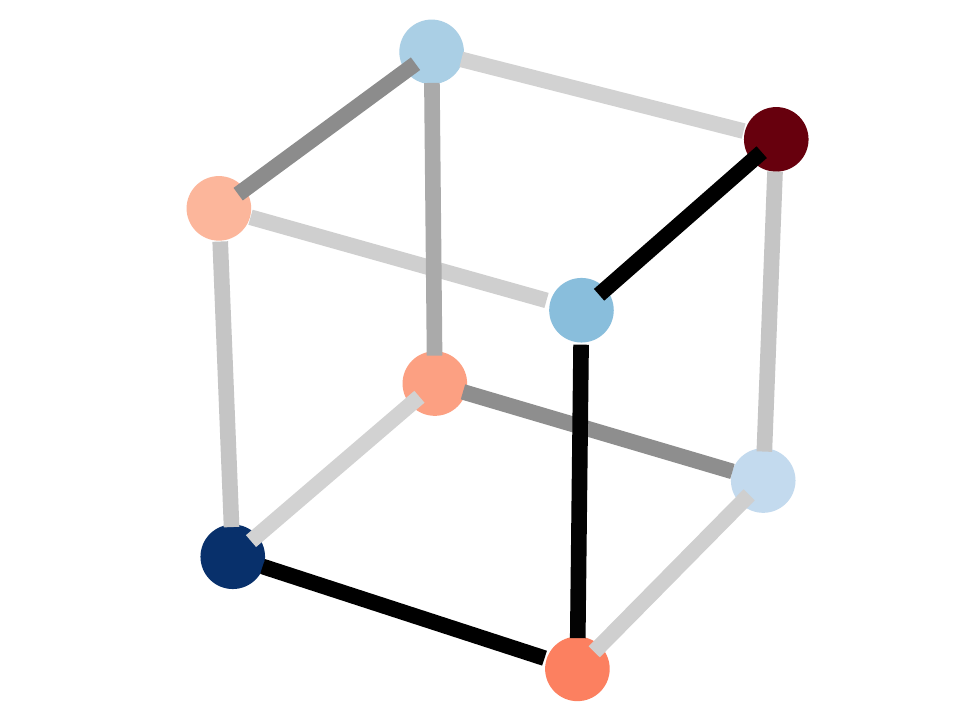}
		\caption{$t=1$}
	\end{subfigure}
	\begin{subfigure}{0.24\textwidth}
		\centering
		\includegraphics[width=\textwidth]{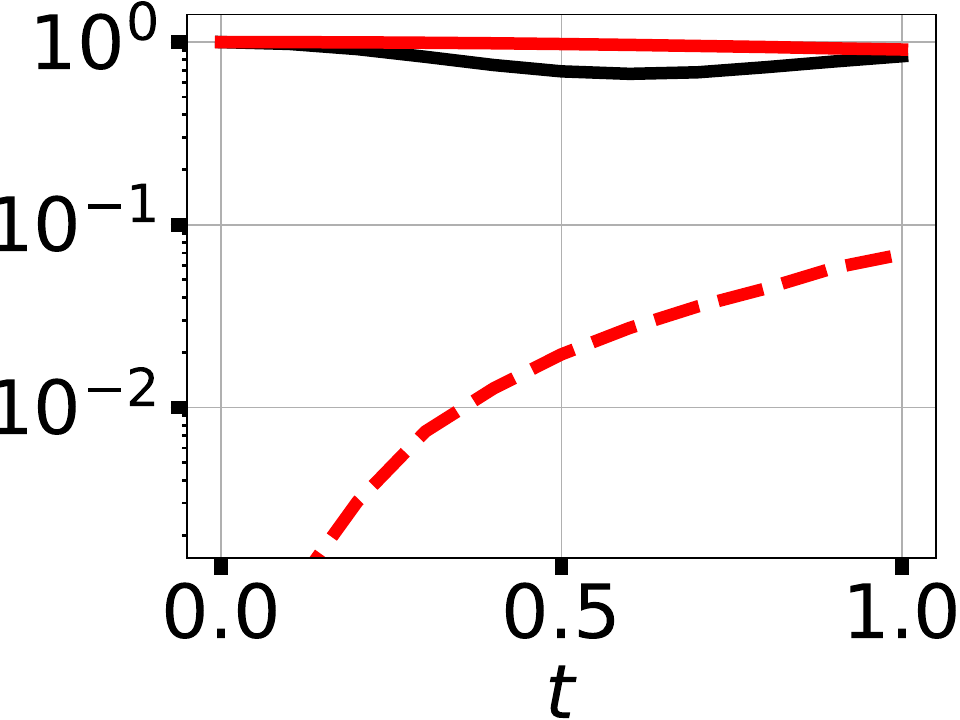}
		\caption{Fermi string}
	\end{subfigure}
	\caption{String breaking for two kinds of strings on a cube with a staggered fermion. The gauge truncation used here is $C_2^* = 2.67$ and the fermion truncation is $M_1$. For these simulations, $10^5$ shots were ran on CUDA-Q's statevector simulator with 32 qubits and up to $\mathcal{O}(10^4)$ untranspiled gates. The simulation parameters used were $\Delta t = 0.1$, $N_f=1$, and $m_f=1$. (a)-(c) and (e)-(g) show snapshots of a string at different times. Darker colors correspond to higher $\Pi^2$ and $N_q$ expected values on links and sites, respectively. (d) and (h) Probabilities of the original string (solid black line), the unbroken string (solid red line), and the string breaking (dashed red line).}
	\label{fig:strings}
\end{figure}

Finally, among the richer dynamics to study in lattice gauge theory is string breaking \cite{CataldiOrlandoHalimeh:2025:nonabelian_string_dynamics, Gupta:2026:LSH_string_breaking, JoshiEtal:2026:genuine_string_breaking, GrieningerSavageZemlevskiy:2026:complexity_string_breaking}. A string can be, for example, a closed loop of chromoelectric flux (loop string), or a particle-antiparticle pair connected through flux (fermi string). These string states are also straightforward to prepare in the local encoding with qubit flips, populating fermions or changing the irreps on links to create flux. To model string breaking, Fig.~\ref{fig:strings} illustrates the evolution of these strings on the cube with a staggered fermion.

The loop string, shown in the top row of Fig.~\ref{fig:strings}, is most easily changed by an application of the kinetic term on one of its links. This either changes the irrep of the link (via $\mathbf{3} \otimes \mathbf{3} \to \mathbf{\bar{3}}$), and excites the string with a particle-antiparticle pair, or it ``breaks'' the string (via $\mathbf{3} \otimes \mathbf{\bar{3}} \to \mathbf{1}$). (The first case is not counted as either an unbroken or broken string.) Over time, the string can be broken into multiple fermi strings, which is reflected by the growing expected $N_q$ values on the sites along the string. At the same time, a single application of the plaquette operator can wiggle the string, and two applications can stretch it, or even shrink it into the space of a single plaquette. Each of these are counted as unbroken configurations of the string.

The fermi string, shown in the bottom row of Fig.~\ref{fig:strings}, only occupies part of the lattice. The other part is quickly excited by one-link fermi strings, as seen from the growing $\langle\Pi^2\rangle$. The original fermi string is most easily broken at its middle ``weakest'' link, as seen by the growing $N_q$ expected value on the sites connected by that link. Similar to the loop string, the fermi string can be wiggled and stretched. Not represented by the plots of Fig.~\ref{fig:strings} are fusions of the fermi string, whereby the original fermi string fuses with one or more one-link fermi string excitations. 

\section{Exact Diagonalization}
\label{sec:exact_diagonalization}

For very small lattices and low truncations, the formalism developed here can also be used to perform classical exact diagonalization (ED) analysis. One application of ED is as a check of the quantum circuits for small simulations. We have performed spot-checks of the noiseless simulation results above and obtained good agreement with ED across the simulated time range, indicating low Trotterization errors.

\begin{figure}[!h]
	\begin{center}
		\includegraphics[width=0.975\textwidth]{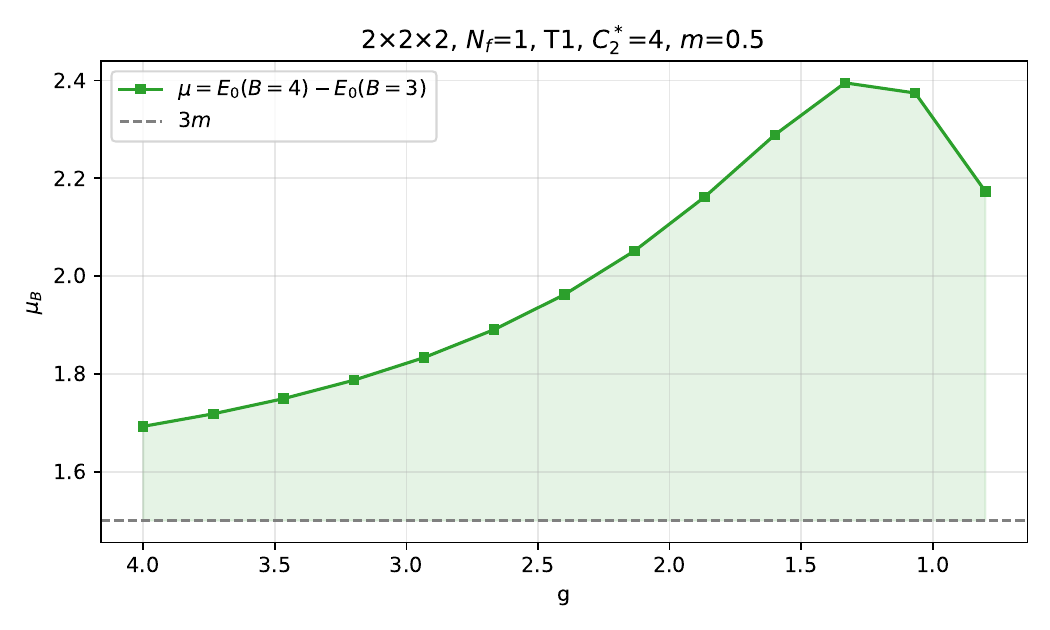}
		\caption{Chemical potential at $B=3-4$ on the single cube, OBC, $T_1$ gauge truncation, $N_f=1$ staggered fermion lattice, obtained via exact diagonalization.}
		\label{fig:muB}
	\end{center}
\end{figure}

A second use of ED is as an exploratory tool for testing other observables. Although we have focused on time evolution circuits in the main body of this work, a second major target of quantum simulations is equilibrium QCD at high baryon density, where classical Monte Carlo suffers a sign problem from the fermion determinant. Fig.~\ref{fig:muB} shows one toy illustration of how one might like to apply our Hamiltonian formulation to this problem. In it we compute the ground state energy cost of adding a baryon to the single-cube $N_f=1$ staggered OBC lattice, starting in the baryon number $=3$ sector, as a function of the bare coupling. The energy cost is always greater than $3m$. This treatment is of course highly contaminated by lattice and truncation artifacts, but it serves as a proof of principle and starting point for more deliberate studies of three-dimensional QCD at high baryon density in the Hamiltonian formalism.

\section{Conclusion}
\label{sec:conclusion}

We hope this work will serve as a useful contribution to the ongoing and long-term effort to develop practical quantum simulations of fundamental physics. The present limitation is on the hardware side; with early fault-tolerant quantum hardware we anticipate that the noiseless simulations performed in this work could be carried out on hardware and pushed to larger lattice sizes and less aggressive truncations. However, there are also clear limitations that will require further algorithmic development. While we wait for the future of quantum hardware, we conclude this paper by suggesting directions for future theoretical work that were not highlighted in the main text.

Developing improved Hamiltonians will be of high value to reach the continuum limit.  The lattice Hamiltonian  contains infinitely many operators that differ from the continuum Hamiltonian at finite lattice spacing. Counterterms will be needed to suppress leading-order lattice artifacts \cite{Carena:2022:improved_hamiltonians, GustafsonVandeWater:2023:improved_fermions}. Furthermore, the irrep truncations for the gauge and fermion degrees of freedom affect the accuracy of simulations. This is pronounced, for example, in state preparation \cite{CiavarellaChernyshev:2022:su3_state_prep, BalajiEtal:2026:ground_state}. A systematic effort could be made to derive Hamiltonian improvement terms that suppress truncation effects \cite{Ciavarella:2023:improved_hamiltonians}. Alternatively, higher truncation cutoffs could become more accessible by improving the efficiency of the time evolution implementation. A complete analysis of lattice QCD time evolution with fault-tolerant algorithms and the efficient representation basis encoding would also be of interest.

The theory can also be modified and extended in a number of other ways. First, it is worth considering how to formulate the theory with other lattice fermions, such as twisted mass fermions \cite{SchwagerlJansenKuhn:2025:twisted_mass} and Ginsparg-Wilson fermions \cite{Singh:2025:ginsparg_wilson}, for their continuum limit and chiral symmetry properties. Of relevance to nuclear physics, introducing a nonzero baryon chemical potential directly in the Hamiltonian,
\begin{equation}
	H_{\mu_\text{B}} = -\frac{\mu_\text{B}}{3} \sum_{\vec{s}} \sum_{f=1}^{N_f} \psi^{(f)\dag}(\vec{s}) \psi^{(f)}(\vec{s}) - \chi^{(f)\dag}(\vec{s}) \chi^{(f)}(\vec{s}),
\end{equation}
would be of interest for preparing states of dense matter, relevant, for example, for the neutron star equation of state~\cite{Nagata:2022:finite_density_QCD}. Finally, quantum electrodynamics must be incorporated to the Hamiltonian by gauging the relevant U(1) symmetry to achieve more realistic simulations of charged matter. Beyond the Standard Model, the theta angle could be promoted to a scalar axion field \cite{Ho:2026:peccei_quinn_schwinger, RouxinolEtal:2026:schwinger_model_axion} to study axion-QCD dynamics through the confinement transition in the early universe.

\section*{Acknowledgments}
\addcontentsline{toc}{section}{Acknowledgments}

We acknowledge the support of the U.S. Department of Energy, Office of Science, Office of High Energy Physics Quantum Information Science Enabled Discovery (QuantISED) program. This work used the DeltaAI system at the National Center for Supercomputing Applications [award OAC 2320345] through allocation PHY260058 from the Advanced Cyberinfrastructure Coordination Ecosystem: Services \& Support (ACCESS) program, which is supported by National Science Foundation grants \#2138259, \#2138286, \#2138307, \#2137603, and \#2138296.

\appendix

\section{Continuum Hamiltonian}
\label{app:continuum_hamiltonian}

Begin with the Lagrangian density
\begin{equation}
	\mathcal{L} = -\frac{1}{4g^2} F_{\mu\nu}^a F^{\mu\nu,a} + \sum_{f=1}^{N_f} \bar{\psi}^{(f)} (i\gamma^\mu \partial_\mu + \gamma^\mu A_\mu^a T^a - m_f) \psi^{(f)} + \frac{\theta}{64\pi^2} \epsilon^{\mu\nu\rho\sigma} F_{\mu\nu}^a F_{\rho\sigma}^a
\end{equation}
The gauge field and the fermion fields have canonical momenta
\begin{equation}
	\Pi^{\mu,a} = \frac{\partial\mathcal{L}}{\partial(\partial_0 A^{\mu,a})} = \frac{1}{g^2} F^{\mu 0,a} + \frac{\theta}{16\pi^2} \epsilon^{0\mu jk} F_{jk}^a \qquad \pi^{(f)} = \frac{\partial\mathcal{L}}{\partial(\partial_0 \psi^{(f)})} = i \bar{\psi}^{(f)} \gamma^0
\end{equation}
The Hamiltonian density is
\begin{equation}
	\mathcal{H} = \Pi^{\mu,a} \partial_0 A_\mu^a + \sum_{f=1}^{N_f} \pi^{(f)} \partial_0 \psi^{(f)} - \mathcal{L}
\end{equation}
and the Hamiltonian is $H = \int d^3x \ \mathcal{H}$.

The first term of $\mathcal{H}$ can be written as
\begin{IEEEeqnarray*}{rCl}
	\Pi^{\mu,a} \partial_0 A_\mu^a & = & \Pi^{i,a} \partial_0 A_i^a \\
	& = & \Pi^{i,a} \left( F_{0i}^a + \partial_i A_0^a - A_0^m A_i^n f^{amn} \right) \\
	& = & \Pi^{i,a} F_{0i}^a - A_0^a \partial_i \Pi^{i,a} - A_0^a A_i^b \Pi^{i,m} f^{bma} \\
	& = & \Pi^{i,a} F_{0i}^a - A_0^a D_i \Pi^{i,a} \\
	& = & -\frac{1}{g^2} F_{i0}^a F^{i 0,a} + \frac{\theta}{16\pi^2} \epsilon^{0ijk} F_{0i}^a F_{jk}^a - A_0^a D_i \Pi^{i,a} \yesnumber
\end{IEEEeqnarray*}
where $\Pi^{i,a} \partial_i A_0^a$ was integrated by parts. Notice that the pure-gauge terms of the Lagrangian density can be rewritten as
\begin{IEEEeqnarray*}{rCl}
	\mathcal{L}_\text{gauge} & = & -\frac{1}{4g^2} F_{\mu\nu}^a F^{\mu\nu,a} + \frac{\theta}{64\pi^2} \epsilon^{\mu\nu\rho\sigma} F_{\mu\nu}^a F_{\rho\sigma}^a \\
	& = & -\frac{1}{2g^2} F_{i0,a} F^{i0,a} - \frac{1}{4g^2} F_{ij}^a F^{ij,a} + \frac{\theta}{16\pi^2} \epsilon^{0ijk} F_{0i}^a F_{jk}^a \yesnumber
\end{IEEEeqnarray*}
Then the pure-gauge Hamiltonian density is
\begin{IEEEeqnarray*}{rCl}
	\mathcal{H}_\text{gauge} & = & \Pi^{\mu,a} \partial_0 A_\mu^a - \mathcal{L}_\text{gauge} \\
	& = & - \frac{1}{2g^2} F_{i0,a}F^{i0,a} + \frac{1}{4g^2} F_{ij}^a F^{ij,a} - A_0^a D_i \Pi^{i,a} \\
	& = & \frac{1}{2g^2} \left( g^2 \Pi^{i,a} - \frac{\theta g^2}{16\pi^2} \epsilon^{0ijk} F_{jk}^a \right) \left( g^2 \Pi^{i,a} - \frac{\theta g^2}{16\pi^2} \epsilon^{0ipq} F_{pq}^a \right) + \frac{1}{4g^2} F_{ij}^a F^{ij,a} - A_0^a D_i \Pi^{i,a} \\
	& = & \frac{g^2}{2} \Pi^{i,a} \Pi^{i,a} + \frac{1}{4g^2} \left( 1 + \frac{\theta^2 g^4}{64\pi^4} \right) F_{ij}^a F^{ij,a} + \frac{\theta g^2}{16\pi^2} \epsilon_{ijk} \Pi^{i,a} F_{jk}^a - A_0^a D_i \Pi^{i,a} \yesnumber
\end{IEEEeqnarray*}
The following properties of the Levi-Civita tensor (for which $\epsilon_{123}=+1$) were used:
\begin{equation}
	\epsilon^{0ijk} = \epsilon^{ijk} \qquad \epsilon^{ijk} = - \epsilon_{ijk} \qquad \epsilon_{ijk} \epsilon^{ipq} = \delta_j^q \delta_k^p - \delta_j^p \delta_k^q
\end{equation}
The complete Hamiltonian density is therefore
\begin{IEEEeqnarray*}{rCl}
	\mathcal{H} = & - & \sum_{f=1}^{N_f} \bar{\psi}^{(f)} (i\gamma^i \partial_i + \gamma^i A_i^a T^a - m_f) \psi^{(f)} - A_0^a \bigg( D_i \Pi^{i,a} + \sum_{f=1}^{N_f} \bar{\psi}^{(f)} \gamma^0 T^a \psi^{(f)} \bigg) \\
	& + & \frac{g^2}{2} \Pi^{i,a} \Pi^{i,a} + \frac{1}{4g^2} \left( 1 + \frac{\theta^2 g^4}{64\pi^4} \right) F_{ij}^a F^{ij,a} + \frac{\theta g^2}{16\pi^2} \epsilon_{ijk} \Pi^{i,a} F_{jk}^a \yesnumber
\end{IEEEeqnarray*}

Promote the fields to field operators in the Schr\"{o}dinger picture. Impose the equal-time canonical (anti)commutation relations
\begin{equation}
	[A_\mu^a(\vec{x}), \Pi^{\nu,b}(\vec{y})] = i\eta_\mu^\nu \delta_{ab} \delta^3(\vec{x} - \vec{y}) \qquad \{ \psi_{\alpha,c}^{(f)}(\vec{x}), \pi_{\beta,c'}^{(f')}(\vec{y}) \} = i \delta_{\alpha\beta} \delta_{cc'} \delta_{ff'} \delta^3(\vec{x}-\vec{y})
\end{equation}
where $\eta_\mu^\nu = \delta_\mu^\nu$. (Anti)commutators of the form $[A,A]$, $[\Pi,\Pi]$, $\{ \psi, \psi \}$, and $\{ \pi, \pi \}$ are zero, and all fermionic operators commute with gauge field operators. The Hamiltonian operator is defined to be
\begin{IEEEeqnarray*}{rCl}
	H = \int d^3x \bigg[ & - & \sum_{f=1}^{N_f} \bar{\psi}^{(f)} (i\gamma^i \partial_i + \gamma^i A_i^a T^a - m_f) \psi^{(f)} - A_0^a \bigg( D_i \Pi^{i,a} + \sum_{f=1}^{N_f} \bar{\psi}^{(f)} \gamma^0 T^a \psi^{(f)} \bigg) \\
	& + & \frac{g^2}{2} \Pi^{i,a} \Pi^{i,a} + \frac{1}{4g^2} \left( 1 + \frac{\theta^2 g^4}{64\pi^4} \right) F_{ij}^a F^{ij,a} + \frac{\theta g^2}{16\pi^2} \epsilon_{ijk} \Pi^{i,a} F_{jk}^a \bigg] \yesnumber
\end{IEEEeqnarray*}

The quantized theory comes with primary constraint $\Pi^{0,a}(\vec{x}) \ket{\text{phys}} = 0$. Defining the Gauss operator
\begin{equation}
	G^a(\vec{x}) = D_i \Pi^{i,a}(\vec{x}) + \sum_{f=1}^{N_f} \bar{\psi}^{(f)}(\vec{x}) \gamma^0 T^a \psi^{(f)}(\vec{x})
\end{equation}
the secondary constraint $G^a(\vec{x}) \ket{\text{phys}} = 0$ follows. The remaining relevant commutators are
\begin{IEEEeqnarray*}{rCl}
	[\Pi^{0,a}(\vec{x}), G^b(\vec{y})] & = & 0 \\
	\left. [G^a(\vec{x}), G^b(\vec{y})] \right. & = & i f^{abc} G^c(\vec{x}) \delta^3(\vec{x}-\vec{y}) \\
	i[H, G^a(\vec{x})] & = & -A_0^b(\vec{x}) G^c(\vec{x}) f^{abc} \yesnumber
\end{IEEEeqnarray*}
Because these are all proportional to the Gauss operator, the constraint algebra closes.

\section{Dirac Fermions}
\label{app:dirac_fermions}

A general solution to the Dirac equation $(i\gamma^\mu \partial_\mu - m) \psi = 0$ is
\begin{equation}
	\psi(x) = \int \frac{d^3p}{(2\pi)^3} \sum_{s=1,2} \left[ u^s(p) e^{-ip_\mu x^\mu} \psi_p^s + v^s(p) e^{ip_\mu x^\mu} \chi_p^s \right]
\end{equation}
where $\psi_p^s$ and $\chi_p^s$ are operators with anticommutation relations
\begin{equation}
	\{ \psi_p^s, \psi_q^{r\dag} \} = \{ \chi_p^s, \chi_q^{r\dag} \} = (2\pi)^3 \delta_{sr} \delta^3(\vec{p}-\vec{q})
\end{equation}
Using $(\gamma^\mu p_\mu - m) u^s(p) = 0 = (\gamma^\mu p_\mu + m) v^s(p)$, and assuming the normalization $u^{r\dag}(p) u^s(p) = v^{r\dag}(p) v^s(p) = \delta_{rs}$ and the orthogonality $u^{r\dag}(\pm\vec{p}) v^s(\mp\vec{p}) = 0$, the Hamiltonian can be written as
\begin{equation}
	H = \int d^3x \ \bar{\psi} ( -i\gamma^i \partial_i + m ) \psi = \int \frac{d^3p}{(2\pi)^3} p^0 \sum_{s=1,2} \psi^{s\dag}_p \psi^s_p - \chi^{s\dag}_p \chi^s_p
\end{equation}
where $p^0 \equiv \sqrt{|\vec{p}|^2 + m^2}$. Likewise, the Noether charge operators are
\begin{equation}
	\mathcal{Q} = \int d^3x \ \psi^\dag \psi = \int \frac{d^3p}{(2\pi)^3} \sum_{s=1,2} \psi^{s\dag}_p \psi^s_p + \chi^{s\dag}_p \chi^s_p
\end{equation}
and
\begin{equation}
	\mathcal{Q}^a = \int d^3x \ \psi^\dag T^a \psi = \int \frac{d^3p}{(2\pi)^3} \sum_{s=1,2} \psi^{s\dag}_{p,c} T^a_{cc'} \psi^s_{p,c'} + \chi^{s\dag}_{p,c} T^a_{cc'} \chi^s_{p,c'}
\end{equation}

A choice must be made to construct the momentum-space Hilbert space; let $\ket{\text{vac}}$ be the state annihilated by $\psi_p^s$ and $\chi_p^s$:
\begin{equation}
	\psi_p^s \ket{\text{vac}} = 0 \qquad \chi_p^s \ket{\text{vac}} = 0
\end{equation}
Then $\ket{\text{vac}}$ is a state of zero energy -- it is the vacuum, or unoccupied state. However, the negative sign in the Hamiltonian suggests that by populating all possible negative-frequency excitations, of which there are an infinite number, the energy can be made arbitrarily low. Therefore, the ground state is
\begin{equation}
	\ket{\Omega} = \prod_{p,s} \chi_p^{s\dag} \ket{\text{vac}}
\end{equation}
It is then also possible is to formulate all states relative to $\ket{\Omega}$. If
\begin{equation}
	a_p^s \equiv \psi_p^s \qquad b^{s\dag}_p \equiv \chi_p^s \qquad \{ a_p^s, a_q^{r\dag} \} = \{ b_p^s, b_q^{r\dag} \} = (2\pi)^3 \delta_{sr} \delta^3(\vec{p}-\vec{q})
\end{equation}
then $a^s_p \ket{\Omega} = 0$ and $b_p^s \ket{\Omega} = 0$. Finally, these new anticommutation relations can be used to normal order all operators so that their expected value in the ground state is zero. The normal ordered Hamiltonian is
\begin{equation}
	:\mathrel{H}: \ = \int \frac{d^3p}{(2\pi)^3} p^0 \sum_{s=1,2} a^{s\dag}_p a^s_p + b^{s\dag}_p b^s_p
\end{equation}
and the charges are (using $b_c T^a_{cc'} b^\dag_{c'} \to -b_{c'}^\dag T^a_{cc'} b_c = -b_c^\dag T_{cc'}^{a*} b_c$)
\begin{equation}
	:\mathrel{\mathcal{Q}}: \ = \int \frac{d^3p}{(2\pi)^3} \sum_{s=1,2} a^{s\dag}_p a^s_p - b^{s\dag}_p b^s_p \qquad :\mathrel{\mathcal{Q}^a}: \ = \int \frac{d^3p}{(2\pi)^3} \sum_{s=1,2} a^{s\dag}_{p,c} T^a_{cc'} a^s_{p,c'} - b^{s\dag}_{p,c} T^{a*}_{cc'} b^s_{p,c'}
\end{equation}
Positive-frequency excitations (particles), $a^{s\dag}_p \ket{\Omega}$, and negative-frequency excitations (antiparticles), $b^{s\dag}_p \ket{\Omega}$, both have positive energy but opposite sign charges.

\subsubsection*{Position-Space Formalism}

Although the field operator
\begin{equation}
	\psi(\vec{x}) = \int \frac{d^3p}{(2\pi)^3} \sum_{s=1,2} \left[ u^s(p) e^{i\vec{p} \cdot \vec{x}} a_p^s + v^s(p) e^{-i\vec{p} \cdot \vec{x}} b_p^{s\dag} \right]
\end{equation}
annihilates the vacuum state $\ket{\text{vac}}$, it does not annihilate the ground state $\ket{\Omega}$. Therefore, position-space operators cannot be normal-ordered with respect to $\ket{\Omega}$. However, it may be useful to work with a state that approximates the ground state better than $\ket{\text{vac}}$. This new zero-point state that is neither the vacuum nor the ground state will be used to construct the Hilbert space.

In particular, consider the zero-momentum modes in the field operator, where $p^0 = m$. For those modes, $(\gamma^0 - 1) u^s(0) = 0$ and $(\gamma^0 + 1) v^s(0) = 0$. Then it is possible to ``project out'' the annihilation and creation operators for particles at rest. Define the particle and antiparticle projectors \cite{Wilson:1977:quarks_strings}
\begin{equation}
	P_+ = \frac{1+\gamma^0}{2} \qquad P_- = \frac{1-\gamma^0}{2}
\end{equation}
Notice that $P_+ \psi(\vec{x}) \ni a_0^s$ (with no $b^{s\dag}_0$) and $\psi^\dag(\vec{x}) P_- \ni b_0^s$ (with no $a^{s\dag}_0$). Restricted to particles at rest, these projected operators do annihilate $\ket{\Omega}$. Now a choice must be made to construct the position-space Hilbert space; write
\begin{equation}
	(P_+ \psi)_\alpha(\vec{x}) \equiv \hat{\psi}_\alpha(\vec{x}) \qquad (\psi^\dag P_-)_\alpha(\vec{x}) \equiv \hat{\chi}_\alpha(\vec{x})
\end{equation}
and let $\ket{0}$ be the zero-point state annihilated by $\hat{\psi}_\alpha$ and $\hat{\chi}_\alpha$:
\begin{equation}
	\hat{\psi}_\alpha \ket{0} = 0 \qquad \hat{\chi}_\alpha \ket{0} = 0
\end{equation}
Note the anticommutators:
\begin{IEEEeqnarray*}{rCl}
	\{ \psi_\alpha(\vec{x}), \psi_\beta^\dag(\vec{y}) \} & = & \{ \hat{\psi}_\alpha(\vec{x}), \hat{\psi}_\beta^\dag(\vec{y}) \} + \{ \hat{\chi}_\alpha^\dag(\vec{x}), \hat{\chi}_\beta(\vec{y}) \} \\
	\{ \hat{\psi}_\alpha(\vec{x}), \hat{\psi}_\beta^\dag(\vec{y}) \} & = & (P_+)_{\alpha\beta} \delta^3(\vec{x}-\vec{y}) \\
	\{ \hat{\chi}_\alpha^\dag(\vec{x}), \hat{\chi}_\beta(\vec{y}) \} & = & (P_-)_{\alpha\beta} \delta^3(\vec{x}-\vec{y}) \yesnumber
\end{IEEEeqnarray*}
While ultimately $\ket{0}$ is not the ground state, it essentially is for particles at rest. If the theory had a finite Hilbert space, then $\ket{0}$ should be finitely closer to the ground state than $\ket{\text{vac}}$.

Now revisit the normal ordering procedure. Using $\gamma^i P_\pm = P_\mp \gamma^i$, the Hamiltonian becomes
\begin{equation}
	::\mathrel{H}:: \ = \int d^3x \ ( -i \hat{\psi}^\dag \gamma^i \partial_i \hat{\chi}^\dag + i \hat{\chi} \gamma^i \partial_i \hat{\psi} ) + m ( \hat{\psi}^\dag \hat{\psi} + \hat{\chi}^\dag \hat{\chi} )
\end{equation}
and the charges are
\begin{equation}
	::\mathrel{\mathcal{Q}}:: \ = \int d^3x \ \hat{\psi}^\dag \hat{\psi} - \hat{\chi}^\dag \hat{\chi} \qquad ::\mathrel{\mathcal{Q}^a}:: \ = \int d^3x \ \hat{\psi}^\dag T^a \hat{\psi} - \hat{\chi}^\dag T^{a*} \hat{\chi}
\end{equation}
These operators are invariant under a change of basis for the gamma matrices. It is convenient to work in the Dirac basis -- where $\gamma^0$ is diagonal -- because $\hat{\psi}$ and $\hat{\chi}$ will have linearly independent components. In this case, the states $\hat{\psi}^\dag_1 \ket{0}$ and $\hat{\psi}^\dag_2 \ket{0}$ contain a particle at rest, and have opposite charge from the states $\hat{\chi}^\dag_3 \ket{0}$ and $\hat{\chi}^\dag_4 \ket{0}$, which have an antiparticle at rest.

\subsection{Lattice Fermions}
\label{app:dirac_fermions:lattice_fermions}

The theory of free Dirac fermions can be discretized with a spatial lattice as done in the main text. In what follows, all operators will appear in their normal-ordered form, and the projected Dirac field operators $\hat{\psi}$ and $\hat{\chi}$ will have their hats dropped. The discrete anticommutation relations are
\begin{equation}
	\{ \psi_\alpha(\vec{s}), \psi_\beta^\dag(\vec{r}) \} = (P_+)_{\alpha\beta} \delta_{\vec{s},\vec{r}} \qquad \{ \chi_\alpha^\dag(\vec{s}), \chi_\beta(\vec{r}) \} = (P_-)_{\alpha\beta} \delta_{\vec{s},\vec{r}}
\end{equation}
On the lattice these operators are dimensionless; they have been rescaled by $\psi \to \frac{1}{\sqrt{a^3}} \psi$ and $\chi \to \frac{1}{\sqrt{a^3}} \chi$ in passing from the continuum to the lattice. For the discretized Hamiltonian, a symmetric finite difference maintains hermiticity:
\begin{IEEEeqnarray*}{rCl}
	H = \sum_{\vec{s}} \bigg[ & - & \frac{i}{2a} \sum_{i=1}^{3} \left[ \psi^\dag(\vec{s}) \gamma^i \chi^\dag(\vec{s}+a\vec{e}_i) - \psi^\dag(\vec{s}) \gamma^i \chi^\dag(\vec{s}-a\vec{e}_i) \right] \\
	& + & \frac{i}{2a} \sum_{i=1}^{3} \left[ \chi(\vec{s}) \gamma^i \psi(\vec{s}+a\vec{e}_i) - \chi(\vec{s}) \gamma^i \psi(\vec{s}-a\vec{e}_i) \right] \\
	& + & m (\psi^\dag(\vec{s}) \psi(\vec{s}) + \chi^\dag(\vec{s}) \chi(\vec{s})) \bigg] \yesnumber
\end{IEEEeqnarray*}
The charge operators are
\begin{equation}
	\mathcal{Q} = \sum_{\vec{s}} \psi^\dag(\vec{s}) \psi(\vec{s}) - \chi^\dag(\vec{s}) \chi(\vec{s}) \qquad \mathcal{Q}^a = \sum_{\vec{s}} \psi^\dag(\vec{s}) T^a \psi(\vec{s}) - \chi^\dag(\vec{s}) T^{a*} \chi(\vec{s})
\end{equation}

The equations of motion are
\begin{IEEEeqnarray*}{rCl}
	\partial_t \psi(\vec{s}) & = & -\frac{1}{2a} \sum_{i=1}^{3} \gamma^i \left[ \chi^\dag(\vec{s}+a\vec{e}_i) - \chi^\dag(\vec{s}-a\vec{e}_i) \right] - im \psi(\vec{s}) \\
	\partial_t \chi^\dag(\vec{s}) & = & \frac{1}{2a} \sum_{i=1}^{3} \gamma^i \left[ \psi(\vec{s}+a\vec{e}_i) - \psi(\vec{s}-a\vec{e}_i) \right] + im \chi^\dag(\vec{s}) \yesnumber
\end{IEEEeqnarray*}
 $\chi^\dag(t,\vec{s})$ can be solved for in terms of $\psi(t,\vec{s})$ by using plane waves $\psi(t,\vec{s}) = e^{-i(\omega t - \vec{p} \cdot \vec{s})} \tilde{\psi}(\vec{p})$ and $\chi^\dag(t,\vec{s}) = e^{-i(\omega t - \vec{p} \cdot \vec{s})} \tilde{\chi}^\dag(\vec{p})$, revealing the dispersion relation
\begin{equation}
	\omega^2 = \frac{1}{a^2} \sum_{i=1}^{3} \sin^2(p^ia) + m^2
\end{equation}
This shows that, on the lattice, the Dirac theory is one of eight kinds of fermion species (called tastes): one for each $p^1,p^2,p^3=0,\frac{\pi}{a}$. This is the fermion doubling problem.

\subsubsection*{Wilson Fermions}

One way to deal with the fermion doubling problem is to modify the dispersion relation -- for instance, making $\omega^2(\frac{\pi}{a}) > m^2$. This can be done by adding operators to the lattice Hamiltonian, granted that these operators vanish in the continuum. Directly adding another (momentum-independent) mass term is ruled out because this simply shifts the minimum of the dispersion relation. Without introducing interaction terms (such as a four-fermion term), only the kinetic term can be modified with higher-order derivatives.

Briefly switching to the non-projected formalism, the Wilson term \cite{Wilson:1977:quarks_strings}
\begin{equation}
	H \supset -\frac{r}{2a} \sum_{\vec{s}} \left[ \bar{\psi}(\vec{s}) \sum_{i=1}^{3} \psi(\vec{s}+a\vec{e}_i) - 2\psi(\vec{s}) + \psi(\vec{s}-a\vec{e}_i) \right]
\end{equation}
is a symmetric finite difference for a Laplacian. The factor of $a$ is added by dimensional analysis and ensures the operator disappears in the continuum at order $\mathcal{O}(a)$:
\begin{equation}
	H_\text{continuum} \supset -\frac{ar}{2} \int d^3x \ \bar{\psi} |\vec{\nabla}|^2 \psi + \mathcal{O}(a^2)
\end{equation}
The factor of $\frac{1}{2}$ is for convenience to match the $\frac{1}{2}$ from the usual kinetic term. Lastly, it will turn out that the factor of $-1$ lets the Wilson parameter $r>0$ be positive.

After inserting projectors and normal ordering, the Hamiltonian is
\begin{IEEEeqnarray*}{rCl}
	H = \sum_{\vec{s}} \bigg[ & - & \frac{i}{2a} \sum_{i=1}^{3} \left[ \psi^\dag(\vec{s}) \gamma^i \chi^\dag(\vec{s}+a\vec{e}_i) - \psi^\dag(\vec{s}) \gamma^i \chi^\dag(\vec{s}-a\vec{e}_i) \right] \\
	& + & \frac{i}{2a} \sum_{i=1}^{3} \left[ \chi(\vec{s}) \gamma^i \psi(\vec{s}+a\vec{e}_i) - \chi(\vec{s}) \gamma^i \psi(\vec{s}-a\vec{e}_i) \right] \\
	& - & \frac{r}{2a} \sum_{i=1}^{3} \left[ \psi^\dag(\vec{s}) \psi(\vec{s}+a\vec{e}_i) + \psi^\dag(\vec{s}) \psi(\vec{s}-a\vec{e}_i) \right] \\
	& - & \frac{r}{2a} \sum_{i=1}^{3} \left[ \chi^\dag(\vec{s}) \chi(\vec{s}+a\vec{e}_i) + \chi^\dag(\vec{s}) \chi(\vec{s}-a\vec{e}_i) \right] \\
	& + & (m + \tfrac{3r}{a}) (\psi^\dag(\vec{s}) \psi(\vec{s}) + \chi^\dag(\vec{s}) \chi(\vec{s})) \bigg] \yesnumber
\end{IEEEeqnarray*}
(If the fermions were massless, then the Wilson term effectively adds a mass; the Wilson term generally breaks chiral symmetry.) The equations of motion
\begin{IEEEeqnarray*}{rCl}
	\partial_t \psi(\vec{s}) & = & -\frac{1}{2a} \sum_{i=1}^{3} \gamma^i \left[ \chi^\dag(\vec{s}+a\vec{e}_i) - \chi^\dag(\vec{s}-a\vec{e}_i) \right] - i(m + \tfrac{3r}{a}) \psi(\vec{s}) \\
	& + & \frac{ir}{2a}  \sum_{i=1}^{3} \left[ \psi(\vec{s}+a\vec{e}_i) + \psi(\vec{s}-a\vec{e}_i) \right] \\
	\partial_t \chi^\dag(\vec{s}) & = & \frac{1}{2a} \sum_{i=1}^{3} \gamma^i \left[ \psi(\vec{s}+a\vec{e}_i) - \psi(\vec{s}-a\vec{e}_i) \right] + i(m + \tfrac{3r}{a}) \chi^\dag(\vec{s}) \\
	& - & \frac{ir}{2a} \sum_{i=1}^{3} \left[ \chi^\dag(\vec{s}+a\vec{e}_i) + \chi^\dag(\vec{s}-a\vec{e}_i) \right] \yesnumber
\end{IEEEeqnarray*}
can be solved with plane waves to obtain the dispersion relation
\begin{equation}
	\omega^2 = \frac{1}{a^2} \sum_{i=1}^{3} \sin^2(p^ia) + \left( m + \frac{2r}{a} \sum_{i=1}^{3} \sin^2\left( \frac{p^ia}{2} \right) \right)^2
\end{equation}
Therefore, by using $r>0$ a positive momentum-dependent shift to the mass has been added. All but one taste decouples in the continuum limit, leaving a lattice theory with a single fermion species.

\subsubsection*{Staggered Fermions}

It is possible to mitigate the fermion doubling problem by staggering components of the Dirac field among sites \cite{Susskind:1977:lattice_fermions}. It will be useful to have explicit gamma matrices. The Dirac basis for gamma matrices is
\begin{IEEEeqnarray*}{rCCCl}
	\gamma^0 = \begin{pmatrix} I & 0 \\ 0 & -I \end{pmatrix} & \qquad & \gamma^i = \begin{pmatrix} 0 & \sigma^i \\ -\sigma^i & 0 \end{pmatrix} & \qquad & \gamma^5 = \begin{pmatrix} 0 & I \\ I & 0 \end{pmatrix} \\ [12pt]
	\sigma^1 = X = \begin{pmatrix} 0 & 1 \\ 1 & 0 \end{pmatrix} & \qquad & \sigma^2 = Y = \begin{pmatrix} 0 & -i \\ i & 0 \end{pmatrix} & \qquad & \sigma^3 = Z =  \begin{pmatrix} 1 & 0 \\ 0 & -1 \end{pmatrix} \yesnumber
\end{IEEEeqnarray*}
The Hamiltonian contains the following bilinears:
\begin{IEEEeqnarray*}{rClCrCl}
	\psi^\dag \psi & = & \psi^\dag_1 \psi_1 + \psi^\dag_2 \psi_2 & \qquad & \chi^\dag \chi & = & \chi^\dag_3 \chi_3 + \chi^\dag_4 \chi_4 \\
	\chi \gamma^1 \psi & = & -\chi_4 \psi_1 - \chi_3 \psi_2 & \qquad & \psi^\dag \gamma^1 \chi^\dag & = & \psi^\dag_1 \chi^\dag_4 + \psi^\dag_2 \chi^\dag_3 \\
	\chi \gamma^2 \psi & = & -i\chi_4 \psi_1 + i\chi_3 \psi_2 & \qquad & \psi^\dag \gamma^2 \chi^\dag & = & -i\psi^\dag_1 \chi^\dag_4 + i\psi^\dag_2 \chi^\dag_3 \\
	\chi \gamma^3 \psi & = & -\chi_3 \psi_1 + \chi_4 \psi_2 & \qquad & \psi^\dag \gamma^3 \chi^\dag & = & \psi^\dag_1 \chi^\dag_3 - \psi^\dag_2 \chi^\dag_4 \yesnumber
\end{IEEEeqnarray*}
The form of the bilinears suggests a procedure for staggering the component operators. Define the quantities $\Sigma_\perp^s = \frac{s_1+s_2}{a}$ and $\Sigma_3^s = \frac{s_3}{a}$ for a site $\vec{s}$. The four components can be defined on sites with specific $\Sigma_\perp^s$ and $\Sigma_3^s$ parities. Without loss of generality, define $\psi_1$ on sites with even $\Sigma_\perp^s$ and $\Sigma_3^s$. This constrains $\chi_4$ to be defined on sites with odd $\Sigma_\perp^s$ and even $\Sigma_3^s$ by inspecting the $\chi \gamma^{1,2} \psi$ bilinears. Likewise, $\chi_3$ is constrained by the $\chi \gamma^3 \psi$ bilinear to be defined on sites with even $\Sigma_\perp^s$ and odd $\Sigma_3^s$. This leaves $\psi_2$ to be defined on sites with odd $\Sigma_\perp^s$ and odd $\Sigma_3^s$. This scheme is summarized in table~\ref{tab:staggering_scheme}. Generally, $\psi_\alpha(\vec{s})$ operators are defined on sites with even $\Sigma^s=\frac{s_1+s_2+s_3}{a}$ (even sites), and $\chi_\alpha(\vec{s})$ operators are defined on sites with odd $\Sigma^s$ (odd sites).
\begin{table}[!ht]
	\centering
	\setlength{\tabcolsep}{12pt}
	\renewcommand{\arraystretch}{1.5}
	\begin{tabular}{c|c|c}
		Operator & $\Sigma_\perp^s$ & $\Sigma_3^s$ \\
		\hline
		$\psi_1(\vec{s})$ & Even & Even \\
		\hline
		$\psi_2(\vec{s})$ & Odd & Odd \\
		\hline
		$\chi_3(\vec{s})$ & Even & Odd \\
		\hline
		$\chi_4(\vec{s})$ & Odd & Even
	\end{tabular}
	\caption{The component operator shown in each row is defined on lattice sites that have specific parities in  $\Sigma_\perp^s = \frac{s_1+s_2}{a}$ and $\Sigma_3^s = \frac{s_3}{a}$, where $s_i$ is the $i$th component of $\protect\vec{s}$.}
	\label{tab:staggering_scheme}
\end{table}

The staggered Hamiltonian is written by expanding the bilinears in the original Hamiltonian and summing over sites of specific parities where now only component operators are defined:
\begin{IEEEeqnarray*}{rCl}
	H = & - & \frac{i}{2a} \sum_{\vec{s} \text{even}} \sum_{i=1}^{3} \gamma^i(\vec{s}) \left[ \psi^\dag(\vec{s}) \chi^\dag(\vec{s}+a\vec{e}_i) - \psi^\dag(\vec{s}) \chi^\dag(\vec{s}-a\vec{e}_i) \right] \\
	& + & \frac{i}{2a} \sum_{\vec{s} \text{odd}} \sum_{i=1}^{3} \gamma^i(\vec{s}) \left[ \chi(\vec{s}) \psi(\vec{s}+a\vec{e}_i) - \chi(\vec{s}) \psi(\vec{s}-a\vec{e}_i) \right] \\
	& + & m \sum_{\vec{s} \text{even}} \psi^\dag(\vec{s}) \psi(\vec{s}) + m \sum_{\vec{s} \text{odd}} \chi^\dag(\vec{s}) \chi(\vec{s}) \yesnumber
\end{IEEEeqnarray*}
Here, $\gamma^i(\vec{s})$ are gamma matrix elements:
\begin{equation}
	\gamma^i(\vec{s}) =
	\begin{cases}
		1 & (i=1 \text{ and } \vec{s} \text{ even}) \text{ or } (i=3 \text{ and } \Sigma^s_3 \text{ even}) \\
		-1 & (i=1 \text{ and } \vec{s} \text{ odd}) \text{ or } (i=3 \text{ and } \Sigma^s_3 \text{ odd}) \\
		i & i=2 \text{ and } \Sigma_3^s \text{ odd} \\
		-i & i=2 \text{ and } \Sigma_3^s \text{ even}
	\end{cases}
\end{equation}
The charge operators should also be written to reflect the staggered field operators:
\begin{IEEEeqnarray*}{rCl}
	\mathcal{Q} & = & \sum_{\vec{s} \text{even}} \psi^\dag(\vec{s}) \psi(\vec{s}) - \sum_{\vec{s} \text{odd}} \chi^\dag(\vec{s}) \chi(\vec{s}) \\
	\mathcal{Q}^a & = & \sum_{\vec{s} \text{even}} \psi^\dag(\vec{s}) T^a \psi(\vec{s}) - \sum_{\vec{s} \text{odd}} \chi^\dag(\vec{s}) T^{a*} \chi(\vec{s}) \yesnumber
\end{IEEEeqnarray*}

The number of continuum degrees of freedom are not obvious in this formalism. It would appear that they are spread out across the lattice. And this is practically the case; the staggering procedure has established a ``unit cell'' that is repeated across the lattice. For $\text{d}=3$, this is a cube. It is then constructive to recombine the eight degrees of freedom on these unit cells into fermionic fields defined on sites of a coarse lattice. These fields will not only have Dirac indices, but also lingering taste indices that reflect a partial solution to the fermion doubling problem. This consolidation is called the spin-taste basis \cite{DeGrandDetar:2006:lattice_methods, CatterallPradhanSamlodia:2026:staggered_fermions}.

Start by defining unit cell coordinate vectors $\vec{e}$. These can be split into
\begin{equation}
	E_\text{even} = \{ (0,0,0), (1,1,0), (1,0,1), (0,1,1) \} \qquad E_\text{odd} = \{ (1,0,0), (0,1,0), (0,0,1), (1,1,1) \}
\end{equation}
The new field operators will be a linear combination of the component operators spread among the unit cell. Matrix elements can account for the taste and Dirac degrees of freedom. In particular, use $\Gamma_e = (-\gamma^1)^{e_1} (i\gamma^2)^{e_2} (-\gamma^3)^{e_3}$. Then define
\begin{equation}
	\Psi_{t,\alpha}(2\vec{s}) = \frac{1}{\sqrt{2}} \sum_{\vec{e} \in E_\text{even}} (\Gamma_e)_{t\alpha} \psi(2\vec{s}+a\vec{e}) \qquad X_{\alpha,t}(2\vec{s}) = \frac{1}{\sqrt{2}} \sum_{\vec{e} \in E_\text{odd}} (\Gamma_e^\dag)_{\alpha t} \chi(2\vec{s}+a\vec{e})
\end{equation}
where taste and Dirac indices $t$ and $\alpha$ have been written suggestively. It turns out that half the nonvanishing components are not unique. Restricting to the use of $t=1,2$, the following constraints hold:
\begin{IEEEeqnarray*}{lClClCl}
	\Psi_{3,3} = \Psi_{1,1} & \quad & \Psi_{3,4} = \Psi_{1,2} & \quad & \Psi_{4,3} = \Psi_{2,1} & \quad & \Psi_{4,4} = \Psi_{2,2} \\
	X_{1,3} = -X_{3,1} & \quad & X_{1,4} = -X_{3,2} & \quad & X_{2,3} = -X_{4,1} & \quad & X_{2,4} = -X_{4,2} \yesnumber
\end{IEEEeqnarray*}
Note that the Hermitian conjugate of the operator $\Psi_{t,\alpha}(\vec{s})$ is $\Psi^\dag_{\alpha,t}(\vec{s}) \equiv (\Psi^\dag)_{\alpha,t}(\vec{s})$.

Trace identities of the gamma matrices imply $\trace[\Gamma^\dag_{e'} \Gamma_e] = 2^{\lceil \frac{\text{d}}{2} \rceil} \delta_{e'e}$. Then it follows that
\begin{equation}
	\psi(2\vec{s}+a\vec{e}) = \frac{1}{\sqrt{2}} \trace[ \Gamma_e^\dag \Psi(2\vec{s}) ] \qquad \chi(2\vec{s}+a\vec{e}) = \frac{1}{\sqrt{2}} \trace[ \Gamma_e X(2\vec{s}) ]
\end{equation}
Moreover, in the Dirac basis, restricted to $t=1,2$, it is possible to find $(\Gamma^\dag_e)_{\beta t'} (\Gamma_e)_{t\alpha} = 2\delta_{tt'} \delta_{\alpha\beta}$. The anticommutators then follow:
\begin{equation}
	\{ \Psi_{t,\alpha}(\vec{s}), \Psi^\dag_{\beta,t'}(\vec{r}) \} = \delta_{tt'} \delta_{\alpha\beta} \delta_{\vec{s},\vec{r}} \qquad \{ X^\dag_{t,\alpha}(\vec{s}), X_{\beta,t'}(\vec{r}) \} = \delta_{tt'} \delta_{\alpha\beta} \delta_{\vec{s},\vec{r}}
\end{equation}

Using the coarse lattice spacing $b=2a$, it is an exercise to find that the staggered Hamiltonian is
\begin{IEEEeqnarray*}{rCl}
	H = \frac{1}{2} \sum_{\vec{s}} \bigg[ & - & \frac{i}{2b} \sum_{i=1}^{3} \trace\left[ \Psi^\dag(\vec{s}) X^\dag(\vec{s}+b\vec{e}_i) \gamma^i - \Psi^\dag(\vec{s}) X^\dag(\vec{s}-b\vec{e}_i) \gamma^i \right] \\
	& + & \frac{i}{2b} \sum_{i=1}^{3} \trace\left[ X(\vec{s}) \Psi(\vec{s}+b\vec{e}_i) \gamma^i - X(\vec{s}) \Psi(\vec{s}-b\vec{e}_i) \gamma^i \right] \\
	& + & \frac{i}{2b} \sum_{i=1}^{3} \trace\left[ \Psi^\dag(\vec{s}) \gamma^i X^\dag(\vec{s}+b\vec{e}_i) - 2\Psi^\dag(\vec{s}) \gamma^i X^\dag(\vec{s}) + \Psi^\dag(\vec{s}) \gamma^i X^\dag(\vec{s}-b\vec{e}_i) \right] \\
	& + & \frac{i}{2b} \sum_{i=1}^{3} \trace\left[ X(\vec{s}) \gamma^i \Psi(\vec{s}+b\vec{e}_i) - 2X(\vec{s}) \gamma^i \Psi(\vec{s}) + X(\vec{s}) \gamma^i \Psi(\vec{s}-b\vec{e}_i) \right] \\
	& + & m \trace\left[ \Psi^\dag(\vec{s}) \Psi(\vec{s}) + X^\dag(\vec{s}) X(\vec{s}) \right] \bigg] \yesnumber
\end{IEEEeqnarray*}
The overall factor of $\frac{1}{2}$ divides out a factor of two that appears from traces with the redundancy constraints. In finite volume, this Hamiltonian resembles a theory of two lattice fermions with a Laplacian-like interaction, albeit with the number of sites per axis cut in half. The first-derivative terms only mix the Dirac indices, as desired. The Laplacian terms generally mix both taste and Dirac indices; however, in the limit $b \to 0$, they vanish.

\section{Aspects of \texorpdfstring{SU($N_c$)}{SU(Nc)} Lattice Gauge Theory}
\label{app:aspects_of_SUNc_LGT}

This appendix reviews some of the representation theory of SU($N_c$) and $\mathfrak{su}(N_c)$ as it pertains to Sec.~\ref{sec:lattice_theory}.

\subsection{Clebsch-Gordan Coefficients}
\label{app:aspects:CGCs}

The direct product of $n_p$ irreps can be written as a direct sum of $n_s$ distinct irreps with multiplicity $m$:
\begin{equation}
	\bigotimes_{i=1}^{n_p} R_i = \bigoplus_{j=1}^{n_s} \bigoplus_{k=1}^{m_j} S_{j,k}
\end{equation}
The product basis states of the direct-product representation are written as
\begin{equation}
	\ket{(R_1, r_1) \otimes \cdots \otimes (R_{n_p}, r_{n_p})} = \ket{\otimes_{i=1}^{n_p} (R_i, r_i)}
\end{equation}
where $r_i = 1,\dots,\dim(R_i)$ enumerates a basis vector of $R_i$. The sum basis states of the direct-sum representation are written as $\ket{(S_{j,k}, s_{j,k})}$, where $s_{j,k} = 1,\dots,\dim(S_{j,k})$ enumerates a basis vector of $S_{j,k}$. These furnish complete, orthonormal bases:
\begin{IEEEeqnarray*}{rCl}
	\sum_{r_1=1}^{\dim(R_1)} \cdots \sum_{r_{n_p}=1}^{\dim(R_{n_p})} \ketbra{\otimes_{i=1}^{n_p} (R_i, r_i)}{\otimes_{i=1}^{n_p} (R_i, r_i)} = 1 & \qquad & \braket{\otimes_{i=1}^{n_p} (R_i, r_i)}{\otimes_{i=1}^{n_p} (R_i, r'_i)} = \prod_{i=1}^{n_p} \delta_{r_i,r'_i} \\ [12pt]
	\sum_{j=1}^{n_s} \sum_{k=1}^{m_j} \sum_{s_{j,k}=1}^{\dim(S_{j,k})} \ketbra{(S_{j,k}, s_{j,k})}{(S_{j,k}, s_{j,k})} = 1 & \qquad & \braket{(S_{j,k}, s_{j,k})}{(S_{j',k'}, s'_{j',k'})} = \delta_{j,j'} \delta_{k,k'} \delta_{s_{j,k},s'_{j',k'}} \\ \yesnumber
\end{IEEEeqnarray*}
Using the completeness relations,
\begin{IEEEeqnarray*}{rCl}
	\ket{\otimes_{i=1}^{n_p} (R_i, r_i)} & = & \sum_{j=1}^{n_s} \sum_{k=1}^{m_j} \sum_{s_{j,k}=1}^{\dim(S_{j,k})} \braket{(S_{j,k}, s_{j,k})}{\otimes_{i=1}^{n_p} (R_i, r_i)} \ \ket{(S_{j,k}, s_{j,k})} \\ [12pt]
	\ket{(S_{j,k}, s_{j,k})} & = & \sum_{r_1=1}^{\dim(R_1)} \cdots \sum_{r_{n_p}=1}^{\dim(R_{n_p})} \braket{\otimes_{i=1}^{n_p} (R_i, r_i)}{(S_{j,k}, s_{j,k})} \ \ket{\otimes_{i=1}^{n_p} (R_i, r_i)} \yesnumber
\end{IEEEeqnarray*}
The coefficients in these expansions are called Clebsch-Gordan coefficients (CGCs) \cite{ArneEtal:2011:cgcs}, and they can be chosen to be real numbers. As a consequence of the orthonormality conditions, the CGCs obey
\begin{IEEEeqnarray*}{rCl}
	\sum_{j=1}^{n_s} \sum_{k=1}^{m_j} \sum_{s_{j,k}=1}^{\dim(S_{j,k})} \braket{\otimes_{i=1}^{n_p} (R_i, r_i)}{(S_{j,k}, s_{j,k})} \ \braket{(S_{j,k}, s_{j,k})}{\otimes_{i=1}^{n_p} (R_i, r'_i)} & = & \prod_{i=1}^{n_p} \delta_{r_i,r'_i} \\ [12pt]
	\sum_{r_1=1}^{\dim(R_1)} \cdots \sum_{r_{n_p}=1}^{\dim(R_{n_p})} \braket{(S_{j,k}, s_{j,k})}{\otimes_{i=1}^{n_p} (R_i, r_i)} \ \braket{\otimes_{i=1}^{n_p} (R_i, r_i)}{(S_{j',k'}, s'_{j',k'})} & = & \delta_{j,j'} \delta_{k,k'} \delta_{s_{j,k},s'_{j',k'}} \\ \yesnumber
\end{IEEEeqnarray*}
This is the statement that the change-of-basis matrix formed by the CGCs is orthogonal.

\subsection{Representation Matrices}
\label{app:aspects:representation_matrices}

Let $D^R(g)$ be the unitary representation matrix for a group element $g \in \text{SU}(N_c)$ in an irrep $R$ of SU($N_c$). $R$ is a group homomorphism; as such,
\begin{equation}
	D^R(g) D^R(h) = D^R(gh)
\end{equation}
for all $g,h \in \text{SU}(N_c)$. Hence,
\begin{equation}
	D^R(g) D^R(g^{-1}) = D^R(1) = 1 \implies D^R(g^{-1}) = (D^R(g))^\dag
\end{equation}
In terms of matrix elements, $D^R_{mn}(g^{-1}) = (D^R_{nm}(g))^*$, where $(D^R(g))^*$ is the complex conjugate of $D^R(g)$. This defines the complex conjugate representation $R^*$ of $R$:
\begin{equation}
	D^{R^*}(g) = (D^R(g))^*
\end{equation}

Let $T^{Ra}$ be a basis element of $\mathfrak{su}(N_c)$ in the irrep $R$. By way of the exponential map,
\begin{equation}
	D^R(g) = e^{i \phi^a_g T^{Ra}} \implies D^{R^*}(g) = e^{-i \phi^a_g T^{Ra*}} \implies T^{R^*a} = -T^{Ra*}
\end{equation}
Basis vectors for the carrier space of an irrep can be distinguished by their weights, or their eigenvalues from the diagonal basis elements $T^{Rk}_z$ ($k=1,\dots,N_c-1$). Although the same carrier space basis can be used in $R$ and $R^*$, the weights of the basis vectors will differ. In particular, the weights of the basis vectors in $R^*$ are negative of those in $R$.

In practice, $R^*$ may be replaced by an isomorphic representation $\bar{R}$. That is,
\begin{equation}
	D^{\bar{R}}(g) = V D^{R^*}(g) V^\dag \implies T^{\bar{R}a} = VT^{R^*a}V^\dag
\end{equation}
$V$ may perform a change -- or a reordering -- of basis vectors, mapping a basis vector $v$ of $R$ to the basis vector $\tilde{v}$ of $\bar{R}$. However, if $w_v$ is the weight of $v$ in $R$, then the weight of $\tilde{v}$ must be $w_{\tilde{v}} = -w_v$ for $\bar{R}$ to remain isomorphic to $R^*$. Therefore, $V$ must be a permutation matrix whose elements are $|V_{\tilde{m}m}| = 1$.

One way to find $V$ for a given $\bar{R}$ is to check the CGCs for $\bar{R} \otimes R \to \mathbf{1}$, where $\mathbf{1}$ denotes the trivial representation. This is because $\sgn\{ \braket{\mathbf{1}}{(R^*,\tilde{r}) \otimes (R,r)} \} = \pm \delta_{\tilde{r}r}$. Therefore, $\sgn\{ \braket{\mathbf{1}}{(\bar{R},\tilde{r}) \otimes (R,r)} \}$ should provide the replacement to $\delta_{\tilde{r}r}$. Thus,
\begin{equation}
	V_{\tilde{r}r} = \sgn\{ \braket{\mathbf{1}}{(\bar{R},\tilde{r}) \otimes (R,r)} \} \equiv \varphi(r)
\end{equation}
Note that $V$ need not be symmetric. But because real-valued CGCs can always be chosen, $V$ can always be an orthogonal matrix. To summarize,
\begin{equation}
	D^R_{mn}(g^{-1}) = D^{R*}_{nm}(g) = V^\text{T}_{n\tilde{n}} D^{\bar{R}}_{\tilde{n}\tilde{m}}(g) V_{\tilde{m}m} = \varphi(n) \varphi(m) D^{\bar{R}}_{\tilde{n}\tilde{m}}(g)
\end{equation}
If $R$ is a real or pseudoreal representation, then $\bar{R}$ becomes a representation isomorphic to $R$. $V$ can still change the basis of $R$ with possible factors of $-1$.

\subsubsection*{Adjoint Representation}

If $R = \text{adj}$ is the adjoint representation, then $T^{\text{adj},a}_{bc} = -if^{abc}$ are a suitable basis for the algebra. Because the structure constants $f^{abc}$ are real, the representation matrices $D^\text{adj}(g)$ can always be written as real-valued matrices. Furthermore, these matrices have a special action on algebra-valued objects. For instance,
\begin{equation}
	D^\text{adj}_{ab}(g) T^{Rb} = D^R(g^{-1}) T^{Ra} D^R(g) = T^{Rb} D^\text{adj}_{ba}(g^{-1})
\end{equation}
where $D^\text{adj}(g^{-1}) = (D^\text{adj}(g))^\text{T}$.

\subsubsection*{Clebsch-Gordan Series}

A useful relation exists between CGCs and representation matrices. Group elements can act on a representation basis state, turning it into a linear combination of basis states, via matrix-vector multiplication:
\begin{equation}
	\ket{(R,r)} \to \sum_{t=1}^{\dim(R)} D^R_{rt}(g) \ket{(R,t)}
\end{equation}
For a general product basis state, the transformation can be viewed in both the product basis and the sum basis (after an insertion of the identity):
\begin{IEEEeqnarray*}{rCl}
	\ket{\otimes_{i=1}^{n_p} (R_i, r_i)} & \to & \sum_{t_1=1}^{\dim(R_1)} \cdots \sum_{t_{n_p}=1}^{\dim(R_{n_p})} D^{R_1}_{r_1 t_1}(g) \cdots D^{R_{n_p}}_{r_{n_p} t_{n_p}}(g) \ \ket{\otimes_{i=1}^{n_p} (R_i, t_i)} \\
	\ket{\otimes_{i=1}^{n_p} (R_i, r_i)} & \to & \sum_{j=1}^{n_s} \sum_{k=1}^{m_j} \sum_{s_{j,k}=1}^{\dim(S_{j,k})} \sum_{\ell_{j,k}=1}^{\dim(S_{j,k})} D^{S_{j,k}}_{s_{j,k} \ell_{j,k}}(g) \ \braket{(S_{j,k}, s_{j,k})}{\otimes_{i=1}^{n_p} (R_i, r_i)} \ \ket{(S_{j,k}, \ell_{j,k})} \\ \yesnumber
\end{IEEEeqnarray*}
Further expanding the sum state in the product basis, and equating both transformations, yields the Clebsch-Gordan series:
\begin{equation}
	\prod_{i=1}^{n_p} D^{R_i}_{r_i t_i}(g) = \sum_{j=1}^{n_s} \sum_{k=1}^{m_j} \sum_{s_{j,k}=1}^{\dim(S_{j,k})} \sum_{\ell_{j,k}=1}^{\dim(S_{j,k})} D^{S_{j,k}}_{s_{j,k} \ell_{j,k}}(g) \ \braket{(S_{j,k}, s_{j,k})}{\otimes_{i=1}^{n_p} (R_i, r_i)} \ \braket{\otimes_{i=1}^{n_p} (R_i, t_i)}{(S_{j,k}, \ell_{j,k})}
\end{equation}

\subsection{Link Hilbert Space Operations}
\label{app:aspects:link_hilbert_space}

This section was inspired by \cite{DAndreaEtal:2024:new_su2_basis}, although different conventions are followed here.

\subsubsection*{Preliminaries}

Each link on the lattice hosts an infinite-dimensional Hilbert space. The group element basis $\{ \ket{g} \}$ for $g \in \text{SU}(N_c)$ serves as a suitable Hilbert basis. Group element basis states are eigenstates of the link operator\footnote{For this section, when representation labels are suppressed in representation-dependent quantities such as $U^R_{mn}$, assume an arbitrary irrep $R$.}
\begin{equation}
	U_{mn} = \int dg \ D_{mn}(g) \ketbra{g}{g}
\end{equation}
where $dg$ is a Haar measure. Translation operators can be defined to transform one group element basis state into another. Because SU($N_c$) is a nonabelian group, this can be accomplished by two such operators, $\Theta_{\mathsf{L}g}$ and $\Theta_{\mathsf{R}g}$, that perform right or left group multiplication:
\begin{equation}
	\Theta_{\mathsf{L}g} \ket{h} = \ket{hg^{-1}} \qquad \Theta_{\mathsf{R}g} \ket{h} = \ket{g^{-1}h}
\end{equation}
The translation operators are group-valued operators and can be written in terms of algebra-valued operators via the exponential map:
\begin{equation}
	\Theta_{\mathsf{L}g} = e^{i\phi_g^a \Pi_\mathsf{L}^a} \qquad \Theta_{\mathsf{R}g} = e^{i\phi_g^a \Pi_\mathsf{R}^a}
\end{equation}
$\Pi_\mathsf{L}^a$ and $\Pi_\mathsf{R}^a$ are the generators of translation, or momentum operators. As will be later shown, these momentum operators furnish two algebras on each link, which informs a representation basis for the link Hilbert space. Representation basis states are written as $\ket{(R,r_\mathsf{R},r_\mathsf{L})} = \ket{(R,r_\mathsf{R})} \otimes \ket{(R,r_\mathsf{L})}$. The remainder of this section will concern identities among the operators and bases discussed so far.

\subsubsection*{Link Operator Transformations}

In the lattice gauge theory, translation operators are used to effect gauge transformations on the link operator. This is possible because, for example,
\begin{IEEEeqnarray*}{rCl}
	\Theta_{\mathsf{L}g} U_{mn} \Theta^\dag_{\mathsf{L}g} & = & \int dh \ D_{mn}(h) \Theta_{\mathsf{L}g} \ketbra{h}{h} \Theta^\dag_{\mathsf{L}g} \\
	& = & \int dh \ D_{mn}(h) \ketbra{hg^{-1}}{hg^{-1}} \\
	& = & \int dh' \ D_{mn}(h'g) \ketbra{h'}{h'} \\
	& = & U_{mm'} D_{m'n}(g) \yesnumber
\end{IEEEeqnarray*}
The left and right translation operators transform the link operator as
\begin{equation}
	\Theta_{\mathsf{L}g} U_{mn} \Theta^\dag_{\mathsf{L}g} = U_{mm'} D_{m'n}(g) \qquad \Theta_{\mathsf{R}g} U_{mn} \Theta^\dag_{\mathsf{R}g} = D_{mm'}(g) U_{m'n}
\end{equation}
Because the translation operators are unitary operators, it follows that $\Theta_{\mathsf{L}g^{-1}} = \Theta_{\mathsf{L}g}^\dag$ and $\Theta_{\mathsf{R}g^{-1}} = \Theta_{\mathsf{R}g}^\dag$. Therefore, transformations with respect to $g^{-1}$ can be made according to
\begin{equation}
	\Theta^\dag_{\mathsf{L}g} U_{mn} \Theta_{\mathsf{L}g} = U_{mm'} D_{m'n}(g^{-1}) \qquad \Theta^\dag_{\mathsf{R}g} U_{mn} \Theta_{\mathsf{R}g} = D_{mm'}(g^{-1}) U_{m'n}
\end{equation}

\subsubsection*{Momentum Operator Commutators}

Link operator transformations close to the identity ($\phi_g^a \approx 0$) contain a commutator. For example, consider that
\begin{IEEEeqnarray*}{rCl}
	\Theta_{\mathsf{L}g} U_{mn} \Theta_{\mathsf{L}g}^\dag & \approx & (1 + i\phi^a_g \Pi^a_\mathsf{L}) U_{mn} (1 - i\phi^a_g \Pi^a_\mathsf{L}) \\
	& \approx & U_{mn} + i\phi^a_g \Pi_\mathsf{L}^a U_{mn} - i\phi^a_g U_{mn} \Pi_\mathsf{L}^a \\
	& = & U_{mn} + i\phi^a_g [\Pi_\mathsf{L}^a, U_{mn}] \yesnumber
\end{IEEEeqnarray*}
and also
\begin{equation}
	U_{mm'} D_{m'n}(g) \approx U_{mn} + i\phi^a_g U_{mm'} T^a_{m'n}
\end{equation}
Hence,
\begin{equation}
	[\Pi_\mathsf{L}^a, U_{mn}] = U_{mm'} T^a_{m'n} \qquad [\Pi_\mathsf{R}^a, U_{mn}] = T^a_{mm'} U_{m'n}
\end{equation}
As an application, by using the antisymmetry of $T^{\text{adj},a}_{bc} = -if^{abc}$, it straightforward to see that
\begin{equation}
	[\Pi_\mathsf{L}^b, U^\text{adj}_{ab}] = 0 \qquad [\Pi_\mathsf{R}^b, U^\text{adj}_{ba}] = 0
\end{equation}

Furthermore, the Jacobi identity implies commutation relations among the momentum operators. For instance,
\begin{IEEEeqnarray*}{rCl}
	& & [\Pi^a_\mathsf{L}, [\Pi^b_\mathsf{L}, U_{mn}]] + [\Pi^b_\mathsf{L}, [U_{mn}, \Pi^a_\mathsf{L}]] + [U_{mn}, [\Pi^a_\mathsf{L}, \Pi^b_\mathsf{L}]] \\
	& = & [\Pi^a_\mathsf{L}, U_{mm'}] T^b_{m'n} - [\Pi^b_\mathsf{L}, U_{mm'}] T^a_{m'n} + [U_{mn}, [\Pi^a_\mathsf{L}, \Pi^b_\mathsf{L}]] \\
	& = & U_{mn'} T^a_{n'm'} T^b_{m'n} - U_{mn'} T^b_{n'm'} T^a_{m'n} + [U_{mn}, [\Pi^a_\mathsf{L}, \Pi^b_\mathsf{L}]] \\
	& = & U_{mn'} [T^a, T^b]_{n'n} + [U_{mn}, [\Pi^a_\mathsf{L}, \Pi^b_\mathsf{L}]] \\
	& = & i f^{abc} U_{mn'} T^c_{n'n} + [U_{mn}, [\Pi^a_\mathsf{L}, \Pi^b_\mathsf{L}]] \\
	& = & 0 \\
	& \implies & [U_{mn}, [\Pi^a_\mathsf{L}, \Pi^b_\mathsf{L}]] = -if^{abc} U_{mn'} T^c_{n'n} = [U_{mn}, if^{abc} \Pi^c_\mathsf{L}] \yesnumber
\end{IEEEeqnarray*}
The calculation can be repeated with other momentum operators to find that
\begin{equation}
	[\Pi^a_\mathsf{L}, \Pi^b_\mathsf{L}] = if^{abc} \Pi^c_\mathsf{L} \qquad [\Pi^a_\mathsf{R}, \Pi^b_\mathsf{R}] = -if^{abc} \Pi^c_\mathsf{R} \qquad [\Pi^a_\mathsf{L}, \Pi^b_\mathsf{R}] = 0
\end{equation}
The first two commutators resemble $[T^a,T^b]=if^{abc}T^c$, which implies the momentum operators are bases for two algebras of operators on each link.

\subsubsection*{Relation Between Left and Right Translations}

The left and right translation operators can be written in terms of one another. For example,
\begin{equation}
	\Theta_{\mathsf{L}g} \ket{h} = \ket{hg^{-1}} = \ket{hg^{-1}h^{-1}h} = \Theta_{\mathsf{R}h^{-1}gh} \ket{h}
\end{equation}
which, in exponential form, shows
\begin{equation}
	e^{i \phi^a_g \Pi^a_\mathsf{L}} \ket{h} = e^{i D^\text{adj}_{ba}(h) \phi^a_g \Pi^b_\mathsf{R}} \ket{h}
\end{equation}
This implies the left and right momentum operators are related. To see this, unpack the exponential, using the fact that $[U_{ba},U_{b'a'}]=0$:
\begin{IEEEeqnarray*}{rCl}
	e^{i D^\text{adj}_{ba}(h) \phi^a_g \Pi^b_\mathsf{R}} \ket{h} & = & \sum_{k=0}^{\infty} \frac{(i D^\text{adj}_{ba}(h) \phi^a_g \Pi^b_\mathsf{R})^k}{k!} \ket{h} \\
	& = & \sum_{k=0}^{\infty} \frac{i^k}{k!} (\textstyle\prod_{i=1}^{k} \Pi^{b_i}_\mathsf{R}) (\textstyle\prod_{i=1}^{k} \phi_g^{a_i} U^\text{adj}_{b_i a_i}) \ket{h} \\
	& = & \sum_{k=0}^{\infty} \frac{i^k}{k!} (\textstyle\prod_{i=1}^{k-1} \Pi^{b_i}_\mathsf{R}) \phi^{a_k}_g \Pi^{b_k}_\mathsf{R} U^\text{adj}_{b_k a_k} (\textstyle\prod_{i=1}^{k-1} \phi_g^{a_i} U^\text{adj}_{b_i a_i}) \ket{h} \yesnumber
\end{IEEEeqnarray*}
It appears this can be written as an exponential of the form $e^{\Pi U}$. However, to proceed requires the commutator
\begin{IEEEeqnarray*}{rCl}
	& & [\phi_g^{a_k} \Pi^{b_k}_\mathsf{R} U^\text{adj}_{b_k a_k}, \phi_g^{a_1} U^\text{adj}_{b_1 a_1} \cdots \phi_g^{a_{k-1}} U^\text{adj}_{b_{k-1} a_{k-1}}] \\
	& = & \sum_{i=1}^{k-1} \phi_g^{a_1} U^\text{adj}_{b_1 a_1} \cdots \phi_g^{a_{i-1}} U^\text{adj}_{b_{i-1} a_{i-1}} [ \phi_g^{a_k} \Pi^{b_k}_\mathsf{R} U^\text{adj}_{b_k a_k}, \phi_g^{a_i} U^\text{adj}_{b_i a_i} ] \phi_g^{a_{i+1}} U^\text{adj}_{b_{i+1} a_{i+1}} \cdots \phi_g^{a_{k-1}} U^\text{adj}_{b_{k-1} a_{k-1}} \\
	& = & \sum_{i=1}^{k-1} \phi_g^{a_1} U^\text{adj}_{b_1 a_1} \cdots \phi_g^{a_{i-1}} U^\text{adj}_{b_{i-1} a_{i-1}} [ \phi_g^{a_k} \Pi^{b_k}_\mathsf{R}, \phi_g^{a_i} U^\text{adj}_{b_i a_i} ] U^\text{adj}_{b_k a_k} \phi_g^{a_{i+1}} U^\text{adj}_{b_{i+1} a_{i+1}} \cdots \phi_g^{a_{k-1}} U^\text{adj}_{b_{k-1} a_{k-1}} \\
	& = & \sum_{i=1}^{k-1} \phi_g^{a_1} U^\text{adj}_{b_1 a_1} \cdots \phi_g^{a_{i-1}} U^\text{adj}_{b_{i-1} a_{i-1}} \phi_g^{a_k} (-if_{b_k b_i m}) U^\text{adj}_{m a_i} \phi_g^{a_i} U^\text{adj}_{b_k a_k} \phi_g^{a_{i+1}} U^\text{adj}_{b_{i+1} a_{i+1}} \cdots \phi_g^{a_{k-1}} U^\text{adj}_{b_{k-1} a_{k-1}} \\
	& = & \sum_{i=1}^{k-1} \phi_g^{a_1} U^\text{adj}_{b_1 a_1} \cdots \phi_g^{a_{i-1}} U^\text{adj}_{b_{i-1} a_{i-1}} \phi_g^{a_i} (+if_{m b_i b_k}) U^\text{adj}_{b_k a_k} \phi_g^{a_k} U^\text{adj}_{m a_i} \phi_g^{a_{i+1}} U^\text{adj}_{b_{i+1} a_{i+1}} \cdots \phi_g^{a_{k-1}} U^\text{adj}_{b_{k-1} a_{k-1}} \\
	& = & 0 \yesnumber
\end{IEEEeqnarray*}
Or, for either momentum operator,
\begin{equation}
	[\phi_g^{a_k} \Pi^{b_k}_\mathsf{R} U^\text{adj}_{b_k a_k}, \textstyle\prod_{i=1}^{k-1} \phi_g^{a_i} U^\text{adj}_{b_i a_i}] = 0 \qquad [\phi_g^{a_k} \Pi^{b_k}_\mathsf{L} U^\text{adj}_{a_k b_k}, \textstyle\prod_{i=1}^{k-1} \phi_g^{a_i} U^\text{adj}_{a_i b_i}] = 0
\end{equation}
By repeated use of this commutator,
\begin{IEEEeqnarray*}{rCl}
	e^{i D^\text{adj}_{ba}(h) \phi^a_g \Pi^b_\mathsf{R}} \ket{h} & = & \sum_{k=0}^{\infty} \frac{i^k}{k!} (\textstyle\prod_{i=1}^{k-1} \Pi^{b_i}_\mathsf{R}) (\textstyle\prod_{i=1}^{k-1} \phi_g^{a_i} U^\text{adj}_{b_i a_i}) \phi_g^{a_k} \Pi^{b_k}_\mathsf{R} U^\text{adj}_{b_k a_k} \ket{h} \\
	& \vdots & \\
	& = & \sum_{k=0}^{\infty} \frac{(i \phi^a_g \Pi^b_\mathsf{R} U^\text{adj}_{ba})^k}{k!} \ket{h} \\
	& = & e^{i \phi^a_g \Pi^b_\mathsf{R} U^\text{adj}_{ba}} \ket{h} \yesnumber
\end{IEEEeqnarray*}
This calculation done for both translation operators shows
\begin{equation}
	e^{i \phi^a_g \Pi^a_\mathsf{L}} \ket{h} = e^{i \phi^a_g \Pi^b_\mathsf{R} U^\text{adj}_{ba}} \ket{h} \qquad e^{i \phi^a_g \Pi^a_\mathsf{R}} \ket{h} = e^{i \phi^a_g U^\text{adj}_{ab} \Pi^b_\mathsf{L}} \ket{h}
\end{equation}
Hence, the momentum operators are related by parallel transport:
\begin{equation}
	\Pi^a_{\mathsf{L}} = \Pi^b_{\mathsf{R}} U^\text{adj}_{ba} \qquad \Pi^a_{\mathsf{R}} = U^\text{adj}_{ab} \Pi^b_{\mathsf{L}}
\end{equation}

There are two useful corollaries. First, $U^\text{adj}_{ba} (U^\text{adj})^\text{T}_{ac} = U^\text{adj}_{ba} U^\text{adj}_{ca} = \delta_{bc}$ implies that both quadratic Casimir operators are equal:
\begin{equation}
	\Pi^a_{\mathsf{L}} \Pi^a_{\mathsf{L}} = \Pi^b_{\mathsf{R}} U^\text{adj}_{ba} \Pi^c_{\mathsf{R}} U^\text{adj}_{ca} = \Pi^a_{\mathsf{R}} \Pi^a_{\mathsf{R}}
\end{equation}
Second, the transformations of $\Pi_{\mathsf{L}} \equiv \Pi_\mathsf{L}^a T^a$ and $\Pi_{\mathsf{R}} \equiv \Pi_\mathsf{R}^a T^a$ can be found. For example,
\begin{IEEEeqnarray*}{rCl}
	\Theta_{\mathsf{L}g} \Pi^a_{\mathsf{L}} T^a \Theta^\dag_{\mathsf{L}g} & = & \Theta_{\mathsf{L}g} \Pi^b_{\mathsf{R}} U^\text{adj}_{ba} T^a \Theta^\dag_{\mathsf{L}g} \\
	& = & \Pi^b_{\mathsf{R}} U^\text{adj}_{bc} D^\text{adj}_{ca}(g) T^a \\
	& = & \Pi^c_{\mathsf{L}} D^\text{adj}_{ca}(g) T^a \\
	& = & D(g^{-1}) \Pi^a_{\mathsf{L}} T^a D(g) \yesnumber
\end{IEEEeqnarray*}
Thus,
\begin{equation}
	\Theta_{\mathsf{L}g} \Pi_{\mathsf{L}} \Theta^\dag_{\mathsf{L}g} = D(g^{-1}) \Pi_{\mathsf{L}} D(g) \qquad \Theta_{\mathsf{R}g} \Pi_{\mathsf{R}} \Theta^\dag_{\mathsf{R}g} = D(g) \Pi_{\mathsf{R}} D(g^{-1})
\end{equation}
which matches the form of a gauge transformation of the canonical momentum operator, or the field strength tensor.

\subsubsection*{Change of Basis}

The relation between the group element and representation bases can be clarified by the great orthogonality theorem,
\begin{equation}
	\int dg \ (D^S_{s_\mathsf{R} s_\mathsf{L}}(g))^* D^R_{r_\mathsf{R} r_\mathsf{L}}(g) = \frac{|G|}{\dim(R)} \delta_{R,S} \delta_{r_\mathsf{R},s_\mathsf{R}} \delta_{r_\mathsf{L},s_\mathsf{L}}
\end{equation}
where $|G|$ is the volume of the group. Then a comparison with
\begin{equation}
	\braket{(R,r_\mathsf{R},r_\mathsf{L})}{(R,r_\mathsf{R},r_\mathsf{L})} = 1 = \int dg \ \braket{(R,r_\mathsf{R},r_\mathsf{L})}{g} \braket{g}{(R,r_\mathsf{R},r_\mathsf{L})}
\end{equation}
suggests
\begin{equation}
	\braket{g}{(R,r_\mathsf{R},r_\mathsf{L})} = \sqrt{\frac{\dim(R)}{|G|}} D^R_{r_\mathsf{R} r_\mathsf{L}}(g)
\end{equation}

In particular, if $R = \mathbf{1}$ is the trivial representation, then $\braket{g}{\mathbf{1}} = \frac{1}{\sqrt{|G|}}$. As a result,
\begin{IEEEeqnarray*}{rCl}
	\ket{(R,r_\mathsf{R},r_\mathsf{L})} & = & \int dg \ \ket{g} \braket{g}{(R,r_\mathsf{R},r_\mathsf{L})} \\
	& = & \sqrt{\frac{\dim(R)}{|G|}} \int dg \ D^R_{r_\mathsf{R} r_\mathsf{L}}(g) \ket{g} \\
	& = & \sqrt{\dim(R)} \int dg \ D^R_{r_\mathsf{R} r_\mathsf{L}}(g) \ket{g} \braket{g}{\mathbf{1}} \\
	& = & \sqrt{\dim(R)} U^R_{r_\mathsf{R} r_\mathsf{L}} \ket{\mathbf{1}} \yesnumber
\end{IEEEeqnarray*}
Therefore, any representation basis state can be created by acting on the trivial state with a link operator.

\subsubsection*{Operators in the Representation Basis}

The trivial state is invariant under translations; for example,
\begin{equation}
	\Theta_{\mathsf{L}g} \ket{\mathbf{1}} = \int dh \ \ketbra{h}{h} \Theta_{\mathsf{L}g} \ket{\mathbf{1}} = \int dh \ \ket{h} \braket{hg}{\mathbf{1}} = \int dh \ \ket{h} \braket{h}{\mathbf{1}} = \ket{\mathbf{1}}
\end{equation}
This implies $\Pi^a_\mathsf{L} \ket{\mathbf{1}} = \Pi^a_\mathsf{R} \ket{\mathbf{1}} = 0$. Then the action of these operators on representation basis states can be computed. The momentum operators act as
\begin{equation}
	\Pi^a_\mathsf{L} \ket{(R,r_\mathsf{R},r_\mathsf{L})} = \sum_{s_\mathsf{L}=1}^{\dim(R)} T^{Ra}_{s_\mathsf{L} r_\mathsf{L}} \ket{(R,r_\mathsf{R},s_\mathsf{L})} \qquad \Pi^a_\mathsf{R} \ket{(R,r_\mathsf{R},r_\mathsf{L})} = \sum_{s_\mathsf{R}=1}^{\dim(R)} T^{Ra}_{r_\mathsf{R} s_\mathsf{R}} \ket{(R,s_\mathsf{R},r_\mathsf{L})}
\end{equation}
This verifies that the representation basis states are eigenstates of the quadratic Casimir operator with the quadratic Casimir eigenvalue of the representation:
\begin{equation}
	\Pi^a_\mathsf{L} \Pi^a_\mathsf{L} \ket{(R,r_\mathsf{R},r_\mathsf{L})} = C_2(R) \ket{(R,r_\mathsf{R},r_\mathsf{L})} \qquad \Pi^a_\mathsf{R} \Pi^a_\mathsf{R} \ket{(R,r_\mathsf{R},r_\mathsf{L})} = C_2(R) \ket{(R,r_\mathsf{R},r_\mathsf{L})}
\end{equation}
which follows from $T^{Ra}_{mm'} T^{Ra}_{m'n} = C_2(R) \delta_{mn}$. The translation operators act as
\begin{equation}
	\Theta_{\mathsf{L}g} \ket{(R,r_\mathsf{R},r_\mathsf{L})} = \sum_{s_\mathsf{L}=1}^{\dim(R)} D^R_{s_\mathsf{L} r_\mathsf{L}}(g) \ket{(R,r_\mathsf{R},s_\mathsf{L})} \qquad \Theta_{\mathsf{R}g} \ket{(R,r_\mathsf{R},r_\mathsf{L})} = \sum_{s_\mathsf{R}=1}^{\dim(R)} D^R_{r_\mathsf{R} s_\mathsf{R}}(g) \ket{(R,s_\mathsf{R},r_\mathsf{L})}
\end{equation}

Finally, it is useful to evaluate the link operator in the representation basis. By definition of the link operator,
\begin{equation}
	\bra{(R',r'_\mathsf{R},r'_\mathsf{L})} U_{mn}^\mathcal{R} \ket{(R,r_\mathsf{R},r_\mathsf{L})} = \frac{\sqrt{\dim(R') \dim(R)}}{|G|} \int dg \ D^\mathcal{R}_{mn}(g) (D^{R'}_{r'_\mathsf{R} r'_\mathsf{L}}(g))^* D^R_{r_\mathsf{R} r_\mathsf{L}}(g)
\end{equation}
The Clebsch-Gordan series can be applied to the unconjugated matrix elements:
\begin{equation}
	\hspace{-0.375in} D^\mathcal{R}_{mn}(g) D^R_{r_\mathsf{R} r_\mathsf{L}}(g) = \sum_{j=1}^{n_s} \sum_{k=1}^{m_j} \sum_{s_{j,k}=1}^{\dim(S_{j,k})} \sum_{\ell_{j,k}=1}^{\dim(S_{j,k})} D^{S_{j,k}}_{s_{j,k} \ell_{j,k}}(g) \ \braket{(S_{j,k}, s_{j,k})}{(\mathcal{R},m) \otimes (R,r_\mathsf{R})} \ \braket{(\mathcal{R},n) \otimes (R,r_\mathsf{L})}{(S_{j,k}, \ell_{j,k})}
\end{equation}
However, the great orthogonality theorem collapses most of these sums, except the one over multiplicities. Therefore,
\begin{equation}
	\hspace{-0.125in} \bra{(R',r'_\mathsf{R},r'_\mathsf{L})} U_{mn}^\mathcal{R} \ket{(R,r_\mathsf{R},r_\mathsf{L})} = \sqrt{\frac{\dim(R)}{\dim(R')}} \sum_{k=1}^{m_{R'}} \braket{(R'_k,r'_\mathsf{R})}{(\mathcal{R},m) \otimes (R,r_\mathsf{R})} \ \braket{(\mathcal{R},n) \otimes (R,r_\mathsf{L})}{(R'_k,r'_\mathsf{L})}
\end{equation}
When $\mathcal{R}$ is the fundamental representation, $m_{R'}=1$, and the sum can be omitted.

\subsection{Site Hilbert Space Operations}
\label{app:aspects:site_hilbert_space}

The content and definitions in the section were inspired by \cite{KasperEtal:2020:jaynes_cummings_model}.

\subsubsection*{Preliminaries}

Each site on the lattice hosts multiple finite-dimensional Hilbert spaces, corresponding to states of particles/antiparticles of a given flavor, spinor index, and color index. These single particle/antiparticle states are spanned by an unoccupied state $\ket{0}$ and an occupied state, $\ket{1}=\psi^{\dag(f)}_{\alpha,c}(\vec{s})\ket{0}$, or $\ket{1}=\chi^{\dag(f)}_{\alpha,c}(\vec{s})\ket{0}$. A tensor product over all such basis states is an occupation number basis state. The particle/antiparticle annihilation and creation operators have nonvanishing anticommutation relations
\begin{equation}
	\{ \psi^{(f)}_{\alpha,c}(\vec{s}), \psi_{\beta,c'}^{\dag(f')}(\vec{r}) \} = (P_+)_{\alpha\beta} \delta_{cc'} \delta_{ff'} \delta_{\vec{s},\vec{r}} \qquad \{ \chi_{\alpha,c}^{\dag(f)}(\vec{s}), \chi^{(f')}_{\beta,c'}(\vec{r}) \} = (P_-)_{\alpha\beta} \delta_{cc'} \delta_{ff'} \delta_{\vec{s},\vec{r}}
\end{equation}
Under gauge transformations, $\psi'(\vec{s}) =\Omega(\vec{s}) \psi(\vec{s})$, whereas,
\begin{equation}
	\chi'(\vec{s}) = \chi(\vec{s}) \Omega^\dag(\vec{s}) = \Omega^*(\vec{s}) \chi(\vec{s})
\end{equation}
Thus, particle states will transform in an irrep $R$ and antiparticle states will transform in the complex conjugate irrep $R^*$.

Color charge $\mathcal{Q}^a$ is conserved in the lattice gauge theory. It contains charge density operators
\begin{equation}
	Q_\alpha^{a(f)}(\vec{s}) = \psi_{\alpha,c}^{\dag(f)}(\vec{s}) T^a_{cc'} \psi_{\alpha,c'}^{(f)}(\vec{s}) \qquad \bar{Q}_\alpha^{a(f)}(\vec{s}) = -\chi_{\alpha,c}^{\dag(f)}(\vec{s}) T^{a*}_{cc'} \chi_{\alpha,c'}^{(f)}(\vec{s})
\end{equation}
where only the color indices $c,c'$ are summed over. Translation operators are defined from these charge operators:
\begin{equation}
	\Theta_{Q_\alpha g}^{(f)}(\vec{s}) = e^{i \phi_g^a Q_\alpha^{a(f)}(\vec{s})} \qquad \bar{\Theta}_{Q_\alpha g}^{(f)}(\vec{s}) = e^{i \phi_g^a \bar{Q}_\alpha^{a(f)}(\vec{s})}
\end{equation}
As will be shown, the charge operators form bases for algebras of operators. These algebras partition the Hilbert spaces at each site into color subspaces of particles/antiparticles of a given flavor and spinor index. Furthermore, the operator algebras inform a representation basis that spans each color subspace. The remainder of this section will concern identities among the operators and bases discussed so far.

\subsubsection*{Annihilation/Creation Operator Transformations}

Commutation relations for the charge operators with the annihilators/creators are\footnote{For the rest of this section, flavor, spinor, and site labels are dropped. Also, representation dependent quantities such as $T^{Ra}$ written without a representation label are understood to be in an arbitrary irrep $R$.}
\begin{IEEEeqnarray*}{rClCrCl}
	[\psi_c, Q^a] & = & T^a_{cc'} \psi_{c'} & \qquad & [\psi^\dag_c, Q^a] & = & -\psi^\dag_{c'} T^a_{c'c} \\ \relax
	[\chi_c, \bar{Q}^a] & = & (-T^{a*})_{cc'} \chi_{c'} & \qquad & [\chi^\dag_c, \bar{Q}^a] & = & -\chi^\dag_{c'} (-T^{a*})_{c'c} \yesnumber
\end{IEEEeqnarray*}
The iterated commutator then applies a power of a basis element; for example,
\begin{IEEEeqnarray*}{rCl}
	[\phi_g^a Q^a, \psi_c]_k & = & [\phi_g^{a_k} Q^{a_k}, \dots, [\phi_g^{a_2} Q^{a_2}, [\phi_g^{a_1} Q^{a_1}, \psi_c]] \cdots ] \\
	& = & -\phi_g^{a_1} T^{a_1}_{cc'} [\phi_g^{a_k} Q^{a_k}, \dots, [\phi_g^{a_2} Q^{a_2}, \psi_{c'}] \cdots ] \\
	& = & (-\phi_g^a T^a)^k_{cc'} \psi_{c'} \yesnumber
\end{IEEEeqnarray*}
Then, by the Campbell identity,
\begin{equation}
	e^{-i\phi_g^a Q^a} \psi_c e^{i\phi_g^a Q^a} = \sum_{k=0}^{\infty} \frac{[-i\phi_g^a Q^a, \psi_c]_k}{k!} = \sum_{k=0}^{\infty} \frac{(i\phi_g^a T^a)^k_{cc'}}{k!} \psi_{c'} = D_{cc'}(g) \psi_{c'}
\end{equation}
Repeating this process for the rest of the commutators yields
\begin{IEEEeqnarray*}{rCl}
	\Theta^\dag_{Q g} \psi_c \Theta_{Q g} = D^R_{cc'}(g) \psi_{c'} & \qquad & \Theta^\dag_{Q g} \psi^\dag_c \Theta_{Q g} = \psi^\dag_{c'} D^R_{c'c}(g^{-1}) \\
	\bar{\Theta}^\dag_{Q g} \chi_c \bar{\Theta}_{Q g} = D^{R^*}_{cc'}(g) \chi_{c'} & \qquad & \bar{\Theta}^\dag_{Q g} \chi^\dag_c \bar{\Theta}_{Q g} = \chi^\dag_{c'} D^{R^*}_{c'c}(g^{-1}) \yesnumber
\end{IEEEeqnarray*}
which shows the translation operators can effect the desired gauge transformations.

\subsubsection*{Charge Operator Commutators}

The commutation relations for the charge operators can be readily calculated. For instance,
\begin{IEEEeqnarray*}{rCl}
	[Q^a, Q^b] & = & T^a_{cc'} T^b_{dd'} \psi^\dag_c \{ \psi_{c'}, \psi^\dag_d \} \psi_{d'} - T^a_{cc'} T^b_{dd'} \{ \psi^\dag_c, \psi^\dag_d \} \psi_{c'} \psi_{d'} \\
	& + & T^a_{cc'} T^b_{dd'} \psi^\dag_d \psi^\dag_c \{ \psi_{c'}, \psi_{d'} \} - T^a_{cc'} T^b_{dd'} \psi^\dag_d \{ \psi^\dag_c, \psi_{d'} \} \psi_{c'} \\
	& = & T^a_{cd} T^b_{dd'} \psi_c^\dag \psi_{d'} - T^a_{cc'} T^b_{dc} \psi_d^\dag \psi_{c'} \\
	& = & \psi^\dag_c \left( T^a_{cc'} T^b_{c'd} - T^b_{cc'} T^a_{c'd} \right) \psi_d \\
	& = & if^{abe} \psi_c^\dag T^e_{cd} \psi_d \\
	& = & if^{abc} Q^c(\vec{s}) \yesnumber
\end{IEEEeqnarray*}
In total,
\begin{equation}
	[Q^a, Q^b] = if^{abc} Q^c \qquad [\bar{Q}^a, \bar{Q}^b] = if^{abc} \bar{Q}^c \qquad [Q^a, \bar{Q}^b] = 0
\end{equation}
and therefore each site contains algebras of operators for particle and antiparticle states. Each charge operator sums over the $N_c$ colors, and so they act on a $2^{N_c}$-dimensional color subspace.

\subsubsection*{Representation Basis State Transformations}

The representation basis for fermions is a change of basis from the color occupation number basis. Let $N_q$ be the color occupation number, defined via
\begin{equation}
	\sum_{c=1}^{N_c} \psi_c^\dag \psi_c \ket{n_1, \dots, n_{N_c}} = \sum_{i=1}^{N_c} n_i \ket{n_1, \dots, n_{N_c}} \equiv N_q \ket{n_1, \dots, n_{N_c}}
\end{equation}
where $\ket{n_1, \dots, n_{N_c}}$ is a color occupation number basis state for a color subspace. Then the representation basis states for a lattice theory in the representation $\mathcal{R}$ can be expressed as
\begin{IEEEeqnarray*}{rCl}
	\ket{(R_{k,N_q},r)} & \coloneq & \frac{1}{\sqrt{N_q!}} \sum_{c_1=1}^{\dim(\mathcal{R})} \cdots \sum_{c_{N_q}=1}^{\dim(\mathcal{R})} \braket{(R_k,r)}{\otimes_{i=1}^{N_q} (\mathcal{R},c_i)} \prod_{i=1}^{N_q} \psi_{c_i}^\dag \ket{0} \\
	\bar{\ket{(R_{k,N_q},r)}} & \coloneq & \frac{1}{\sqrt{N_q!}} \sum_{c_1=1}^{\dim(\mathcal{R})} \cdots \sum_{c_{N_q}=1}^{\dim(\mathcal{R})} \braket{\otimes_{i=1}^{N_q} (\mathcal{R},c_i)}{(\bar{R}_k,\tilde{r})} \prod_{i=1}^{N_q} \chi_{c_i}^\dag \ket{0} \yesnumber
\end{IEEEeqnarray*}
The $\ket{(R_{k,N_q},r)}$ states correspond to particle states, while the $\bar{\ket{(R_{k,N_q},r)}}$ states correspond to antiparticle states. The CGCs present in these definitions are contracted with an antisymmetric product of creation operators. As such, the CGCs must manifest antisymmetry. This forms the restriction that $R_k$ or $\bar{R}_k$ is an antisymmetric irrep found in the direct-sum decomposition of $\mathcal{R}^{\otimes N_q}$. (Here, $k$ is a multiplicity label.) Generally, the representation basis states are superpositions of color occupation number basis states. However, when $\mathcal{R}$ is the fundamental representation, they are equal to color occupation number basis states, and, additionally, the $k,N_q$ labels can be dropped.

The zero-point state $\ket{0}$ is invariant under gauge transformations because $Q^a \ket{0} = 0$ and $\bar{Q}^a \ket{0} = 0$. Hence, it is associated with a trivial representation. All other representation basis states may be acted on nontrivially by the translation operators:
\begin{IEEEeqnarray*}{rCl}
	\Theta_{Qg} \ket{(R_{k,N_q},r)} & = & \frac{1}{\sqrt{N_q!}} \sum_{c_1,d_1=1}^{\dim(\mathcal{R})} \cdots \sum_{c_{N_q},d_{N_q}=1}^{\dim(\mathcal{R})} \braket{(R_k,r)}{\otimes_{i=1}^{N_q} (\mathcal{R},c_i)} \prod_{i=1}^{N_q} \psi_{d_i}^\dag D^\mathcal{R}_{d_i c_i}(g) \ket{0} \\
	\bar{\Theta}_{Qg} \bar{\ket{(R_{k,N_q},r)}} & = & \frac{1}{\sqrt{N_q!}} \sum_{c_1,d_1=1}^{\dim(\mathcal{R})} \cdots \sum_{c_{N_q},d_{N_q}=1}^{\dim(\mathcal{R})} \braket{\otimes_{i=1}^{N_q} (\mathcal{R},c_i)}{(\bar{R}_k,\tilde{r})} \prod_{i=1}^{N_q} D^\mathcal{R}_{c_i d_i}(g^{-1}) \chi_{d_i}^\dag \ket{0} \\ \yesnumber
\end{IEEEeqnarray*}
This includes a product of representation matrix elements; it can be expanded with the Clebsch-Gordan series:
\begin{IEEEeqnarray*}{rClll}
	\Theta_{Qg} \ket{(R_{k,N_q},r)} & = & \frac{1}{\sqrt{N_q!}} \sum_{c_1,d_1=1}^{\dim(\mathcal{R})} & \cdots & \sum_{c_{N_q},d_{N_q}=1}^{\dim(\mathcal{R})} \sum_{j=1}^{n_s} \sum_{k'=1}^{m_j} \sum_{s_{j,k'}=1}^{\dim(S_{j,k'})} \sum_{\ell_{j,k'}=1}^{\dim(S_{j,k'})} D^{S_{j,k'}}_{s_{j,k'} \ell_{j,k'}}(g) \ \braket{(R_k,r)}{\otimes_{i=1}^{N_q} (\mathcal{R},c_i)} \\
	& & & \times & \braket{(S_{j,k'}, s_{j,k'})}{\otimes_{i=1}^{N_q} (\mathcal{R},d_i)} \ \braket{\otimes_{i=1}^{N_q} (\mathcal{R},c_i)}{(S_{j,k'}, \ell_{j,k'})} \prod_{i=1}^{N_q} \psi_{d_i}^\dag \ket{0} \\
	& = & \frac{1}{\sqrt{N_q!}} \sum_{d_1=1}^{\dim(\mathcal{R})} & \cdots & \sum_{d_{N_q}=1}^{\dim(\mathcal{R})} \sum_{s=1}^{\dim(R)} D^{R}_{s r}(g) \ \braket{(R_k, s)}{\otimes_{i=1}^{N_q} (\mathcal{R},d_i)} \prod_{i=1}^{N_q} \psi_{d_i}^\dag \ket{0} \\
	\bar{\Theta}_{Qg} \bar{\ket{(R_{k,N_q},r)}} & = & \frac{1}{\sqrt{N_q!}} \sum_{c_1,d_1=1}^{\dim(\mathcal{R})} & \cdots & \sum_{c_{N_q},d_{N_q}=1}^{\dim(\mathcal{R})} \sum_{j=1}^{n_s} \sum_{k'=1}^{m_j} \sum_{s_{j,k'}=1}^{\dim(S_{j,k'})} \sum_{\ell_{j,k'}=1}^{\dim(S_{j,k'})} D^{S_{j,k'}}_{s_{j,k'} \ell_{j,k'}}(g^{-1}) \ \braket{\otimes_{i=1}^{N_q} (\mathcal{R},c_i)}{(\bar{R}_k,\tilde{r})} \\
	& & & \times & \braket{(S_{j,k'}, s_{j,k'})}{\otimes_{i=1}^{N_q} (\mathcal{R},c_i)} \ \braket{\otimes_{i=1}^{N_q} (\mathcal{R},d_i)}{(S_{j,k'}, \ell_{j,k'})} \prod_{i=1}^{N_q} \chi_{d_i}^\dag \ket{0} \\
	& = & \frac{1}{\sqrt{N_q!}} \sum_{d_1=1}^{\dim(\mathcal{R})} & \cdots & \sum_{d_{N_q}=1}^{\dim(\mathcal{R})} \sum_{\tilde{s}=1}^{\dim(R)} D^{\bar{R}}_{\tilde{r} \tilde{s}}(g^{-1}) \ \braket{\otimes_{i=1}^{N_q} (\mathcal{R},d_i)}{(\bar{R}_k, \tilde{s})} \prod_{i=1}^{N_q} \chi_{d_i}^\dag \ket{0} \yesnumber
\end{IEEEeqnarray*}
which follows from the orthogonality of the CGCs. The final result is
\begin{equation}
	\Theta_{Qg} \ket{(R_{k,N_q},r)} = \sum_{s=1}^{\dim(R)} D^R_{s r}(g) \ket{(R_{k,N_q},s)} \qquad \bar{\Theta}_{Qg} \bar{\ket{(R_{k,N_q},r)}} = \sum_{\tilde{s}=1}^{\dim(R)} D^{\bar{R}}_{\tilde{r}\tilde{s}}(g^{-1}) \bar{\ket{(R_{k,N_q},s)}}
\end{equation}

\subsection{Construction of Physical Hilbert Space}
\label{app:aspects:physical_hilbert_space}

Gauge transformations on sites and links can be made with the application of translation operators. Operators that make only $\Omega(\vec{s})$ modifications are $\Theta_{Q_\alpha\Omega}(\vec{s})$, $\bar{\Theta}_{Q_\alpha\Omega}(\vec{s})$, $\Theta_{\mathsf{L}\Omega}(\vec{s},\vec{e}_i)$, and $\Theta_{\mathsf{R}\Omega}(\vec{s}-a\vec{e}_i,\vec{e}_i)$. This motivates the definition
\begin{equation}
	\Theta_\Omega(\vec{s}) = \prod_{f=1}^{N_f} \left( \prod_{\alpha=1}^{2} \Theta_{Q_\alpha\Omega}^{(f)}(\vec{s}) \prod_{\alpha=3}^{4} \bar{\Theta}_{Q_\alpha\Omega}^{(f)}(\vec{s}) \right) \prod_{i=1}^{3} \Theta_{\mathsf{L}\Omega}(\vec{s},\vec{e}_i) \Theta^\dag_{\mathsf{R}\Omega}(\vec{s}-a\vec{e}_i,\vec{e}_i)
\end{equation}
An arbitrary operator $O$ undergoes the transformation $\Theta_\Omega^\dag O \Theta_\Omega$. By using the exponential definitions of each translation operator,
\begin{equation}
	\Theta_\Omega(\vec{s}) = e^{i\phi_\Omega^a [ \sum_{f=1}^{N_f} \sum_{\alpha=1}^{4} Q_\alpha^{a(f)}(\vec{s}) + \bar{Q}_\alpha^{a(f)}(\vec{s}) + \sum_{i=1}^{3} \Pi_\mathsf{L}^{i,a}(\vec{s}) - \Pi_\mathsf{R}^{i,a}(\vec{s}-a\vec{e}_i) ]} \equiv e^{i\phi_\Omega^a G^a(\vec{s})}
\end{equation}
Now recall Gauss's law. Physical states are those invariant under a gauge transformation: $\Theta_\Omega(\vec{s}) \ket{\text{phys}} = \ket{\text{phys}}$. In what follows, a basis for solutions to this equation will be constructed in the representation basis, and assuming $\mathcal{R}$ is the fundamental representation.

Each physical basis state $\ket{\Lambda}$ will be such that the irreps of the fermion and link states are fixed. What remains is a superposition of representation basis states in fixed irreps. Within $\ket{\Lambda}$, there exist tensor products $\ket{\text{site tensor}}$ of all particle/antiparticle and half-link states meeting at a site, such that $\ket{\Lambda} = \bigotimes_{\vec{s}} \ket{\text{site tensor}}$ and
\begin{equation}
	\ket{\text{site tensor}} = \sum_{\{ r_x \}} C(r_x) \bigotimes_{\psi \in \mathbf{P}} \ket{(R_\psi, r_\psi)} \bigotimes_{\chi \in \mathbf{A}} \bar{\ket{(R_\chi, r_\chi)}} \bigotimes_{\ell_\mathsf{L} \in \mathbf{L}} \ket{(R_{\ell_\mathsf{L}}, r_{\ell_\mathsf{L}})} \bigotimes_{\ell_\mathsf{R} \in \mathbf{R}} \ket{(R_{\ell_\mathsf{R}}, r_{\ell_\mathsf{R}})}
\end{equation}
Here, $\mathbf{P}$ is the set of all particles, $\mathbf{A}$ is the set of all antiparticles, $\mathbf{L}$ is the set of all left half-links, and $\mathbf{R}$ is the set of all right half-links meeting at a site $\vec{s}$. The label ``$x$'' is used as a catch-all for link and fermion data. In particular, let $\mathbf{S} = \mathbf{P} \cup \mathbf{A} \cup \mathbf{L} \cup \mathbf{R}$. Then the sum over $\{ r_x \} = \{ r_x \, | \, x \in \mathbf{S} \}$ runs through all basis vectors of each irrep. Lastly, $C(r_x)$ is some coefficient in the superposition.

Under a gauge transformation, $\ket{\text{site tensor}}$ transforms as\footnote{It is a conventional choice whether to write the representation matrices in terms of $\Omega$ or $\Omega^{-1}$. The choice determines which conjugate fermion and link state appear in the CGCs.}
\begin{IEEEeqnarray*}{rCl}
	\Theta_\Omega \ket{\text{site tensor}} & = & \sum_{\{ r_x,s_x \}} C(r_x) \bigotimes_{\psi \in \mathbf{P}} D^{R_\psi}_{s_\psi r_\psi}(\Omega) \ket{(R_\psi, s_\psi)} \bigotimes_{\chi \in \mathbf{A}} \varphi(r_\chi) \varphi(s_\chi) D^{R_\chi}_{s_\chi r_\chi}(\Omega) \bar{\ket{(R_\chi, s_\chi)}} \\
	& & \bigotimes_{\ell_\mathsf{L} \in \mathbf{L}} D^{R_{\ell_\mathsf{L}}}_{s_{\ell_\mathsf{L}} r_{\ell_\mathsf{L}}}(\Omega) \ket{(R_{\ell_\mathsf{L}}, s_{\ell_\mathsf{L}})} \bigotimes_{\ell_\mathsf{R} \in \mathbf{R}} \varphi(s_{\ell_\mathsf{R}}) \varphi(r_{\ell_\mathsf{R}}) D^{\bar{R}_{\ell_\mathsf{R}}}_{\tilde{s}_{\ell_\mathsf{R}} \tilde{r}_{\ell_\mathsf{R}}}(\Omega) \ket{(R_{\ell_\mathsf{R}}, s_{\ell_\mathsf{R}})} \yesnumber
\end{IEEEeqnarray*}
Now the Clebsch-Gordan series can be employed. There is a choice here whether to expand the entire product of representation matrix elements at once. The alternative of applying the Clebsch-Gordan series to batches of the matrix elements resembles a procedure called point splitting \cite{AnishettySreeraj:2018:point_splitting, RaychowdhuryStryker:2020:solving_gauss_law}, which could trade the cost of computing certain CGCs with the introduction of virtual degrees of freedom. This work proceeds without point splitting, yielding
\begin{IEEEeqnarray*}{rCl}
	\Theta_\Omega \ket{\text{site tensor}} & = & \sum_{\{ r_x,s_x \}} \Big( {\textstyle\prod\limits_{x \in \mathbf{C}}} \varphi(r_x) \varphi(s_x) \Big) C(r_x) \ket{\underset{x \in \mathbf{S}}{\otimes} (R_x, s_x)} \\
	& \times & \sum_{j=1}^{n_s} \sum_{k=1}^{m_j} \sum_{s_{j,k}=1}^{\dim(S_{j,k})} \sum_{\ell_{j,k}=1}^{\dim(S_{j,k})} D^{S_{j,k}}_{s_{j,k} \ell_{j,k}}(\Omega) \\
	& \times & \braket{(S_{j,k}, s_{j,k})}{\underset{f \in \mathbf{F}}{\otimes} (R_f, s_f) \underset{\ell_\mathsf{L} \in \mathbf{L}}{\otimes} (R_{\ell_\mathsf{L}}, s_{\ell_\mathsf{L}}) \underset{\ell_\mathsf{R} \in \mathbf{R}}{\otimes} (\bar{R}_{\ell_\mathsf{R}}, \tilde{s}_{\ell_\mathsf{R}})} \\
	& \times & \braket{\underset{f \in \mathbf{F}}{\otimes} (R_f, r_f) \underset{\ell_\mathsf{L} \in \mathbf{L}}{\otimes} (R_{\ell_\mathsf{L}}, r_{\ell_\mathsf{L}}) \underset{\ell_\mathsf{R} \in \mathbf{R}}{\otimes} (\bar{R}_{\ell_\mathsf{R}}, \tilde{r}_{\ell_\mathsf{R}})}{(S_{j,k}, \ell_{j,k})} \yesnumber
\end{IEEEeqnarray*}
where $\mathbf{C} = \mathbf{A} \cup \mathbf{R}$, $\mathbf{F} = \mathbf{P} \cup \mathbf{A}$, and
\begin{equation}
	\ket{\underset{x \in \mathbf{S}}{\otimes} (R_x, s_x)} \equiv \bigotimes_{\psi \in \mathbf{P}} \ket{(R_\psi, s_\psi)} \bigotimes_{\chi \in \mathbf{A}} \bar{\ket{(R_\chi, s_\chi)}} \bigotimes_{\ell_\mathsf{L} \in \mathbf{L}} \ket{(R_{\ell_\mathsf{L}}, s_{\ell_\mathsf{L}})} \bigotimes_{\ell_\mathsf{R} \in \mathbf{R}} \ket{(R_{\ell_\mathsf{R}}, s_{\ell_\mathsf{R}})}
\end{equation}

The $C(r_x)$ coefficients must be tuned such that the original $\ket{\text{site tensor}}$ is recovered. Note that these coefficients cannot be $\Omega$-dependent because $\ket{\text{site tensor}}$ should be invariant under any gauge transformation. The only way the representation matrix elements can be eliminated is if the $C(r_x)$ coefficients collapse the $r_x$ summations in such a way that $S_{j,k} = \mathbf{1}$ to force $D^{S_{j,k}}_{s_{j,k} \ell_{j,k}}(\Omega) \to 1$. This implies the solution:
\begin{equation}
	C^{(\Gamma)}(r_x) = \Big( {\textstyle\prod\limits_{x \in \mathbf{C}}} \varphi(r_x) \Big) \sum_{\kappa=1}^{m_\mathbf{1}} \Xi^{(\Gamma)}_\kappa \braket{\mathbf{1},\kappa}{\underset{f \in \mathbf{F}}{\otimes} (R_f, r_f) \underset{\ell_\mathsf{L} \in \mathbf{L}}{\otimes} (R_{\ell_\mathsf{L}}, r_{\ell_\mathsf{L}}) \underset{\ell_\mathsf{R} \in \mathbf{R}}{\otimes} (\bar{R}_{\ell_\mathsf{R}}, \tilde{r}_{\ell_\mathsf{R}})}
\end{equation}
which incorporates the orthogonality of CGCs and the fact $\varphi(r_x)^2=1$. $m_\mathbf{1}$ is the multiplicity of the direct-sum trivial representation in the direct product of all irreps meeting at the site. $\Xi$ is an $m_\mathbf{1} \times m_\mathbf{1}$ unitary matrix. In this paper, $\Xi$ is chosen to be the identity matrix, so that $\Gamma$ is the multiplicity index of a direct-sum trivial representation. Hence, a physical basis state is given by
\begin{equation}
	\ket{\Lambda} = \bigotimes_{\vec{s}} \sum_{\{ r_x \}} \Big( {\textstyle\prod\limits_{x \in \mathbf{C}}} \varphi(r_x) \Big) \braket{\mathbf{1},\Gamma_s}{\underset{f \in \mathbf{F}}{\otimes} (R_f, r_f) \underset{\ell_\mathsf{L} \in \mathbf{L}}{\otimes} (R_{\ell_\mathsf{L}}, r_{\ell_\mathsf{L}}) \underset{\ell_\mathsf{R} \in \mathbf{R}}{\otimes} (\bar{R}_{\ell_\mathsf{R}}, \tilde{r}_{\ell_\mathsf{R}})} \ \ket{\underset{x \in \mathbf{S}}{\otimes} (R_x, r_x)}
\end{equation}

\section{Hamiltonian Matrix Elements}
\label{app:hamiltonian_matrix_elements}

The local operators that appear in the lattice Hamiltonian generate transitions of the site singlets. When $\mathcal{R}$ is the fundamental representation, these transitions occur with independent amplitudes called site factors.

\begin{figure}[ht]
	\centering
	\includegraphics[width=\textwidth]{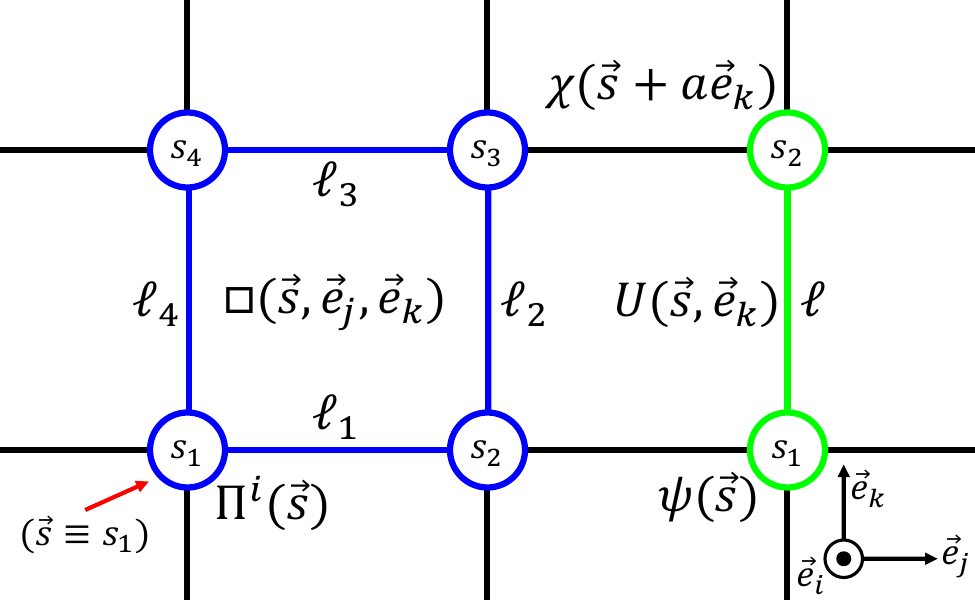}
	\caption{Cross section of a lattice highlighting where Hamiltonian operators act. The plaquette operator acts nontrivially on four links and four sites. Sites are labeled $s_1,\dots,s_4$ as shown, where the site $\protect\vec{s}$ on which the plaquette is defined coincides with $s_1$. Links are also labeled $\ell_1,\dots,\ell_4$ counterclockwise starting from $s_1$. The theta term adds to this a momentum operator acting on the perpendicular ``$\perp$'' link extending from $s_1$. Meanwhile, fermion kinetic terms act nontrivially on a link and two sites. Again, the sites are labeled $s_1$ and $s_2$, such that $s_1$ coincides with the site $\protect\vec{s}$ from which the link $\ell$ is defined. These labeling schemes are used for site factor formulas.}
	\label{fig:lattice_diagram}
\end{figure}

\subsubsection*{Plaquette Operator}

The matrix elements of the plaquette operator in the gauge invariant basis are
\begin{equation}
	\bra{\Lambda_n} \trace[ \Box(\vec{s},\vec{e}_i,\vec{e}_j) ] \ket{\Lambda_m} = \bra{\Lambda_n} \trace[ U^\dag(\vec{s}, \vec{e}_j) U^\dag(\vec{s} + a\vec{e}_j, \vec{e}_i) U(\vec{s} + a\vec{e}_i, \vec{e}_j) U(\vec{s}, \vec{e}_i) ] \ket{\Lambda_m}
\end{equation}
Note that the Hermitian conjugate of a link operator can be written in terms of a link operator in the conjugate representation:
\begin{equation}
	U^\dag = (e^{-iA_i^a T^{a*}})^\text{T} = (U^{R^*})^\text{T} \implies (U^\dag)_{mn} = \varphi(m) \varphi(n) U^{\bar{R}}_{\tilde{n}\tilde{m}}
\end{equation}
Here, $\tilde{n}$ is a basis vector of $\bar{R}$ conjugate to the basis vector $n$ of $R$, and $\varphi(n) = \pm 1$ is phase incurred when translating $n$ to $\tilde{n}$. Hence,
\begin{equation}
	\trace[\Box] = \varphi(\sigma_1) \varphi(\sigma_3) U^{\bar{\mathcal{R}}}_{\tilde{\sigma}_4 \tilde{\sigma}_1} U^{\bar{\mathcal{R}}}_{\tilde{\sigma}_3 \tilde{\sigma}_4} U^\mathcal{R}_{\sigma_3 \sigma_2} U^\mathcal{R}_{\sigma_2 \sigma_1}
\end{equation}

In the representation basis, $U^\mathcal{R}_{\sigma_2 \sigma_1}$ and $U^{\bar{\mathcal{R}}}_{\tilde{\sigma}_4 \tilde{\sigma}_1}$ yield four CGCs, two of which contain the left half-link states from $\ell_1$ and $\ell_4$ meeting at $s_1$ (refer to Fig.~\ref{fig:lattice_diagram} for these labels). Therefore, the $\sigma_1$ summation corresponds to a quantity that contains all $s_1$ data. This is the $s_1$ site factor,
\begin{IEEEeqnarray*}{rCl}
	\begin{Bmatrix}
		R_1 & \Gamma_{s_1} & R_4 \\ \mathcal{R} & \vec{C}_{s_1} & \bar{\mathcal{R}} \\ R'_{1,\gamma_1} & \Gamma'_{s_1} & R'_{4,\gamma_4}
	\end{Bmatrix} & = & \sum_{\sigma_1=1}^{\dim(\mathcal{R})} \sum_{r_1=1}^{\dim(R_1)} \sum_{r'_1=1}^{\dim(R'_1)} \sum_{r_4=1}^{\dim(R_4)} \sum_{r'_4=1}^{\dim(R'_4)} \sum_{\vec{c} \in \vec{C}_{s_1}} \varphi(\sigma_1) \\
	& \times & \braket{(\mathcal{R},\sigma_1) \otimes (R_1,r_1)}{(R'_{1,\gamma_1},r'_1)} \ \braket{(\bar{\mathcal{R}},\tilde{\sigma}_1) \otimes (R_4,r_4)}{(R'_{4,\gamma_4},r'_4)} \\
	& \times & \braket{(R'_1,r'_1) \otimes (R'_4,r'_4) \otimes (\vec{C}_{s_1},\vec{c})}{\mathbf{1},\Gamma'_{s_1}} \ \braket{\mathbf{1},\Gamma_{s_1}}{(R_1,r_1) \otimes (R_4,r_4) \otimes (\vec{C}_{s_1},\vec{c})} \\ \yesnumber
\end{IEEEeqnarray*}
It is the transition amplitude from the initial physical basis state -- labeled by irreps $R_{1,4}$ on links $\ell_{1,4}$, irreps $\vec C_{s_1}$ on links external to the plaquette and the fermions at $s_1$, and a site-singlet multiplicity index $\Gamma_{s_1}$ -- to the final physical basis state -- labeled by the same $\vec C_{s_1}$, and the primed quantities $R'_{1,4}$ and $\Gamma'_{s_1}$. Furthermore, note that the CGCs from the link operators contain a provisional multiplicity index $\gamma_k$ that is only necessary when $\mathcal{R}$ is not the fundamental representation, and is ultimately summed over.

\begin{figure}
	\centering
	\begin{subfigure}{\textwidth}
		\centering
		\includegraphics[width=\textwidth]{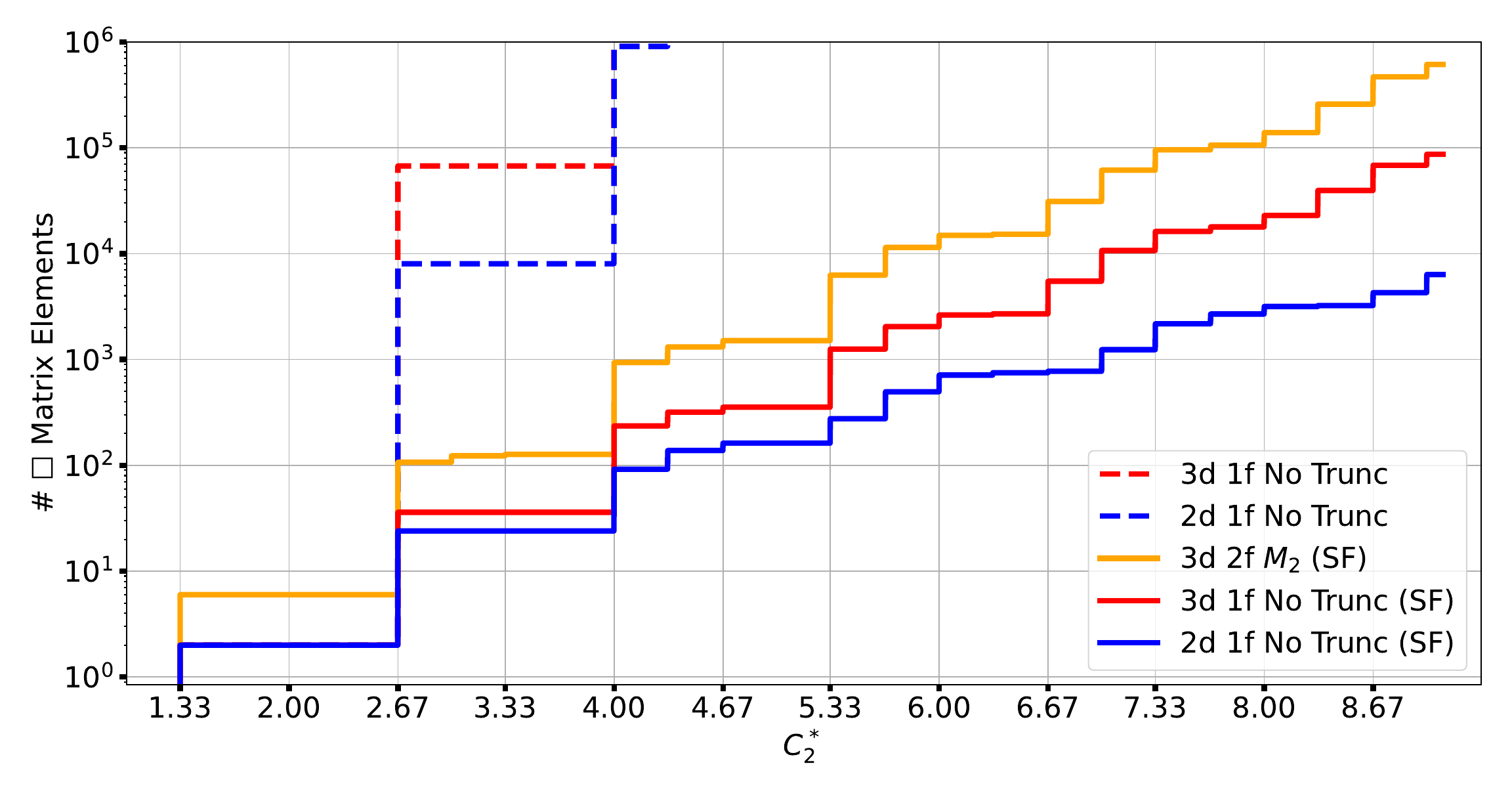}
		\caption{Staggered fermions}
	\end{subfigure}
	
	\begin{subfigure}{\textwidth}
		\centering
		\includegraphics[width=\textwidth]{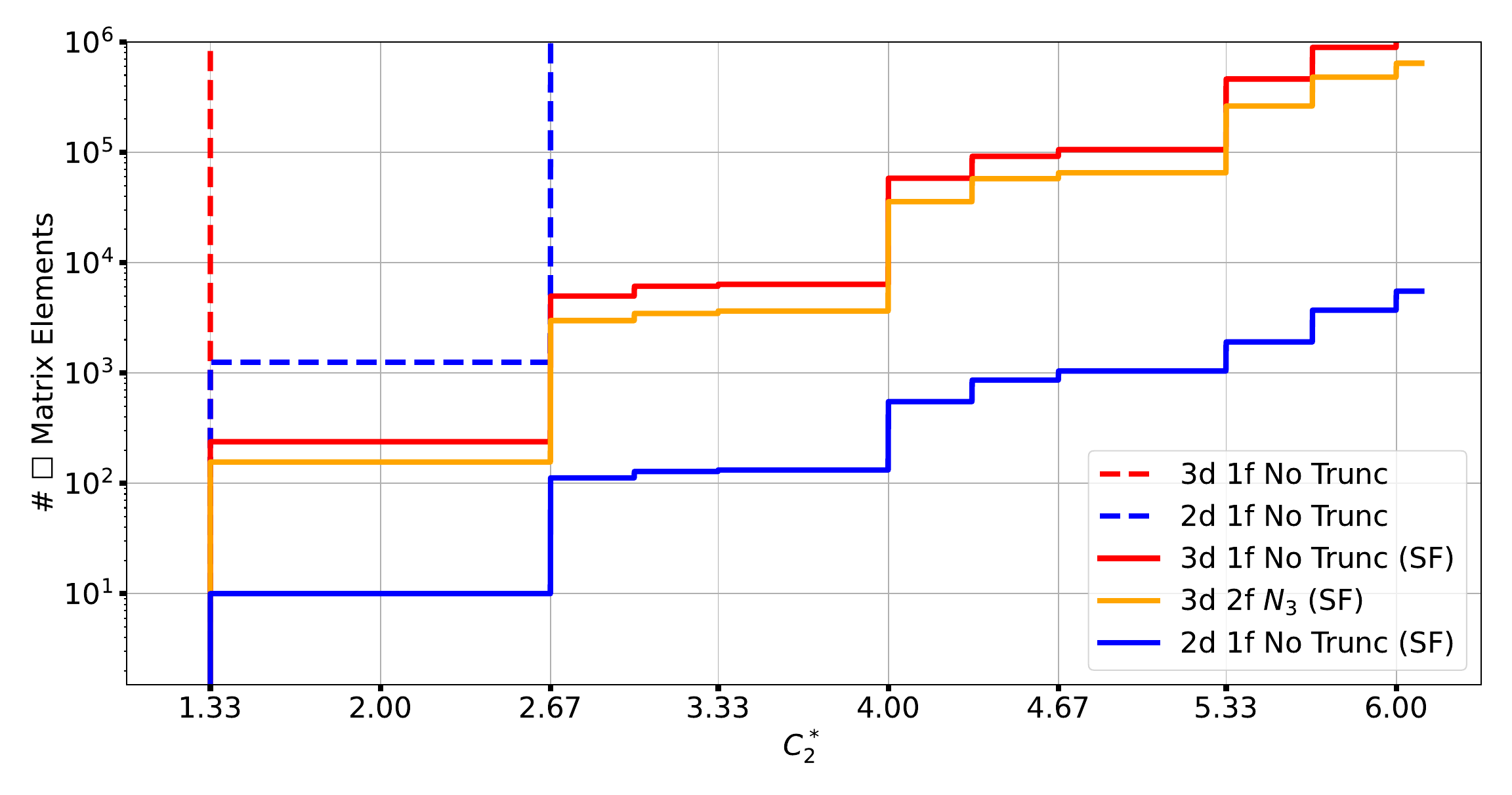}
		\caption{Wilson fermions}
	\end{subfigure}
	\caption{SU(3) plaquette matrix element and site factor (SF) counts. ``\#f'' refers to the number of flavors and ``No Trunc'' means no truncation was performed on the fermion Hilbert space. In both plots, the gauge field is truncated at various $C_2^*$ truncation values. The truncations shown here allow, at certain fermion truncations, irreps including $\{ \mathbf{1}, \mathbf{3}, \mathbf{6}, \mathbf{8}, \mathbf{10}, \mathbf{15} \}$ (plus their conjugates) and maximum singlet multiplicities of 23. (a) The $M_2$ truncation allows each flavor to be at most in a double occupied state. (b) The $N_3$ truncation allows each flavor to have at most color occupation number three, summed over all spinor components.}
	\label{fig:plaq_matrix_elements}
\end{figure}

The cases are similar for $s_2$, $s_3$, and $s_4$. $U^\mathcal{R}_{\sigma_3 \sigma_2}$ and $U^\mathcal{R}_{\sigma_2 \sigma_1}$ contain CGCs for the right half-link of $\ell_1$ and the left half-link of $\ell_2$. This gives the $s_2$ site factor,
\begin{IEEEeqnarray*}{rCl}
	\begin{Bmatrix}
		R_1 & \Gamma_{s_2} & R_2 \\ \mathcal{R} & \vec{C}_{s_2} & \mathcal{R} \\ R_{1,\gamma_1}' & \Gamma_{s_2}' & R_{2,\gamma_2}'
	\end{Bmatrix} & = & \sum_{\sigma_2=1}^{\dim(\mathcal{R})} \sum_{r_1=1}^{\dim(R_1)} \sum_{r'_1=1}^{\dim(R'_1)} \sum_{r_2=1}^{\dim(R_2)} \sum_{r'_2=1}^{\dim(R'_2)} \sum_{\vec{c} \in \vec{C}_{s_2}} \varphi(r_1) \varphi(r'_1) \\
	& \times & \braket{(R'_{1,\gamma_1},r'_1)}{(\mathcal{R},\sigma_2) \otimes (R_1,r_1)} \ \braket{(\mathcal{R},\sigma_2) \otimes (R_2,r_2)}{(R'_{2,\gamma_2},r'_2)} \\
	& \times & \braket{(\bar{R}'_1,\tilde{r}'_1) \otimes (R'_2,r'_2) \otimes (\vec{C}_{s_2},\vec{c})}{\mathbf{1},\Gamma'_{s_2}} \ \braket{\mathbf{1},\Gamma_{s_2}}{(\bar{R}_1,\tilde{r}_1) \otimes (R_2,r_2) \otimes (\vec{C}_{s_2},\vec{c})} \\ \yesnumber
\end{IEEEeqnarray*}
Now the right half-link state of $\ell_1$ contributes $\varphi(r)$ phases by the definition of $\ket{\Lambda}$. $U^{\bar{\mathcal{R}}}_{\tilde{\sigma}_3 \tilde{\sigma}_4}$ and $U^\mathcal{R}_{\sigma_3 \sigma_2}$ give CGCs with the right half-link states of $\ell_2$ and $\ell_3$. The $s_3$ site factor is
\begin{IEEEeqnarray*}{rCl}
	\begin{Bmatrix}
		R_3 & \Gamma_{s_3} & R_2 \\ \bar{\mathcal{R}} & \vec{C}_{s_3} & \mathcal{R} \\ R_{3,\gamma_3}' & \Gamma_{s_3}' & R_{2,\gamma_2}'
	\end{Bmatrix} & = & \sum_{\sigma_3=1}^{\dim(\mathcal{R})} \sum_{r_3=1}^{\dim(R_3)} \sum_{r'_3=1}^{\dim(R'_3)} \sum_{r_2=1}^{\dim(R_2)} \sum_{r'_2=1}^{\dim(R'_2)} \sum_{\vec{c} \in \vec{C}_{s_3}} \varphi(\sigma_3) \varphi(r_3) \varphi(r'_3) \varphi(r_2) \varphi(r'_2) \\
	& \times & \braket{(R'_{3,\gamma_3},r'_3)}{(\bar{\mathcal{R}},\tilde{\sigma}_3) \otimes (R_3,r_3)} \ \braket{(R'_{2,\gamma_2},r'_2)}{(\mathcal{R},\sigma_3) \otimes (R_2,r_2)} \\
	& \times & \braket{(\bar{R}'_3,\tilde{r}'_3) \otimes (\bar{R}'_2,\tilde{r}'_2) \otimes (\vec{C}_{s_3},\vec{c})}{\mathbf{1},\Gamma'_{s_3}} \ \braket{\mathbf{1},\Gamma_{s_3}}{(\bar{R}_3,\tilde{r}_3) \otimes (\bar{R}_2,\tilde{r}_2) \otimes (\vec{C}_{s_3},\vec{c})} \\ \yesnumber
\end{IEEEeqnarray*}
In addition to $\varphi(\sigma_3)$, both right half-links meeting at $s_3$ contribute their phases as well. Finally, $U^{\bar{\mathcal{R}}}_{\tilde{\sigma}_4 \tilde{\sigma}_1}$ and $U^{\bar{\mathcal{R}}}_{\tilde{\sigma}_3 \tilde{\sigma}_4}$ contribute CGCs with the left half-link state of $\ell_3$ and the right half-link state of $\ell_4$. The $s_4$ site factor is
\begin{IEEEeqnarray*}{rCl}
	\begin{Bmatrix}
		R_3 & \Gamma_{s_4} & R_4 \\ \bar{\mathcal{R}} & \vec{C}_{s_4} & \bar{\mathcal{R}} \\ R_{3,\gamma_3}' & \Gamma_{s_4}' & R_{4,\gamma_4}'
	\end{Bmatrix} & = & \sum_{\tilde{\sigma}_4=1}^{\dim(\mathcal{R})} \sum_{r_3=1}^{\dim(R_3)} \sum_{r'_3=1}^{\dim(R'_3)} \sum_{r_4=1}^{\dim(R_4)} \sum_{r'_4=1}^{\dim(R'_4)} \sum_{\vec{c} \in \vec{C}_{s_4}} \varphi(r_4) \varphi(r'_4) \\
	& \times & \braket{(\bar{\mathcal{R}},\tilde{\sigma}_4) \otimes (R_3,r_3)}{(R'_{3,\gamma_3},r'_3)} \ \braket{(R'_{4,\gamma_4},r'_4)}{(\bar{\mathcal{R}},\tilde{\sigma}_4) \otimes (R_4,r_4)} \\
	& \times & \braket{(R'_3,r'_3) \otimes (\bar{R}'_4,\tilde{r}'_4) \otimes (\vec{C}_{s_4},\vec{c})}{\mathbf{1},\Gamma'_{s_4}} \ \braket{\mathbf{1},\Gamma_{s_4}}{(R_3,r_3) \otimes (\bar{R}_4,\tilde{r}_4) \otimes (\vec{C}_{s_4},\vec{c})} \\ \yesnumber
\end{IEEEeqnarray*}
The product of the four site factors, along with the dimension coefficients from the link operators, gives the matrix element
\begin{IEEEeqnarray*}{rCl}
	\bra{\Lambda_n} \trace[\Box] \ket{\Lambda_m} = \left( \prod_{k=1}^{4} \sqrt{\frac{\dim(R_k^m)}{\dim(R_k^n)}} \right) \sum_{\vec{\gamma}} & & \begin{Bmatrix} R_1^m & \Gamma_{s_1}^m & R_4^m \\ \mathcal{R} & \vec{C}_{s_1} & \bar{\mathcal{R}} \\ R_{1,\gamma_1}^n & \Gamma_{s_1}^n & R_{4,\gamma_4}^n \end{Bmatrix} \begin{Bmatrix} R_1^m & \Gamma_{s_2}^m & R_2^m \\ \mathcal{R} & \vec{C}_{s_2} & \mathcal{R} \\ R_{1,\gamma_1}^n & \Gamma_{s_2}^n & R_{2,\gamma_2}^n \end{Bmatrix} \\
	& \times & \begin{Bmatrix} R_3^m & \Gamma_{s_3}^m & R_2^m \\ \bar{\mathcal{R}} & \vec{C}_{s_3} & \mathcal{R} \\ R_{3,\gamma_3}^n & \Gamma_{s_3}^n & R_{2,\gamma_2}^n \end{Bmatrix} \begin{Bmatrix} R_3^m & \Gamma_{s_4}^m & R_4^m \\ \bar{\mathcal{R}} & \vec{C}_{s_4} & \bar{\mathcal{R}} \\ R_{3,\gamma_3}^n & \Gamma_{s_4}^n & R_{4,\gamma_4}^n \end{Bmatrix} \\ \yesnumber
\end{IEEEeqnarray*}
The number of matrix elements is shown in Fig.~\ref{fig:plaq_matrix_elements}.

\subsubsection*{Theta Term}

\begin{figure}
	\centering
	\includegraphics[width=\textwidth]{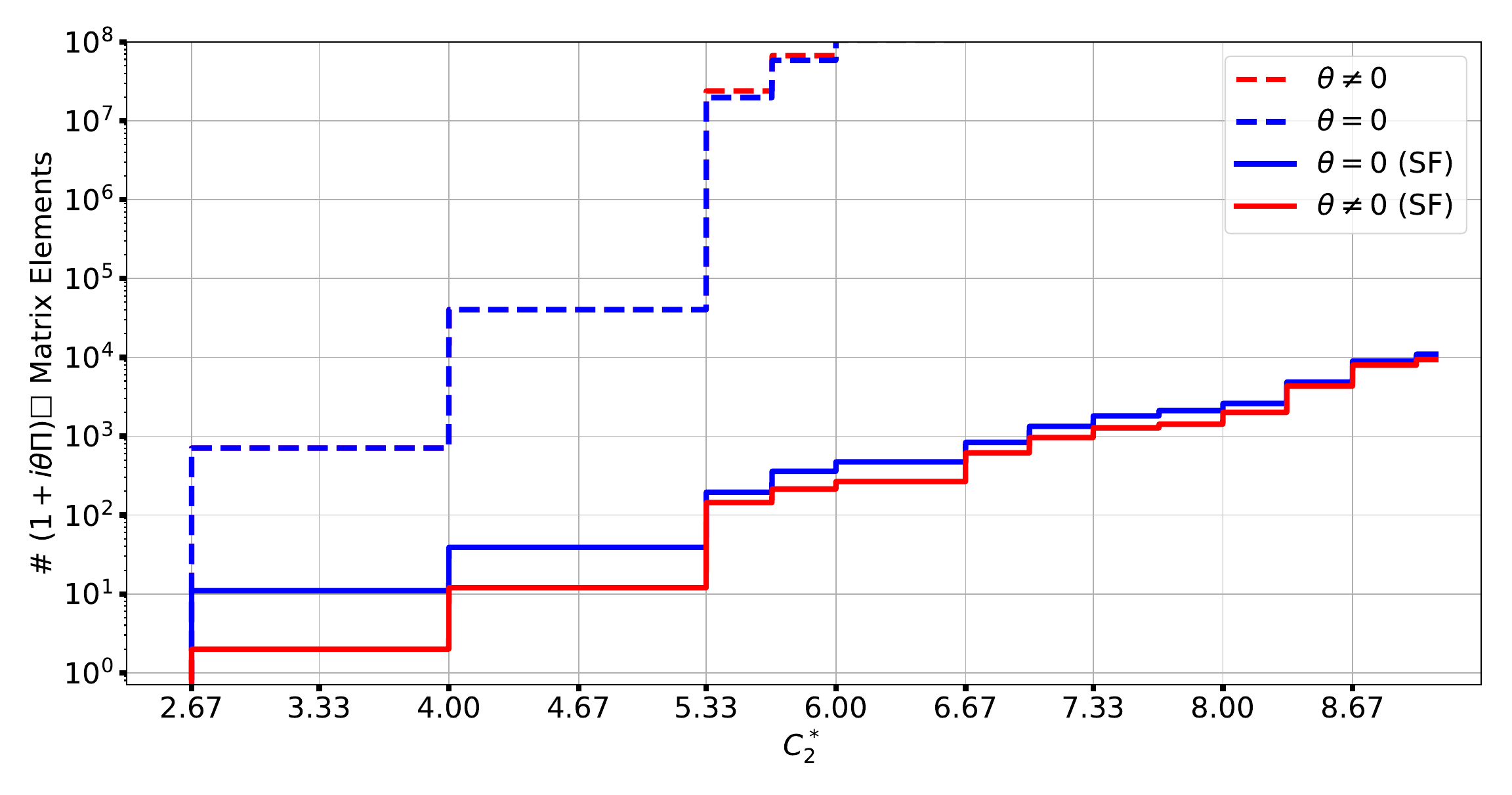}
	\caption{SU(3) theta term matrix element and site factor (SF) counts for a pure gauge theory.  Each curve scans over various $C_2^*$ gauge truncation values. The $\theta \neq 0$ matrix element counts add together plaquette and theta term matrix element counts, hence ``$(1+i\theta\Pi)\Box$.'' Because the number of theta term site factors is typically on the order of the number of plaquette site factors, the matrix element counts are comparable.}
	\label{fig:theta_matrix_elements}
\end{figure}

The matrix elements of the theta term in the gauge invariant basis are
\begin{equation}
	\bra{\Lambda_n} \trace[\Pi^i(\vec{s}) \Box(\vec{s},\vec{e}_j,\vec{e}_k)] \ket{\Lambda_m} = \bra{\Lambda_n} \trace[ \Pi^i(\vec{s}) U^\dag(\vec{s}, \vec{e}_k) U^\dag(\vec{s} + a\vec{e}_k, \vec{e}_j) U(\vec{s} + a\vec{e}_j, \vec{e}_k) U(\vec{s}, \vec{e}_j) ] \ket{\Lambda_m}
\end{equation}
Here, $\Pi^i(\vec{s}) = \Pi^{i,a}_\mathsf{L}(\vec{s}) T^{\mathcal{R}a}$. Therefore,
\begin{equation}
	\trace[\Pi \Box] = \varphi(\tau) \varphi(\sigma_3) \Pi^{i,a}_\mathsf{L} T^{\mathcal{R}a}_{\sigma_1 \tau} U^{\bar{\mathcal{R}}}_{\tilde{\sigma}_4 \tilde{\tau}} U^{\bar{\mathcal{R}}}_{\tilde{\sigma}_3 \tilde{\sigma}_4} U^\mathcal{R}_{\sigma_3 \sigma_2} U^\mathcal{R}_{\sigma_2 \sigma_1}
\end{equation}
The matrix element for this operator has the same site factors of $\trace[\Box]$ for $s_2$, $s_3$, and $s_4$, which can be read off by comparing the appearances of $\sigma_2$, $\sigma_3$, and $\sigma_4$ in $\trace[\Pi\Box]$ with those in $\trace[\Box]$. However, now $\sigma_1$ appears contracted with an algebra basis element rather than a link operator. The operator $U^{\bar{\mathcal{R}}}_{\tilde{\sigma}_4 \tilde{\tau}}$ will produce a CGC with the left half-link state of $\ell_4$ and the index $\tilde{\tau}$, while $U^\mathcal{R}_{\sigma_2 \sigma_1}$ will produce a CGC with the left half-link state of $\ell_1$ and the index $\sigma_1$. Both $\sigma_1$ and $\tau$ are summed with a factor of $T^{\mathcal{R}a}_{\sigma_1\tau}$. This leads to the special theta term $s_1$ site factor,
\begin{IEEEeqnarray*}{rCl}
	\begin{bmatrix}
		R_1 & \Gamma_{s_1} & R_4 \\ \mathcal{R} & \vec{C}_{s_1} & \bar{\mathcal{R}} \\ R_{1,\gamma_1}' & \Gamma_{s_1}' & R_{4,\gamma_4}'
	\end{bmatrix}_{R_\perp} & = & \sum_{\tau=1}^{\dim(\mathcal{R})} \sum_{\sigma_1=1}^{\dim(\mathcal{R})} \sum_{r_1=1}^{\dim(R_1)} \sum_{r'_1=1}^{\dim(R'_1)} \sum_{r_4=1}^{\dim(R_4)} \sum_{r'_4=1}^{\dim(R'_4)} \sum_{r_\perp=1}^{\dim(R_\perp)} \sum_{r'_\perp=1}^{\dim(R_\perp)} \sum_{\vec{c} \in \vec{C}_{s_1}} \varphi(\tau) T^{R_\perp a}_{r'_\perp r_\perp} T^{\mathcal{R}a}_{\sigma_1 \tau} \\
	& \times & \braket{(\mathcal{R},\sigma_1) \otimes (R_1,r_1)}{(R'_{1,\gamma_1},r'_1)} \ \braket{(\bar{\mathcal{R}},\tilde{\tau}) \otimes (R_4,r_4)}{(R'_{4,\gamma_4},r'_4)} \\
	& \times & \braket{(R'_1,r'_1) \otimes (R'_4,r'_4) \otimes (R_\perp,r'_\perp) \otimes (\vec{C}_{s_1},\vec{c})}{\mathbf{1},\Gamma_{s_1}'} \\
	& \times & \braket{\mathbf{1},\Gamma_{s_1}}{(R_1,r_1) \otimes (R_4,r_4) \otimes (R_\perp,r_\perp) \otimes (\vec{C}_{s_1},\vec{c})} \yesnumber
\end{IEEEeqnarray*}
This takes into account $\Pi^a_\mathsf{L}$ acting on a representation basis state, which generates the factor of $T^{R_\perp a}_{r'_\perp r_\perp}$. (Note that each $T^a$ is either real or imaginary, so these site factors can be made real-valued.) This site factor also has an additional label for $R_\perp$ for the state of the link extending in the $i$th direction, which is perpendicular to the plaquette's plane. The irrep $R_\perp$ does not change because it is not acted on by a link operator, but $\Pi^a_\mathsf{L}$ does change the half-link state $\ket{(R_\perp,r_\perp)}$. As a result, $\vec{C}_{s_1}$ does not contain this perpendicular state. All together, the theta term matrix element is
\begin{IEEEeqnarray*}{rCl}
	\bra{\Lambda_n} \trace[\Pi \Box] \ket{\Lambda_m} = \left( \prod_{k=1}^{4} \sqrt{\frac{\dim(R_k^m)}{\dim(R_k^n)}} \right) \sum_{\vec{\gamma}} & & \begin{bmatrix} R_1^m & \Gamma_{s_1}^m & R_4^m \\ \mathcal{R} & \vec{C}_{s_1} & \bar{\mathcal{R}} \\ R_{1,\gamma_1}^n & \Gamma_{s_1}^n & R_{4,\gamma_4}^n \end{bmatrix}_{R_\perp} \begin{Bmatrix} R_1^m & \Gamma_{s_2}^m & R_2^m \\ \mathcal{R} & \vec{C}_{s_2} & \mathcal{R} \\ R_{1,\gamma_1}^n & \Gamma_{s_2}^n & R_{2,\gamma_2}^n \end{Bmatrix} \\
	& \times & \begin{Bmatrix} R_3^m & \Gamma_{s_3}^m & R_2^m \\ \bar{\mathcal{R}} & \vec{C}_{s_3} & \mathcal{R} \\ R_{3,\gamma_3}^n & \Gamma_{s_3}^n & R_{2,\gamma_2}^n \end{Bmatrix} \hspace{0.11in} \begin{Bmatrix} R_3^m & \Gamma_{s_4}^m & R_4^m \\ \bar{\mathcal{R}} & \vec{C}_{s_4} & \bar{\mathcal{R}} \\ R_{3,\gamma_3}^n & \Gamma_{s_4}^n & R_{4,\gamma_4}^n \end{Bmatrix} \\ \yesnumber
\end{IEEEeqnarray*}
The number of matrix elements is shown in Fig.~\ref{fig:theta_matrix_elements}.

\subsubsection*{Fermion Kinetic Terms}

The matrix elements for all possible fermion kinetic terms are
\begin{IEEEeqnarray*}{rCl}
	\bra{\Lambda_n} \psi_c^\dag(\vec{s}+a\vec{e}_i) U_{cc'}(\vec{s},\vec{e}_i) \chi_{c'}^\dag(\vec{s}) \ket{\Lambda_m} & \qquad & \bra{\Lambda_n} \chi_c(\vec{s}+a\vec{e}_i) U_{cc'}(\vec{s},\vec{e}_i) \psi_{c'}(\vec{s}) \ket{\Lambda_m} \\
	\bra{\Lambda_n} \psi_c^\dag(\vec{s}+a\vec{e}_i) U_{cc'}(\vec{s},\vec{e}_i) \psi_{c'}(\vec{s}) \ket{\Lambda_m} & \qquad & \bra{\Lambda_n} \chi_{c'}^\dag(\vec{s}) U_{cc'}(\vec{s},\vec{e}_i) \chi_c(\vec{s}+a\vec{e}_i) \ket{\Lambda_m} \\ \yesnumber
\end{IEEEeqnarray*}
where the spinor and flavor indices have been suppressed. To calculate these matrix elements, it is convenient to use the occupation number basis; each fermionic representation basis state is equal to an occupation number basis state when $\mathcal{R}$ is the fundamental representation. Then each fermionic operator can be written as a conventional tensor-product operator via the Jordan-Wigner transformation, containing $Z$, $\sigma^\pm$, and $I$. These operators must first be multiplied together to track where the nontrivial $Z$ and $\sigma^\pm$ act.

For example, consider the bilinear $\psi^\dag(s_2) \psi(s_1)$. Suppose $s_1$ ``comes before'' $s_2$. Then in a Jordan-Wigner transformation, this operator may appear as
\begin{IEEEeqnarray*}{rCl}
	\psi^\dag(s_2) \psi(s_1) & = & \left( Z \otimes \cdots \otimes Z \otimes Z \hspace{1ex} \otimes Z \otimes Z \otimes \sigma^+ \otimes I \otimes \cdots \otimes I \right) \\
	& \times & \left( Z \otimes \cdots \otimes Z \otimes \sigma^- \otimes I \; \otimes I \: \otimes I \hspace{1.5ex} \otimes I \otimes \cdots \otimes I \right) \\
	& = & I \otimes \cdots \otimes I \otimes Z\sigma^- \otimes Z \otimes Z \otimes \sigma^+ \otimes I \otimes \cdots \otimes I \yesnumber
\end{IEEEeqnarray*}
Two ladder operators remain ($Z \sigma^\pm = \mp \sigma^\pm$ and $\sigma^\pm Z = \pm \sigma^\pm$), and many $Z$ operators became $Z^2 = I$. However, some $Z$ operators linger between the ladder operators. In the context of the kinetic term, these $Z$ operators act on color occupation number basis states, including possibly the states being raised or lowered. When the $Z$ operators act on the raised/lowered states, they may either act on the colors ``to the left'' or ``to the right'' of the color being raised/lowered. This will be represented in the matrix elements by sub-operators $\tilde{\psi}$ and $\tilde{\chi}$, which indicate to apply $Z$ operators to the color occupation number state being raised/lowered. The remaining $Z$ operators acting on other states accumulate a product of $(-1)^{N_q}$ phases. These phases may come from states on either $s_1$ and $s_2$ (for instance, from different spinor or flavor states not being raised/lowered), which can be included in the matrix element site factors. However, the phases can also come from other lattice sites that are not $s_1$ or $s_2$. These cannot be immediately calculated in the matrix elements without encoding whole-lattice states, which is not the approach taken in this paper. Instead, the phases will have to be calculated in real time during a simulation, such that a $-1$ phase is applied if a color occupation number basis state has odd color occupation. In the matrix elements, this will appear as a product of $Z$ operators over the color occupation number states between $s_1$ and $s_2$.

Once again, the matrix elements can be written in terms of site factors. The contraction $U_{cc'} \chi_{c'}^\dag$ yields two CGCs from the link operator, one of which contains the left half-link state of $\ell$ (joined at $s_1$). The operator $\chi^\dag$ will change the state of an antiparticle at $s_1$, and so the $c'$ summation will be associated with a quantity that contains all $s_1$ data. This is the $s_1$ site factor for operators with $\chi^\dag$:
\begin{IEEEeqnarray*}{rCl}
	\begin{pmatrix}
		R & \Gamma_{s_1} & S \\ \mathcal{R} & \vec{C}_{s_1} & \chi^\dag \\ R' & \Gamma'_{s_1} & S'
	\end{pmatrix} & = & (-1)^{\zeta_{s_1}} \sum_{\sigma=1}^{\dim(\mathcal{R})} \sum_{r=1}^{\dim(R)} \sum_{r'=1}^{\dim(R')} \sum_{s=1}^{\dim(S)} \sum_{s'=1}^{\dim(S')} \sum_{\vec{c} \in \vec{C}_{s_1}} \varphi(s) \varphi(s') \\
	& \times & \braket{(\mathcal{R},\sigma) \otimes (R,r)}{(R',r')} \ \bra{(S',s')} \tilde{\chi}^\dag_\sigma \ket{(S,s)} \\
	& \times & \braket{(R',r') \otimes (S',s') \otimes (\vec{C}_{s_1},\vec{c})}{\mathbf{1},\Gamma'_{s_1}} \ \braket{\mathbf{1},\Gamma_{s_1}}{(R,r) \otimes (S,s) \otimes (\vec{C}_{s_1},\vec{c})} \\ \yesnumber
\end{IEEEeqnarray*}
Here, $\varphi(r)$ phases appear for the antiparticle state by the definition of $\ket{\Lambda}$. No such phase is necessary for the left half-link state being modified here, and all other would-be phases from right half-link and antiparticle control states have squared to one. Note the $\tilde{\chi}^\dag$ matrix element that appears, meaning to account for $Z$ operators in a way consistent with whether $s_1$ comes before $s_2$. Moreover, $\zeta_{s_1}$ is a sum over color occupation numbers for those fermions in $\vec{C}_{s_1}$ that ``come between'' the antiparticle in $s_1$ and the particle in $s_2$ being created:
\begin{equation}
	\zeta_{s_1} = \sum_{\substack{C \in \vec{C}_{s_1} \\ \text{between}}} N_q(C)
\end{equation}

\begin{figure}
	\centering
	\begin{subfigure}{\textwidth}
		\centering
		\includegraphics[width=\textwidth]{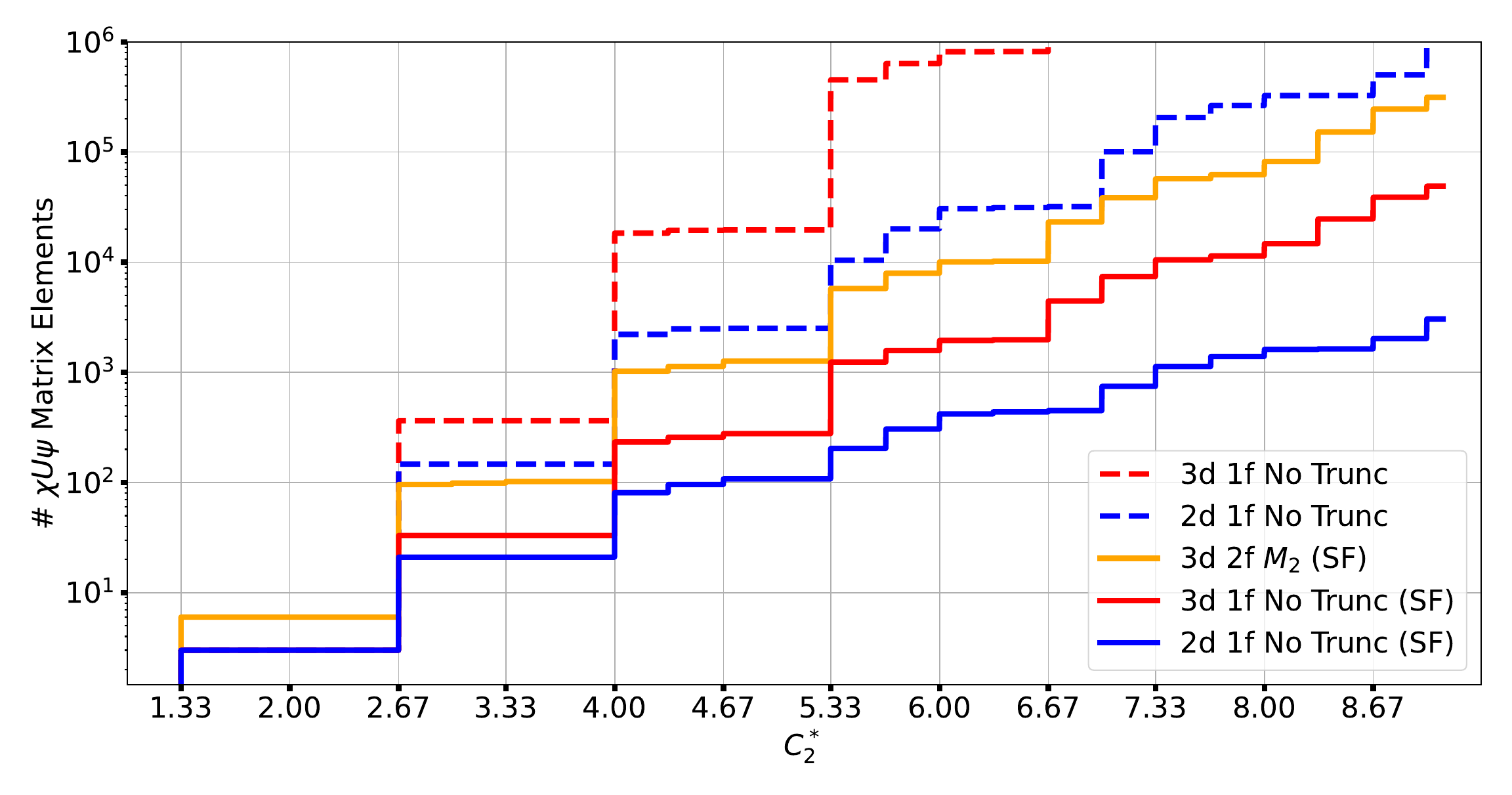}
		\caption{Staggered fermions}
	\end{subfigure}
	
	\begin{subfigure}{\textwidth}
		\centering
		\includegraphics[width=\textwidth]{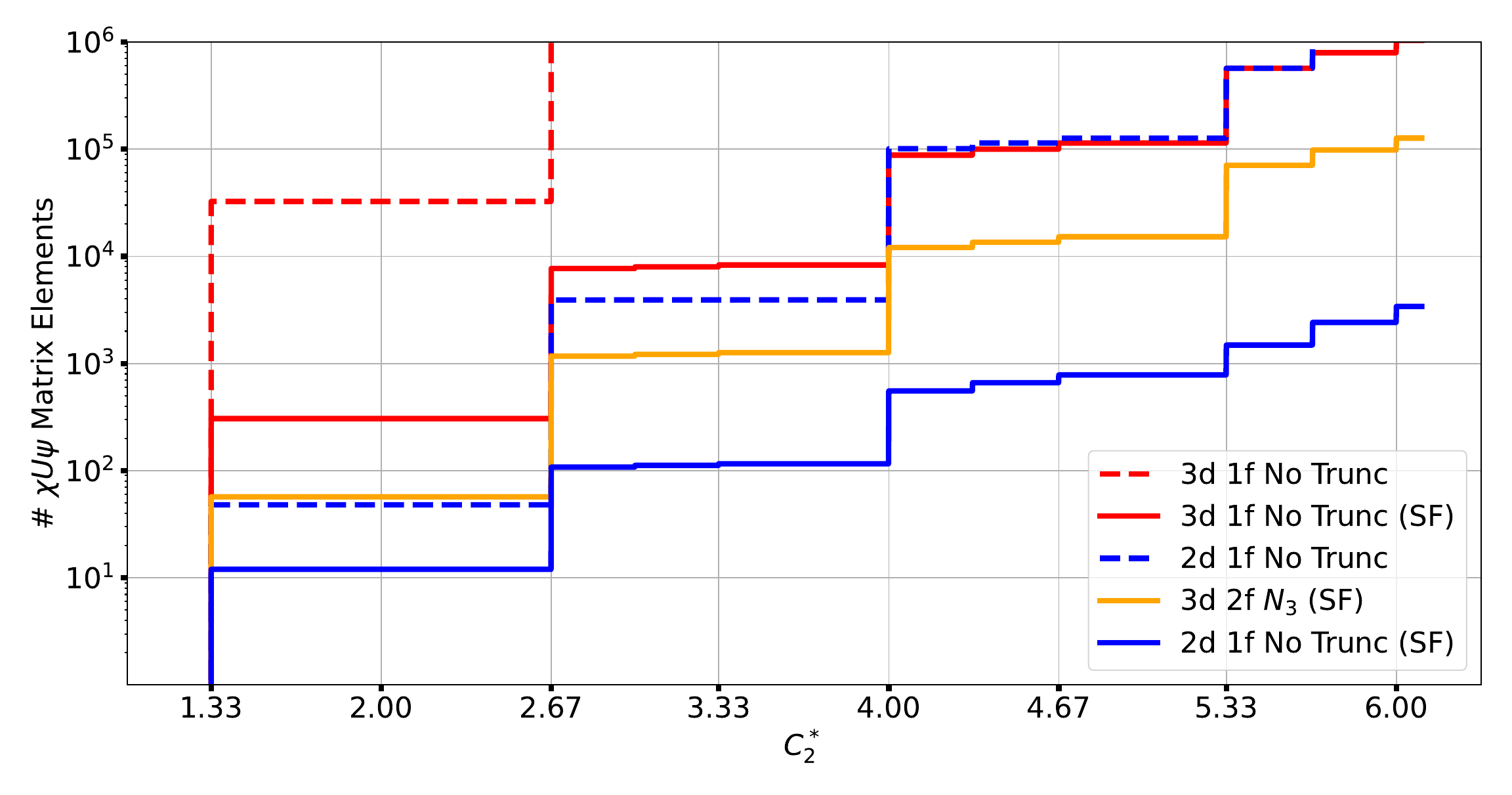}
		\caption{Wilson fermions}
	\end{subfigure}
	\caption{SU(3) kinetic term matrix element and site factor (SF) counts. ``\#f'' refers to the number of flavors and ``No Trunc'' means no truncation was performed on the fermion Hilbert space. In both plots, the gauge field is truncated at various $C_2^*$ truncation values. (a) The $M_2$ truncation allows each flavor to be at most in a double occupied state. (b) The $N_3$ truncation allows each flavor to have at most color occupation number three summed over all spinor components. (Counts for Wilson terms versus ordinary kinetic terms are comparable.)}
	\label{fig:link_matrix_elements}
\end{figure}

Three site factors remain. The contraction $\psi_c^\dag U_{cc'}$ will contain quantities defined on $s_2$. The $s_2$ site factor for operators with $\psi^\dag$ is
\begin{IEEEeqnarray*}{rCl}
	\begin{pmatrix}
		R & \Gamma_{s_2} & S \\ \mathcal{R} & \vec{C}_{s_2} & \psi^\dag \\ R' & \Gamma'_{s_2} & S'
	\end{pmatrix} & = & (-1)^{\zeta_{s_2}} \sum_{\sigma=1}^{\dim(\mathcal{R})} \sum_{r=1}^{\dim(R)} \sum_{r'=1}^{\dim(R')} \sum_{s=1}^{\dim(S)} \sum_{s'=1}^{\dim(S')} \sum_{\vec{c} \in \vec{C}_{s_2}} \varphi(r) \varphi(r') \\
	& \times & \braket{(R',r')}{(\mathcal{R},\sigma) \otimes (R,r)} \ \bra{(S',s')} \tilde{\psi}^\dag_\sigma \ket{(S,s)} \\
	& \times & \braket{(\bar{R}',\tilde{r}') \otimes (S',s') \otimes (\vec{C}_{s_2},\vec{c})}{\mathbf{1},\Gamma'_{s_2}} \ \braket{\mathbf{1},\Gamma_{s_2}}{(\bar{R},\tilde{r}) \otimes (S,s) \otimes (\vec{C}_{s_2},\vec{c})} \\ \yesnumber
\end{IEEEeqnarray*}
Note the $\varphi(r)$ phases that appear for the right half-link state of $\ell$ meeting at $s_2$. Moving on to undaggered operators, the contraction $U_{cc'} \psi_{c'}$ picks up $s_1$ quantities. The $s_1$ site factor for operators with $\psi$ is
\begin{IEEEeqnarray*}{rCl}
	\begin{pmatrix}
		R & \Gamma_{s_1} & S \\ \mathcal{R} & \vec{C}_{s_1} & \psi \\ R' & \Gamma'_{s_1} & S'
	\end{pmatrix} & = & (-1)^{\zeta_{s_1}} \sum_{\sigma=1}^{\dim(\mathcal{R})} \sum_{r=1}^{\dim(R)} \sum_{r'=1}^{\dim(R')} \sum_{s=1}^{\dim(S)} \sum_{s'=1}^{\dim(S')} \sum_{\vec{c} \in \vec{C}_{s_1}} \\
	& \times & \braket{(\mathcal{R},\sigma) \otimes (R,r)}{(R',r')} \ \bra{(S',s')} \tilde{\psi}_\sigma \ket{(S,s)} \\
	& \times & \braket{(R',r') \otimes (S',s') \otimes (\vec{C}_{s_1},\vec{c})}{\mathbf{1},\Gamma'_{s_1}} \ \braket{\mathbf{1},\Gamma_{s_1}}{(R,r) \otimes (S,s) \otimes (\vec{C}_{s_1},\vec{c})} \\ \yesnumber
\end{IEEEeqnarray*}
No $\varphi(r)$ phases appear because only left half-link and particle states are modified. Finally, the contraction $\chi_c U_{cc'}$ has quantities defined at $s_2$. The $s_2$ site factor for operators with $\chi$ is
\begin{IEEEeqnarray*}{rCl}
	\begin{pmatrix}
		R & \Gamma_{s_2} & S \\ \mathcal{R} & \vec{C}_{s_2} & \chi \\ R' & \Gamma'_{s_2} & S'
	\end{pmatrix} & = & (-1)^{\zeta_{s_2}} \sum_{\sigma=1}^{\dim(\mathcal{R})} \sum_{r=1}^{\dim(R)} \sum_{r'=1}^{\dim(R')} \sum_{s=1}^{\dim(S)} \sum_{s'=1}^{\dim(S')} \sum_{\vec{c} \in \vec{C}_{s_2}} \varphi(r) \varphi(r') \varphi(s) \varphi(s') \\
	& \times & \braket{(R',r')}{(\mathcal{R},\sigma) \otimes (R,r)} \ \bra{(S',s')} \tilde{\chi}_\sigma \ket{(S,s)} \\
	& \times & \braket{(\bar{R}',\tilde{r}') \otimes (S',s') \otimes (\vec{C}_{s_2},\vec{c})}{\mathbf{1},\Gamma'_{s_2}} \ \braket{\mathbf{1},\Gamma_{s_2}}{(\bar{R},\tilde{r}) \otimes (S,s) \otimes (\vec{C}_{s_2},\vec{c})} \\ \yesnumber
\end{IEEEeqnarray*}
Here, $\varphi(r)$ phases appear for right half-link and antiparticle states. The product of two site factors, together with a dimension coefficient from the link operator and a contribution of $Z$ operators from the Jordan-Wigner transformation, gives the matrix elements. The usual kinetic terms are
\begin{equation}
	\bra{\Lambda_n} \psi_c^\dag(s_2) U_{cc'} \chi_{c'}^\dag(s_1) \ket{\Lambda_m} = \left( \prod_\text{between} Z \right) \sqrt{\frac{\dim(R_\ell^m)}{\dim(R_\ell^n)}} \begin{pmatrix} R_\ell^m & \Gamma^m_{s_1} & R_\chi^m \\ \mathcal{R} & \vec{C}_{s_1} & \chi^\dag \\ R_\ell^n & \Gamma^n_{s_1} & R_\chi^n \end{pmatrix} \begin{pmatrix} R_\ell^m & \Gamma^m_{s_2} & R_\psi^m \\ \mathcal{R} & \vec{C}_{s_2} & \psi^\dag \\ R_\ell^n & \Gamma^n_{s_2} & R_\psi^n \end{pmatrix}
\end{equation}
and
\begin{equation}
	\bra{\Lambda_n} \chi_c(s_2) U_{cc'} \psi_{c'}(s_1) \ket{\Lambda_m} = \left( \prod_\text{between} Z \right) \sqrt{\frac{\dim(R_\ell^m)}{\dim(R_\ell^n)}} \begin{pmatrix} R_\ell^m & \Gamma^m_{s_1} & R_\psi^m \\ \mathcal{R} & \vec{C}_{s_1} & \psi \\ R_\ell^n & \Gamma^n_{s_1} & R_\psi^n \end{pmatrix} \begin{pmatrix} R_\ell^m & \Gamma^m_{s_2} & R_\chi^m \\ \mathcal{R} & \vec{C}_{s_2} & \chi \\ R_\ell^n & \Gamma^n_{s_2} & R_\chi^n \end{pmatrix}
\end{equation}
These two are necessary for both staggered and Wilson fermions. Wilson fermions also require the matrix elements
\begin{equation}
	\bra{\Lambda_n} \psi_c^\dag(s_2) U_{cc'} \psi_{c'}(s_1) \ket{\Lambda_m} = \left( \prod_\text{between} Z \right) \sqrt{\frac{\dim(R_\ell^m)}{\dim(R_\ell^n)}} \begin{pmatrix} R_\ell^m & \Gamma^m_{s_1} & R_\psi^m \\ \mathcal{R} & \vec{C}_{s_1} & \psi \\ R_\ell^n & \Gamma^n_{s_1} & R_\psi^n \end{pmatrix} \begin{pmatrix} R_\ell^m & \Gamma^m_{s_2} & R_\psi^m \\ \mathcal{R} & \vec{C}_{s_2} & \psi^\dag \\ R_\ell^n & \Gamma^n_{s_2} & R_\psi^n \end{pmatrix}
\end{equation}
and
\begin{equation}
	\bra{\Lambda_n} \chi_{c'}^\dag(s_1) U_{cc'} \chi_c(s_2) \ket{\Lambda_m} = \left( \prod_\text{between} Z \right) \sqrt{\frac{\dim(R_\ell^m)}{\dim(R_\ell^n)}} \begin{pmatrix} R_\ell^m & \Gamma^m_{s_1} & R_\chi^m \\ \mathcal{R} & \vec{C}_{s_1} & \chi^\dag \\ R_\ell^n & \Gamma^n_{s_1} & R_\chi^n \end{pmatrix} \begin{pmatrix} R_\ell^m & \Gamma^m_{s_2} & R_\chi^m \\ \mathcal{R} & \vec{C}_{s_2} & \chi \\ R_\ell^n & \Gamma^n_{s_2} & R_\chi^n \end{pmatrix}
\end{equation}
from the Wilson term. The number of matrix elements is shown in Fig.~\ref{fig:link_matrix_elements}.

\section{The Theta Angle}
\label{app:the_theta_angle}

The Lagrangian density ``$\theta$ term'' $\epsilon^{\mu\nu\rho\sigma} F_{\mu\nu}^a F_{\rho\sigma}^a$ is not invariant under parity (P) and charge-conjugation-parity (CP) transformations, unlike the kinetic term \cite{Weinberg:1996:qft_vol_2}. Also, because of the axial anomaly, in QCD the $\theta$ term is not field-redefinition invariant and combines with a phase from the fermion mass matrix to make a physical CP-violating parameter.

A general Lagrangian density for an SU($N_c$) gauge theory with Dirac fermions can be written with a complex fermion mass term:
\begin{equation}
	\mathcal{L} = -\frac{1}{4g^2} F_{\mu\nu}^a F^{\mu\nu,a} + \sum_{f=1}^{N_f} \bar{\psi}^{(f)} (i\gamma^\mu \partial_\mu + \gamma^\mu A_\mu^a T^a - m_f e^{i\varphi\gamma^5}) \psi^{(f)} + \frac{\vartheta}{64\pi^2} \epsilon^{\mu\nu\rho\sigma} F_{\mu\nu}^a F_{\rho\sigma}^a
\end{equation}
where
\begin{equation}
	e^{i\varphi\gamma^5} = e^{i\varphi}P_\mathsf{R} + e^{-i\varphi}P_\mathsf{L} \qquad P_\mathsf{R} = \frac{1+\gamma^5}{2} \qquad P_\mathsf{L} = \frac{1-\gamma^5}{2}
\end{equation}
for chiral projectors $P_\mathsf{R},P_\mathsf{L}$ and free parameters $\varphi,\vartheta$. Because $\psi \to \gamma^0 \psi$ under parity transformations, and $\{ \gamma^0, \gamma^5 \} = 0$, the complex fermion mass term is not invariant under P and CP transformations -- it gets conjugated. In the Standard Model, however, fermion masses originate from the electroweak sector, which explicitly violates CP through the Jarlskog invariant. Therefore, there are no immediate theoretical reasons to constrain $\varphi=\vartheta=0$ \cite{BonannoBonatiDElia:2026:strong_cp_problem}.

\clearpage

Only a linear combination of $\vartheta$ and $\varphi$ is invariant under axial field redefinitions. A chiral transformation of the fermionic path integral measure has the effect
\begin{equation}
	\mathcal{D} \bar{\psi} \ \mathcal{D} \psi \to \mathcal{D} \bar{\psi} \ \mathcal{D} \psi \ e^{i\frac{2\phi N_f}{32\pi^2} \int d^4x \ \epsilon^{\mu\nu\rho\sigma} \trace[ F_{\mu\nu} F_{\rho\sigma}]}
\end{equation}
as illustrated by the Fujikawa method \cite{Fujikawa:1979:fujikawa_method}. The Lagrangian density then transforms as
\begin{IEEEeqnarray*}{rCl}
	\mathcal{L} \to & - & \frac{1}{4g^2} F_{\mu\nu}^a F^{\mu\nu,a} + \frac{1}{32\pi^2}\left( \frac{\vartheta}{2} + 2\phi I_\mathcal{R} N_f \right) \epsilon^{\mu\nu\rho\sigma} F_{\mu\nu}^a F_{\rho\sigma}^a \\
	& + & \sum_{f=1}^{N_f} \bar{\psi}^{(f)} (i\gamma^\mu \partial_\mu + \gamma^\mu A_\mu^a T^a - m_f e^{i(\varphi+2\phi)} P_\mathsf{R} - m_f e^{-i(\varphi+2\phi)} P_\mathsf{L}) \psi^{(f)} \yesnumber
\end{IEEEeqnarray*}
Now it can be seen that eliminating the CP violating pure-gauge term does not eliminate the CP violating fermion mass term and vice versa:
\begin{IEEEeqnarray*}{rCl}
	\phi = -\frac{\vartheta}{4I_\mathcal{R}N_f} \implies \mathcal{L} \to & - & \frac{1}{4g^2} F_{\mu\nu}^a F^{\mu\nu,a} \\
	& + & \sum_{f=1}^{N_f} \bar{\psi}^{(f)} (i\gamma^\mu \partial_\mu + \gamma^\mu A_\mu^a T^a - m_f e^{-i\tilde{\theta}} P_\mathsf{R} - m_f e^{i\tilde{\theta}} P_\mathsf{L}) \psi^{(f)} \\
	\phi = -\frac{\arg(\det[M])}{2N_f} \implies \mathcal{L} \to & - & \frac{1}{4g^2} F_{\mu\nu}^a F^{\mu\nu,a} + \frac{\theta}{64\pi^2} \epsilon^{\mu\nu\rho\sigma} F_{\mu\nu}^a F_{\rho\sigma}^a \\
	& + & \sum_{f=1}^{N_f} \bar{\psi}^{(f)} (i\gamma^\mu \partial_\mu + \gamma^\mu A_\mu^a T^a - m_f) \psi^{(f)} \yesnumber
\end{IEEEeqnarray*}
where
\begin{equation}
	M = \text{diag}(m_1 e^{i\varphi}, \dots, m_{N_f} e^{i\varphi}) \qquad \arg(\det[M]) = \varphi N_f
\end{equation}
and
\begin{equation}
	\theta = \vartheta - 2I_\mathcal{R} \arg(\det[M]) \qquad \tilde{\theta} = \frac{\theta}{2I_\mathcal{R}N_f}
\end{equation}
In either attempt, the theta angle $\theta$ parameterizes CP violation.

Hence, there are two special ways to include the theta angle. One way is to have it as a coefficient to a pure-gauge term, keeping the fermion masses real. This is the presentation used throughout the paper. The other way is to have it as part of the complex fermion masses. This latter presentation will now be used for the remainder of the appendix. With this choice, the Hamiltonian can be found to be
\begin{IEEEeqnarray*}{rCl}
	H = \int d^3x \bigg[ & - & \sum_{f=1}^{N_f} \bar{\psi}^{(f)} (i\gamma^i \partial_i + \gamma^i A_i^a T^a - m_f e^{-i\tilde{\theta}} P_\mathsf{R} - m_f e^{i\tilde{\theta}} P_\mathsf{L}) \psi^{(f)} \\
	& + & \frac{g^2}{2} \Pi^{i,a} \Pi^{i,a} + \frac{1}{4g^2} F_{ij}^a F^{ij,a} - A_0^a \bigg( D_i \Pi^{i,a} + \sum_{f=1}^{N_f} \psi^{\dag(f)} T^a \psi^{(f)} \bigg) \bigg] \yesnumber
\end{IEEEeqnarray*}
following the canonical quantization procedure in App.~\ref{app:continuum_hamiltonian}. Note that the same Gauss law constrains the physical states of the Hilbert space. For the discussion of the lattice theory that follows, it will be convenient to have the identity
\begin{equation}
	\bar{\psi} (e^{-i\tilde{\theta}} P_\mathsf{R} + e^{i\tilde{\theta}} P_\mathsf{L}) \psi = \psi^\dag ( \cos(\tilde{\theta}) P_+ - \cos(\tilde{\theta}) P_- - i\sin(\tilde{\theta}) P_+ \gamma^5 P_- + i\sin(\tilde{\theta}) P_- \gamma^5 P_+) \psi
\end{equation}
where $P_\pm$ are the projectors $\frac{1\pm\gamma^0}{2}$.

\subsection{Lattice Theory}

The theory can be discretized on a cubic spatial lattice as is done in the main text. The lattice Hamiltonian will share the same $H_E$, $H_B$, $H_\text{kin}$ terms. The difference lies in the fermion mass term.

The complex mass term will be split into a diagonal term $H_\text{mass}$ and a kinetic-term-like operator $H_\theta$. For Wilson fermions, these terms are
\begin{IEEEeqnarray*}{rCl}
	H^\text{W}_\text{mass} & = & (m\cos(\tilde{\theta}) + \tfrac{3r}{a}) \sum_{\vec{s}} \psi^\dag(\vec{s}) \psi(\vec{s}) + \chi^\dag(\vec{s}) \chi(\vec{s}) \\
	H^\text{W}_\theta & = & -im \sin(\tilde{\theta}) \sum_{\vec{s}} \psi^\dag(\vec{s}) \gamma^5 \chi^\dag(\vec{s}) - \chi(\vec{s}) \gamma^5 \psi(\vec{s}) \yesnumber
\end{IEEEeqnarray*}
Note the bilinears in the Dirac basis of gamma matrices:
\begin{equation}
	\chi \gamma^5 \psi = \chi_3 \psi_1 + \chi_4 \psi_2 \qquad \psi^\dag \gamma^5 \chi^\dag = \psi_1^\dag \chi_3^\dag + \psi_2^\dag \chi_4^\dag
\end{equation}
For staggered fermions, $H^\text{stag}_\theta$ will contain an interaction between fermions on two sites. Therefore a link operator must be inserted to maintain gauge invariance:
\begin{IEEEeqnarray*}{rCl}
	H^\text{stag}_\text{mass} & = & m \cos(\tilde{\theta}) \sum_{\vec{s} \text{even}} \psi^\dag(\vec{s}) \psi(\vec{s}) + m \cos(\tilde{\theta}) \sum_{\vec{s} \text{odd}} \chi^\dag(\vec{s}) \chi(\vec{s}) \\
	H^\text{stag}_\theta & = & -\frac{im}{2} \sin(\tilde{\theta}) \sum_{\vec{s} \text{even}} \gamma^5(\vec{s}) \left[ \psi^\dag(\vec{s}) U^\dag(\vec{s},\vec{e}_3) \chi^\dag(\vec{s}+a\vec{e}_3) + \psi^\dag(\vec{s}) U(\vec{s}-a\vec{e}_3, \vec{e}_3) \chi^\dag(\vec{s}-a\vec{e}_3) \right] \\
	& + & \frac{im}{2} \sin(\tilde{\theta}) \sum_{\vec{s} \text{odd}} \gamma^5(\vec{s}) \left[ \chi(\vec{s}) U^\dag(\vec{s},\vec{e}_3) \psi(\vec{s}+a\vec{e}_3) + \chi(\vec{s}) U(\vec{s}-a\vec{e}_3, \vec{e}_3) \psi(\vec{s}-a\vec{e}_3) \right] \\ \yesnumber
\end{IEEEeqnarray*}
Here, an average was used to maintain an $\mathcal{O}(a^2)$ discretization error in the staggered theory. Moreover, in the Dirac basis, $\gamma^5(\vec{s}) = 1$, because the nonzero components of $\gamma^5$ are equal to one.

\begin{figure}[!ht]
	\centering
	\includegraphics[width=0.375\textwidth]{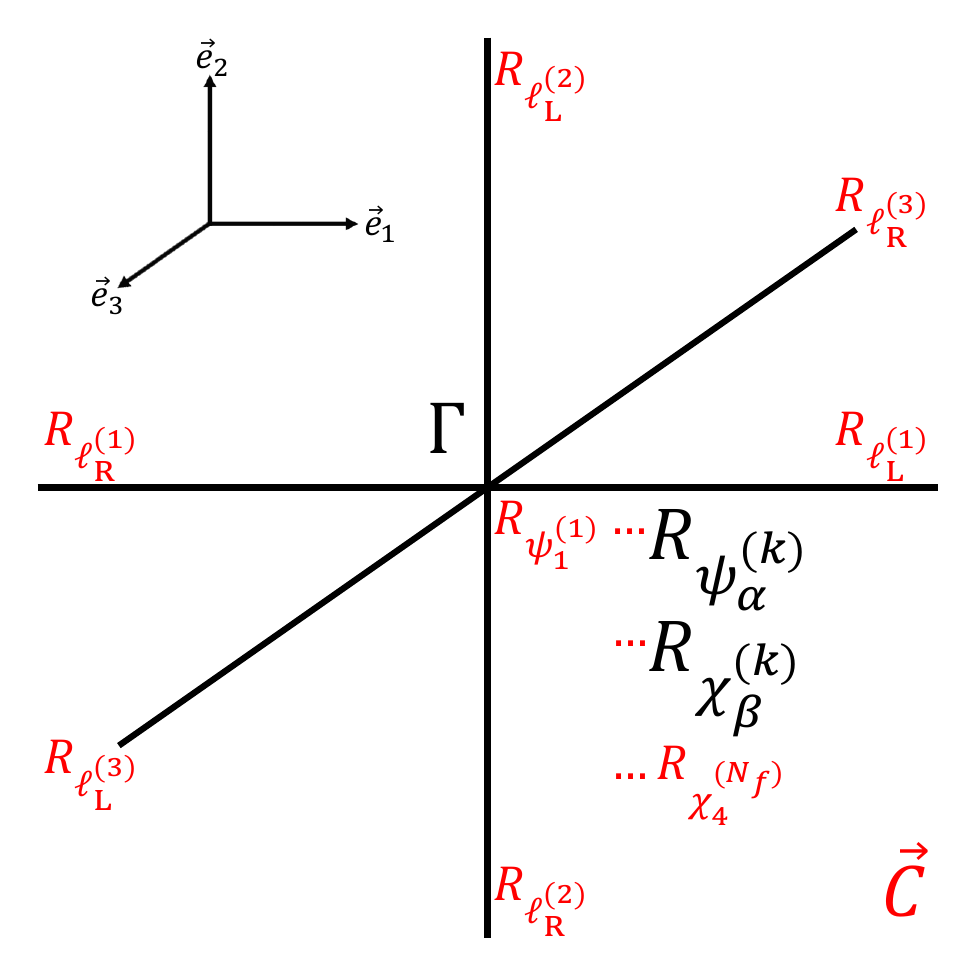}
	\caption{Physical site states. These are specified by assignments of irreps to the degrees of freedom meeting at a lattice site, as well as a singlet multiplicity index. $H^\text{W}_\theta$ generates transitions between physical site states. In particular, it changes the two fermion irreps and the multiplicity index $\Gamma$, all written in black font. All other irreps, written in colored font, remain unchanged upon action of the Hamiltonian. These irreps can be grouped into the vector $\protect\vec{C}_k$ comprised of ``control links'' and other fermions.}
	\label{fig:site_state}
\end{figure}

\begin{figure}
	\centering
	\includegraphics[width=\textwidth]{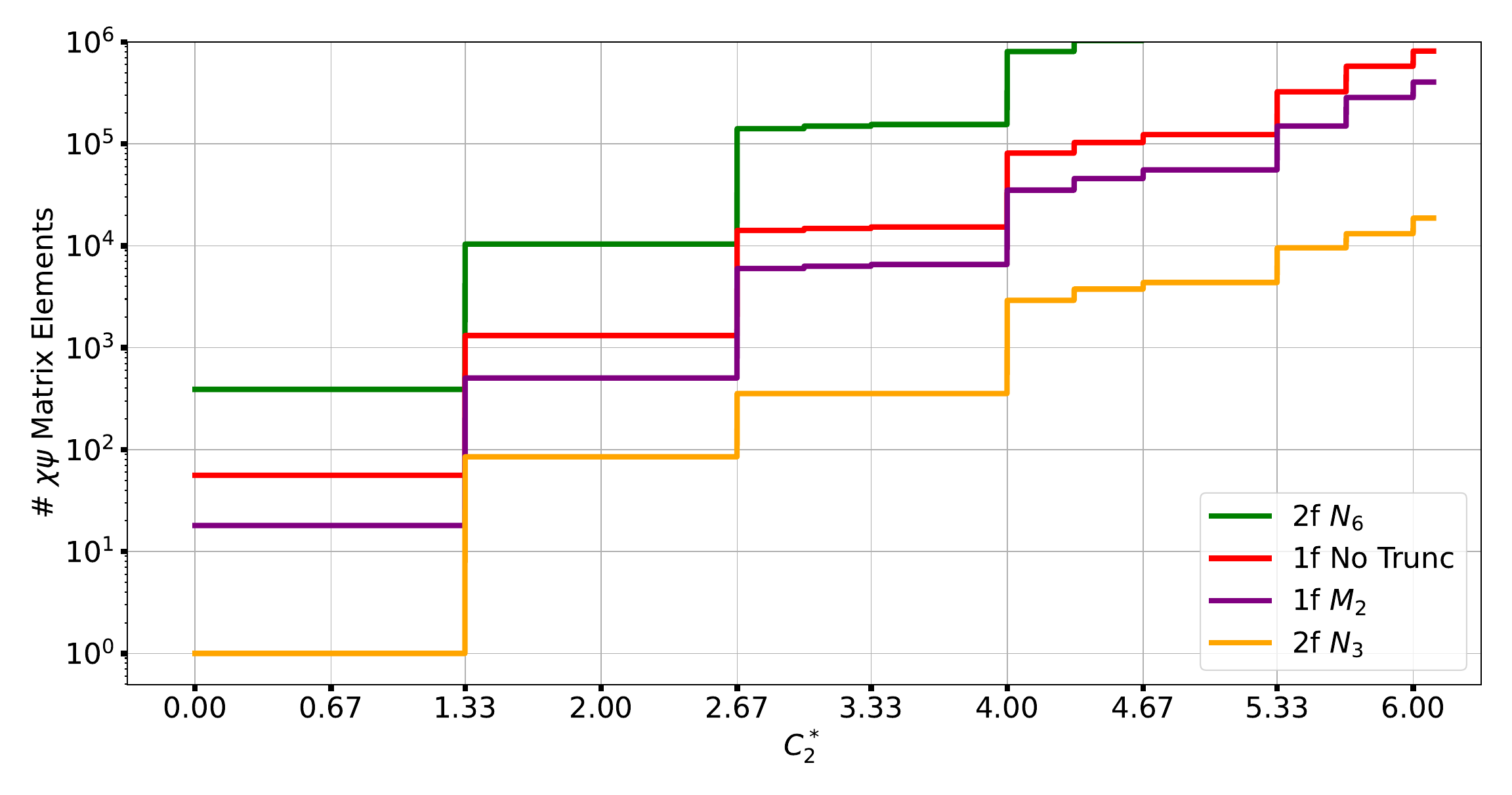}
	\caption{SU(3) site factor counts for theta-induced singlet transitions. ``\#f'' refers to the number of flavors and ``No Trunc'' means no truncation was performed on the fermion Hilbert space. The gauge field is truncated at various $C_2^*$ truncation values. The $M_2$ truncation means each spinor component can be at most in a doubly occupied state. The $N_3$ ($N_6$) truncation means each flavor can have at most color occupation number three (six) summed over all spinor components.}
	\label{fig:site_matrix_elements}
\end{figure}

Thus, $H^\text{stag}_\theta$ is a shift to $H^\text{stag}_\text{kin}$ along the $\vec{e}_3$ axis. On the other hand, with Wilson fermions, $H^\text{W}_\theta$ features an on-site singlet transition; it requires a matrix element calculation unlike one found in the main text. The relevant site factor is
\begin{IEEEeqnarray*}{rCl}
	\left\langle \begin{matrix}
		R & \Gamma & S \\ \chi & \vec{C} & \psi \\ R' & \Gamma' & S'
	\end{matrix} \right\rangle & = & (-1)^\zeta \sum_{\sigma=1}^{\dim(\mathcal{R})} \sum_{r=1}^{\dim(R)} \sum_{r'=1}^{\dim(R')} \sum_{s=1}^{\dim(S)} \sum_{s'=1}^{\dim(S')} \sum_{\vec{c} \in \vec{C}} \varphi(r) \varphi(r') \\
	& \times & \bra{(R',r')} \tilde{\chi}_\sigma \ket{(R,r)} \ \bra{(S',s')} \tilde{\psi}_\sigma \ket{(S,s)} \\
	& \times & \braket{(R',r') \otimes (S',s') \otimes (\vec{C},\vec{c})}{\mathbf{1},\Gamma'} \ \braket{\mathbf{1},\Gamma}{(R,r) \otimes (S,s) \otimes (\vec{C},\vec{c})} \\ \yesnumber
\end{IEEEeqnarray*}
where Dirac indices are suppressed, and $\tilde{\chi}$ and $\tilde{\psi}$ indicate to apply $Z$ operator consistent with the Jordan-Wigner transformation. (See Fig.~\ref{fig:site_state} for an illustration of the degrees of freedom modified by this transition.) $(-1)^\zeta$ is the contribution from $Z$ operators that measure color occupation number parity of the particles and antiparticles of different spinor and flavor indices between those being acted on by $\tilde{\chi}$ and $\tilde{\psi}$. The $\varphi(r)$ phases appear for the antiparticle state by definition of the physical basis states $\ket{\Lambda}$. The complete matrix element is
\begin{equation}
	\bra{\Lambda_n} \chi_c(\vec{s}) \psi_c(\vec{s}) \ket{\Lambda_m} = \left\langle \begin{matrix}
		R^m_\chi & \Gamma^m & R^m_\psi \\ \chi & \vec{C} & \psi \\ R^n_\chi & \Gamma^n & R^n_\psi
	\end{matrix} \right\rangle
\end{equation}
The number of matrix elements is shown in Fig.~\ref{fig:site_matrix_elements}.

\subsection{Quantum Simulation}

Gauss's law and the formulation of the physical Hilbert space are unchanged from what is described in the main text. For quantum simulation, the same local encoding and Hilbert space truncations can be used. The Trotterization formula is slightly modified because of the different Hamiltonian. For staggered fermions,
\begin{equation}
	U(t) = \left( \prod_{k} e^{-iH^\text{stag}_{\text{mass},k} \Delta t} \prod_{k} e^{-iH_{E,k} \Delta t} \prod_{k} e^{-iH_{B,k} \Delta t} \prod_{k} e^{-iH^\text{stag}_{\text{kin}+\theta,k} \Delta t} \right)^{n_t}
\end{equation}
where $H^\text{stag}_{\text{kin}+\theta,k}$ reflects the fact that the kinetic term is merely shifted by the theta angle. For Wilson fermions,
\begin{equation}
	U(t) = \left( \prod_{k} e^{-iH^\text{W}_{\text{mass},k} \Delta t} \prod_{k} e^{-iH_{E,k} \Delta t} \prod_{k} e^{-iH_{B,k} \Delta t} \prod_{k} e^{-iH^\text{W}_{\theta,k} \Delta t} \prod_{k} e^{-iH^\text{W}_{\text{kin},k} \Delta t} \right)^{n_t}
\end{equation}
where $H^\text{W}_{\theta,k}$ reflects the fact that the theta angle gives rise to a distinct on-site operator.

\begin{figure}[!h]
	\centering
	\includegraphics[width=\textwidth]{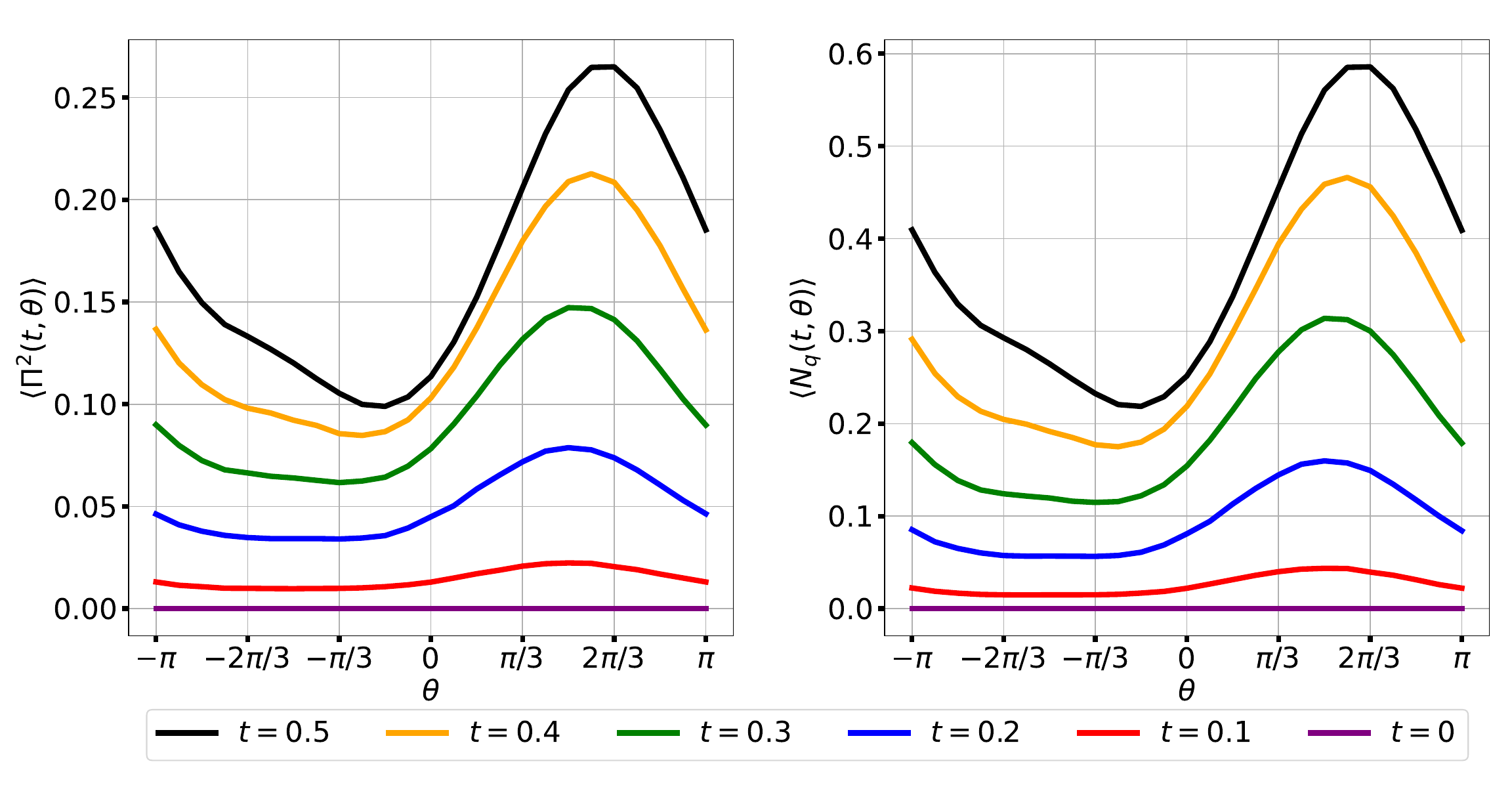}
	\caption{Measurements of $\expect{\Pi^2(t)}$ and $\expect{N_q(t)}$ on the cube with varying theta angle. The gauge truncation used here is $C_2^*=2.67$ and the fermion truncation is $M_1$. For these simulations, $10^5$ shots were run on CUDA-Q's statevector simulator with 32 qubits and up to $\mathcal{O}(10^4)$ untranspiled gates. The simulation parameters used were $\Delta t = 0.1$, $N_f=1$, and $m_f=1$. Each curve is labeled by a different time $t$.}
	\label{fig:theta_fermion}
\end{figure}

The complex fermion mass changes the dynamics of the theory \cite{Mueller:2023:phase_transitions_tomography}. To showcase the effects of the complex fermion mass, the simulations performed in the main text with the theta angle are repeated here. Fig.~\ref{fig:theta_fermion} plots the expectations values $\expect{\Pi^2(t)}$ and $\expect{N_q(t)}$ defined in Sec.~\ref{sec:quantum_simulation} for a staggered fermion on the cube. These results are dramatically different from Fig.~\ref{fig:theta_gauge}, which used the same truncations. The Hamiltonian used in this section is manifestly periodic in $\theta \to \theta + 2\pi N_f$, unlike the Hamiltonian used in the main text. (In the continuum theory both expressions are periodic as $\theta \to \theta + 2\pi$ due to topological charge quantization, which does not hold on the lattice.) The observables respect this periodicity, as can be seen by comparing the data at $\theta = \pm\pi$. However, similar to Fig.~\ref{fig:theta_gauge}, these results are not symmetric in $\theta \to -\theta$. An interesting asymmetry grows over time, where for $\theta > 0$, the lattice is more quickly excited with matter than for $\theta < 0$. Because $H^\text{stag}_\text{mass}$ is symmetric in $\theta \to -\theta$, this asymmetry is accounted for by the kinetic term shift $H^\text{stag}_\theta$; the actual symmetry must compose $\theta \to -\theta$ with time reversal.

Fig.~\ref{fig:theta_gauge} and Fig.~\ref{fig:theta_fermion} differ greatly due to unsuppressed lattice artifacts at strong coupling ($g=2$ in the plots) and the aggressive Hilbert space truncation. For example, at weak coupling (small lattice spacing)  the quantization of topological charge is better respected by the lattice theory because unit-cell-sized topological objects that violate charge quantization are heavily suppressed. These contributions become unsuppressed at large coupling, which is roughly what is driving the form of Fig.~\ref{fig:theta_gauge}.

\bibliographystyle{compact}
\bibliography{references}
\addcontentsline{toc}{section}{References}

\end{document}